\documentclass{aa}  

\usepackage{graphicx}
\usepackage{multicol}
\usepackage{txfonts}
\setcitestyle{numbers}
\begin{document}

   \title{Electromagnetic cascade masquerade:\\ a way to mimic $\gamma$-axion-like particle mixing effects in blazar spectra}
   \titlerunning{Electromagnetic cascade masquerade}
   \author{T.A. Dzhatdoev\inst{1,*}, E.V. Khalikov\inst{1,**}, A.P. Kircheva\inst{1,2} \and A.A. Lyukshin\inst{2}}
   \institute{Federal State Budget Educational Institution of Higher Education M.V. Lomonosov Moscow State University, Skobeltsyn Institute of Nuclear Physics (SINP MSU), 1(2), Leninskie gory, GSP-1, Moscow 119991, Russian Federation \\
   $^{*}$\email{timur1606@gmail.com}, $^{**}$\email{nanti93@mail.ru}
    \and
Federal State Budget Educational Institution of Higher Education M.V. Lomonosov Moscow State University, Department of Physics, 1(2), Leninskie gory, GSP-1, Moscow 119991, Russian Federation}

  \abstract
   {Most of the studies on extragalactic $\gamma$-ray propagation performed up to now only accounted for primary gamma-ray absorption and adiabatic losses (``absorption-only model''). However, there is growing evidence that this model is oversimplified and must be modified in some way. In particular, it was found that the intensity extrapolated from the optically-thin energy range of some blazar spectra is insufficient to explain the optically-thick part of these spectra. This effect was interpreted as an indication for $\gamma$-axion-like particle (ALP) oscillation. On the other hand, there are many hints that a secondary component from electromagnetic cascades initiated by primary $\gamma$-rays or nuclei may be observed in the spectra of some blazars.}
   {We study the impact of electromagnetic cascades from primary $\gamma$-rays or protons on the physical interpretation of blazar spectra obtained with imaging Cherenkov telescopes.}
   {We use the publicly-available code ELMAG to compute observable spectra of electromagnetic cascades from primary $\gamma$-rays. For the case of primary proton, we develop a simple, fast and reasonably accurate hybrid method to calculate the observable spectrum. We perform the fitting of the observed spectral energy distributions (SEDs) with various physical models: the absorption-only model, the ``electromagnetic cascade model'' (for the case of primary $\gamma$-rays), and several versions of the hadronic cascade model (for the case of primary proton). We distinguish the following species of hadronic cascade models: 1) ``basic hadronic model'', where it is assumed that the proton beam travels undisturbed by extragalactic magnetic field and that all observable $\gamma$-rays are produced by primary protons through photohadronic processes with subsequent development of electromagnetic cascades 2) ``intermediate hadronic model'', the same as the basic hadronic model, but the primary beam is terminated at some redshift $z_{c}$ 3) ``modified hadronic model'' that includes the contribution from primary (redshifted and partially absorbed) $\gamma$-rays.}
   {Electromagnetic cascades show at least two very distinct regimes labeled by the energy of the primary $\gamma$-ray ($E_{0}$): the one-generation regime for the case of $E_{0}$<10 $TeV$ and the universal regime for $E_{0}$>100 $TeV$ and redshift to the source $z_{s}$>0.02. Spectral signatures of the observable spectrum for the case of the basic hadronic model, $z_{s}$= 0.186 and low energy ($E$<200 $GeV$) are nearly the same as for purely electromagnetic cascade, but for $E$>200 $GeV$ the spectrum is much harder for the case of the basic hadronic model. In the framework of the intermediate hadronic model, the observable spectrum depends only slightly on the primary proton energy, but it strongly depends on $z_{c}$ at $E$>500 $GeV$. As a rule, both electromagnetic and hadronic cascade models provide acceptable fits to the observed SEDs. We show that the best-fit model intensity in the multi-$TeV$ region of the spectrum in the framework of the electromagnetic cascade model is typically greater than the one for the case of the absorption-only model. Finally, for the case of blazar 1ES 0229+200 we provide strong constraints on the intermediate hadronic model assuming the blazar emission model of Tavecchio (2013) and the model of magnetic field around the source according to Meyer et al. (2013).}
   {}
   \keywords{astroparticle physics -- radiation mechanisms: non-thermal -- methods: numerical -- BL Lacertae objects: individual -- cosmic background radiation}

   \maketitle

\section{Introduction}

The development of ground-based $\gamma$-ray astronomy with imaging Cherenkov telescopes has been very fast during the last two decades (for a review see, e.g., Hillas \cite{hillas}). Indeed, only 7 years after the detection of $TeV$ photons from the active galactic nucleus (AGN) Mkn 421 in Punch et al. \cite{punch}, the first detailed study of a blazar (AGN with the jet presumably pointed towards the observer) spectrum was made by Aharonian et al. \cite{aharonian99}. Almost immediately, these observations were utilized to put some constraints on the intensity of extragalactic background light (EBL) (in Stecker \& de Jager \cite{stecker}, de Jager et al. \cite{dejager} for the former observations, and in Aharonian et al. \cite{aharonian99} itself for the latter).

Indeed, primary very high energy (VHE, $E$>100 $GeV$) $\gamma$-rays are absorbed on EBL photons by means of the $\gamma \gamma \rightarrow e^{+}e^{-}$ process (Nikishov \cite{nikishov}, Gould \& Shroeder \cite{gould}), and for the case of primary energy $E\sim100$ $TeV$ and higher --- on the cosmic microwave background (CMB) as well (Jelley \cite{jelley}). More recently, the signatures of this absorption process were observed with the Fermi LAT instrument (Ackermann et al. \cite{ackermann}) and the H.E.S.S. Cherenkov telescope (Abramowski et al. (The H.E.S.S. Collaboration) \cite{abramowski}) with high statistical significance ($\sim6\sigma$ and 8.8 $\sigma$, respectively).

However, Horns \& Meyer \cite{horns12} found that the strength of absorption at high optical depth ($\tau_{\gamma\gamma}$>2, hereafter simply $\tau$) appears to be lower than expected. This result was obtained on a sample of blazar spectra measured with imaging Cherenkov telescopes by comparing the distributions of the flux points for the $\tau$ regions 1<$\tau$<2 and $\tau$>2 around the intensity extrapolated from the optically-thin regime $\tau$<1. The statistical significance of this effect, according to Horns \& Meyer \cite{horns12}, is 4.2 $\sigma$. While this result was not confirmed by Biteau \& Williams \cite{biteau}, very recently Horns \cite{horns16} again found an indication for this anomaly with another analysis method. Such an anomaly, in fact, closely resembles the so-called ``$TeV$-IR crisis'' (Protheroe \& Meyer \cite{protheroe}) that was derived from the already mentioned observations of Mkn 501 (Aharonian et al. \cite{aharonian99}). The ``$TeV$-IR crisis'', however, was later found to be less severe after the development of more advanced EBL models. On the contrary, the ``new'' anomaly of Horns \& Meyer \cite{horns12}, Horns \cite{horns16} persists for most of these contemporary models of EBL intensity.

The authors of Horns \& Meyer \cite{horns12} interpreted their result as an indication for the existence of some non-conventional physical effect, for instance, the process of oscillations of $\gamma$-rays into axion-like particles (ALPs) and back into photons in magnetic field on the way from the source to the observer ($\gamma \rightarrow ALP$). Indeed, a part of photons reconverted from ALPs near the observer can significantly enhance the observed intensity in the $\tau$>2 region. Moreover, once the anomaly is well established, it is possible to put constraints on the gamma-ALP mixing parameters $m_{a}$ (the mass of ALP) and $g_{a\gamma}$ (the photon-ALP coupling constant). In Meyer et al. \cite{meyer} a lower limit on the $g_{a\gamma}$ was found, depending on the $m_{a}$ value. For any fixed $m_{a}$ in the range considered in Meyer et al. \cite{meyer}, some values of $g_{a\gamma}$ greater than the lower limit $g_{a\gamma-min}(m_{a})$ could explain the observed anomaly. Together with the upper limits from the CAST experiment (Andriamonje et al. (CAST Collaboration) \cite{andriamonje}), this result allowed to construct a confidence interval for $g_{a\gamma}$.

Besides the anomaly at $\tau$>2, there exists another signature of $\gamma \rightarrow ALP$ oscillation, namely, a step-like irregularity that is situated at the energy lower than the starting energy of the VHE anomaly (e.g, Sanchez-Conde et al. \cite{sanchez-conde}). The drop of intensity associated with this spectral feature is usually about 1/3 as photons (two polarization states) attain equipartition with ALPs (one polarization state). A very recent analysis of Fermi LAT data (Ajello et al. \cite{ajello}) (observations of the NGC 1275 Seyfert galaxy were used), however, did not find this signature. Moreover, other bounds (Ayala et al. \cite{ayala}, Abramowski et al. (H.E.S.S. Collaboration) \cite{abramowski13}, Payez et al. \cite{payez}, Wouters \& Brun \cite{wouters}) on the parameters of $\gamma$-ALP mixing allowed to strongly constrain the scenario considered in Meyer et al. \cite{meyer}. Therefore, the hypothesis that the VHE anomaly is explained by $\gamma \rightarrow ALP$ oscillation appears to be less attractive than before; one needs to search for some other physical mechanism of the anomaly.

In this respect, we note that most of extragalactic $\gamma$-ray propagation studies were performed with account of only two elementary processes: the absorption of primary photons (by means of pair-production) and their adiabatic losses. This model (hereafter ``the absorption-only model'') rests on the assumption that the secondary electrons and positrons (hereafter simply ``electrons'' unless otherwise stated) are deflected and delayed by the extragalactic magnetic field (EGMF), thus the cascade photons produced by these electrons by means of the inverse Compton (IC) scattering do not contribute to the observed spectrum.

However, the EGMF strength $B$ in voids of the Large Scale Structure (LSS) may be small enough to violate this assumption. The existing constraints (Blasi et al. \cite{blasi}, Pshirkov et al. \cite{pshirkov}, Dolag et al. \cite{dolag}, Neronov \& Vovk \cite{neronov}, Dermer et al. \cite{dermer}, Taylor et al. \cite{taylor}, Vovk et al. \cite{vovk}, Abramowski et al. (H.E.S.S. Collaboration) \cite{abramowski14}, Takahashi et al. \cite{takahashi12}, Takahashi et al. \cite{takahashi13}) on the EGMF strength on characteristic coherence scale 1 $Mpc$ (Akahori \& Ryu \cite{akahori}) (some of them were summarized, e.g., in Fig. 4 of Dzhatdoev \cite{dzhatdoev15a}) do not exclude the probability that cascade emission may contribute to the observed spectrum.

The recent work Finke et al. \cite{finke}, using a more conservative method than the above-mentioned references, excluded $B<10^{-19}$ $G$, again on characteristic coherence scale 1 $Mpc$, with statistical significance $5\sigma$. Arlen et al. \cite{arlen} did not find any reason to reject the null hypothesis of $B=0$ at all. Tashiro \& Vachaspati \cite{tashiro} found $B\sim10^{-14}$ $G$. This study was made using the angular correlation pattern of Fermi LAT (Atwood et al. \cite{atwood}) diffuse $\gamma$-rays; however, it is not clear how much this result may be affected by the existence of comparatively strong magnetic field in galaxy clusters.

Moreover, there are some hints that the cascade component does contribute to the observed spectrum of blazars at energies $E<300$ $GeV$. Neronov et al. \cite{neronov12}, using observations of Mkn 501 in a flaring state by the Fermi LAT instrument and the VERITAS Cherenkov telescope (Abdo et al. \cite{abdo}) during the 2009 multiwavelength campaign, obtained the energy spectrum $dN/dE$ ($dN$ is the number of photons per energy bin $dE$) from 300 $MeV$ up to 5 $TeV$. It was found that the Fermi LAT spectrum in the 10--200 $GeV$ energy range had a power-law ($dN/dE=C\cdot E^{-\gamma}$) index $\gamma= 1.1\pm0.2$, while in the 300 $GeV$ -- 5 $TeV$ $\gamma \approx 2$. It is interesting that the Fermi LAT lightcurves in the 0.3-3 $GeV$ and 3-30 $GeV$ energy bins do not show any evidence for strong, fast variability, while in the 30-300 $GeV$ energy bin the flare is readily identified on the lightcurve. Neronov et al. \cite{neronov12} found that such a behaviour of the spectral and timing characteristics is typical for intergalactic cascade, assuming $B\sim10^{-16}-10^{-17}$ $G$ on the maximum spatial scale 1 $Mpc$.

Another result supporting the incompleteness of the absorption-only model was also obtained with the Fermi LAT telescope. Namely, Furniss et al. \cite{furniss} found a correlation between 10-500 $GeV$ energy flux of blazars with relatively hard ($\gamma<3$) observed spectra above 10 $GeV$ and the fraction along the line of sight occupied by voids in the LSS. This effect may be explained by the same physical mechanism: the cascade emission that is likely to be angle-broadened by the EGMF at some energy below 200-300 $GeV$. Finally, Chen et al. \cite{chen}, again with Fermi LAT data, found the evidence for the existence of ``pair halos'' (extended emission) around the positions of various blazars, thus giving additional support to the hypothesis that the cascade component may contribute to the observed spectra.

These works motivated us to study how the inclusion of the cascade component into the fitting of the VHE spectra would influence the data interpretation, especially in the optically thick ($\tau$>1) region. Our paper is by far not the first study of intergalactic cascade spectra; besides the already mentioned work (Neronov et al. \cite{neronov12}), there are many others that included more or less detailed duscussions of these, namely: Aharonian et al. \cite{aharonian99}, Aharonian et al. \cite{aharonian02}, d'Avesac et al. \cite{davezac}, Murase et al. \cite{murase}, Takami et al. \cite{takami}.

However, very recently in a conference paper Dzhatdoev \cite{dzhatdoev15b} it was shown that if the primary spectrum is hard enough ($\gamma \approx 1$) and the cutoff in the spectral energy distribution (SED) of the source is situated at a sufficiently high energy ($E_{c}>>1$ $TeV$), then the intersection of a low-energy cascade component and a high-energy primary (absorbed) component forms a kind of an ``ankle'' that to some extent may mimic the $\gamma-ALP$ mixing effect mentioned above. One of the main aims of the present paper is to discuss this effect in more depth. We will see that the last model is qualitatively different from other ``electromagnetic cascade models'', i.e. the models that involve intergalactic cascades initiated by primary gamma-rays.

There exists another class of intergalactic cascade models of blazar spectra dealing with primary protons or nuclei that initiate secondary photons by means of photopion losses and pair production with subsequent development of electromagnetic cascades (``hadronic cascade models'') (e.g. Uryson \cite{uryson}, Essey \& Kusenko \cite{essey10a}, Essey et al. \cite{essey10b}, Essey et al. \cite{essey11}, Murase et al. \cite{murase}, Takami et al. \cite{takami}, Essey \& Kusenko \cite{essey14}, Zheng et al. \cite{zheng}). We will compare hadronic models with the electromagnetic cascade model of Dzhatdoev \cite{dzhatdoev15b} to reveal their advantages and diffuculties.

The paper is organized as follows. In Sec.~\ref{sec:casc} we discuss some general properties of cascades from primary photon or proton developing on EBL/CMB and give a detailed description of our calculation method for the case of primary proton. In Section 3 we define the sample of blazar spectra to be used in the analysis (in this paper we focus on the observations made with Cherenkov telescopes). In Section 4 we present the fits to observed SEDs. Section 5 contains discussion, where our main results are recalled and discussed in the broader context; and, finally, in Section 6 the conclusions to our work are drawn.

\section{Electromagnetic cascades from primary gamma-rays and primary nuclei \label{sec:casc}}

After a primary gamma-ray, travelling through the Universe, is absorbed by an EBL photon, secondary electrons may upscatter CMB/EBL photons and produce the new generation of ``cascade'' photons by means of the IC process. If the energy of the primary photon is high enough and the source is sufficiently distant, the number of generations in such a cascade may be considerably higher than unity. In this case the spectrum of the cascade takes a universal form. On the contrary, if the energy of the primary photon is sufficiently low (but still high enough to be absorbed on EBL), the cascade takes a ``degenerate'' form and may be well described by a one-generation approximation (see. e.g., Dermer et al. \cite{dermer}). In the next subsection we discuss both regimes, as well as the transition between them. 

When the calculations presented in this section were almost finished, we noticed an insightful paper Berezinsky \& Kalashev \cite{berezinsky16} that discusses the universal regime in depth. We will express our results in terms of this work, when applicable. All results presented in this section are calculated for the case of a cascade particle threshold of 10 $GeV$, unless otherwise stated, and the redshift to the source $z_{s}$= 0.186. We find it convenient as we will see that 3 out of 6 blazars studied in this paper (see Section 3) have redshifts very close to 0.186. In the present paper we use the ROOT (Brun \& Rademakers \cite{brun}) analysis framework for drawing figures and performing a part of analysis.

\subsection{Cascades from primary gamma-rays}

Throughout this paper we use the publicly-available code \mbox{ELMAG} (Kachelriess et al., \cite{kachelriess}) version 2.02 to calculate the observable spectra of $\gamma$-rays for the case of primary photon. In what follows, when running the ELMAG code, we assume the Kneiske \& Dole \cite{kneiske10} (hereafter KD10) EBL model. The KD10 model was the first to be implemented in the ELMAG code, therefore this code is best tested (and most reliable) with this model. All calculations with ELMAG presented in this paper are full direct MC simulations (with parameter $\alpha_{smp}=0$).

There was a considerable amount of discussion whether pair beams resulting from the development of electromagnetic cascades are subject to plasma instabilities and additional (with respect to the IC process) energy losses. The conclusions of various studies are conflicting: while Broderick et al. \cite{broderick}, Schlickeiser et al. \cite{schlickeiser}, Chang et al. \cite{chang}, Menzler \& Schlickeiser \cite{menzler} found that these effects are considerable and must be accounted for, at least in the Fermi LAT region of the spectrum, Miniati \& Elyiv \cite{miniati}, Venters \& Pavlidou \cite{venters}, Sironi \& Giannios \cite{sironi}, Kempf et al. \cite{kempf}) argue that plasma losses are subdominant with respect to IC losses. The last work from this list (Kempf et al. \cite{kempf})), running a detailed numerical simulation, found that the growth rate of the instability is likely not sufficient to cause an appreciable effect. Therefore, we do not include such a process in our calculations.

\begin{figure}
\centering
\includegraphics[width=8cm]{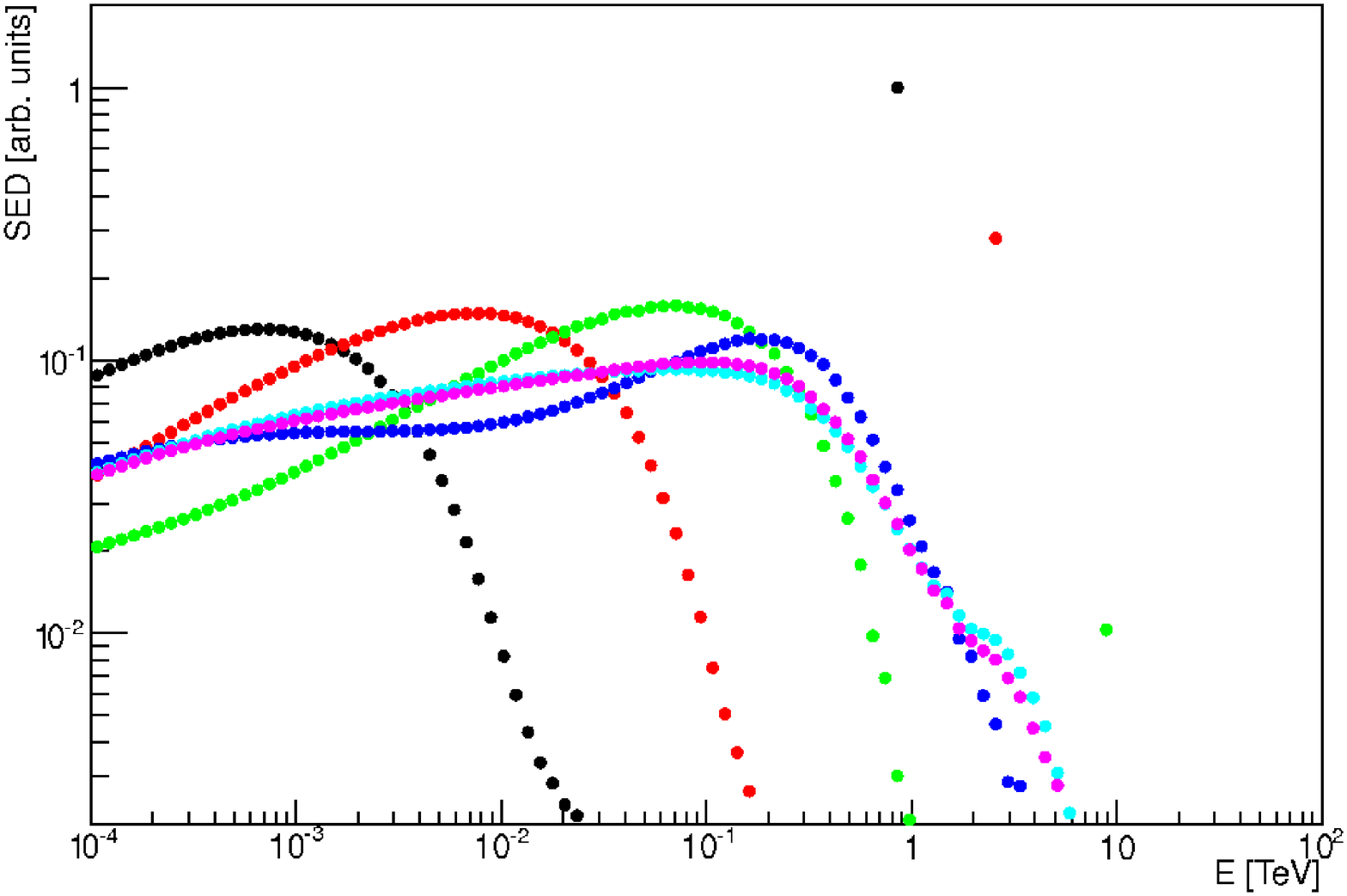}
\caption{Primary and secondary components (observable spectra) of $\gamma$-rays from primary monoenergetic injection. Black circles --- primary energies $E_{0}$= 1 $TeV$, red --- 3 $TeV$, green --- 10 $TeV$, blue --- 30 $TeV$, cyan --- 100 $TeV$, and magenta --- 1000 $TeV$= 1 $PeV$.}
\label{fig1}
\end{figure}

To demonstrate the basic regimes of electromagnetic cascade development on CMB/EBL, Fig.~\ref{fig1} presents observable spectra for different primary energies for the case of monoenergetic primary injection. Two principal components may be readily identified in Fig.~\ref{fig1} --- the cascade component itself and the primary (absorbed) component that is seen in the right part of the graph for the case of $E_{0}\le$ 10 $TeV$. The higher the energy, the more the primary component is attenuated due to a rapid rise of the $\tau$ value with increasing energy. A slight shift of the primary energy to the lower values is due to adiabatic losses. For the case of the primary energy above 10 $TeV$, the primary photons are almost completely absorbed due to very large value of $\tau$ for these energies.

Fig.~\ref{fig1} reveals two very different regimes: for the case of $E_{0}\le$ 10 $TeV$ the energy of the maximum in the spectral energy distribution (SED=$E^{2}dN/dE$) of the cascade component is approximately $\propto E_{0}^{2}$, while, on the contrary, for the case of $E_{0}\ge$ 100 $TeV$ the cascade spectrum is practically independent of energy. The first case may be well described by the one-generation approximation; the second one is essentially the universal regime. As we will see, the universality of the cascade spectrum is a ``weak'' one (Berezinsky \& Kalashev \cite{berezinsky16}), i.e. the spectrum of observable $\gamma$-rays is practically independent of the energy and type (photon/electron) of the primary particle, but depends on $z_{s}$. The transition between these regimes appears to be very sharp; the cascade completely changes its mode of propagation on only one decade of the primary energy. The physical reason for such a fast transition is clear: the rapid rise of the $\gamma\gamma$ interaction rate in the energy range of 10--100 $TeV$ (see, e.g., Kachelriess et al., \cite{kachelriess}, Fig. 2, left panel). Up to our knowledge, no study before us has remarked such a fast transition between these completely different regimes of cascade development. 

Additional graphs of electromagnetic cascade spectra in the universal regime are presented in Appendix A. They demonstrate that the property of cascade universality is fulfilled in the observable energy range 10 $GeV$-- 30 $TeV$ within 4 decades of the primary energy (from 100 $TeV$ to 1 $EeV$) with an accuracy \mbox{$\le$30 \%} independently of the primary particle type for $z_{s}\ge$0.02.

To be able to distinquish between two categories of observable $\gamma$-rays: primary (redshifted) (type 1) and cascade (type 2) photons, we modified the output of the ELMAG code and created a database that contains observable spectrum for every primary photon. For every array that contains observable spectrum we performed the following classification procedure. If this array contains only one non-zero entry and the redshifted value of the primary energy fits the energy range of the corresponding bin with the non-zero enrty, this array was classified as a type 1 event, otherwise --- as a type 2 event. Appendix B contains several graphs that show primary, absorbed and cascade components for various primary spectra. These figures demonstrate that the contribution of the cascade component to the total intensity is appreciable at 100 $GeV$ (roughly the threshold for Cherenkov telescope observations) and above only if the primary spectrum is hard enough so that $\gamma<2$ and that the cutoff energy of the spectrum $E_{c}>$10 $TeV$.

\subsection{Cascades from primary proton}

For the case of primary proton we developed an original code based on a hybrid approach. First of all, we propagate primary protons from the source to the observer ($z$= 0) with a small step $\delta z= 10^{-5}$, updating their energy at every step, and compute their mean energy loss according to Berezinsky et al. \cite{berezinsky}, equation (8). In this work we are interested in the case of proton energy losses on CMB. Full energy losses of a proton with energy $E_{p}$ per unit time are:
\begin{eqnarray}
-dE_{p}/dt= E_{p}\beta_{Ad}+b, \\
\beta_{Ad}= H_{0}\sqrt{\Omega_{m}(1+z)^{3}+\Omega_{\Lambda}}, \\
b= (1+z)^{2}b_{0}((1+z)E_{p}),
\end{eqnarray}
where $H_{0}$= 67.8 $km\cdot s^{-1}Mpc^{-1}$, $\Omega_{m}$= 0.308 (Ade et al. (The Planck Collaboration) \cite{ade15}), $\Omega_{\Lambda}\approx 1-\Omega_{m}$, and $b_{0}= -(dE_{p}/dt)_{pair+pion}$ --- pair production and pion production energy losses at $z=0$.

\begin{figure}
\centering
\includegraphics[width=8cm]{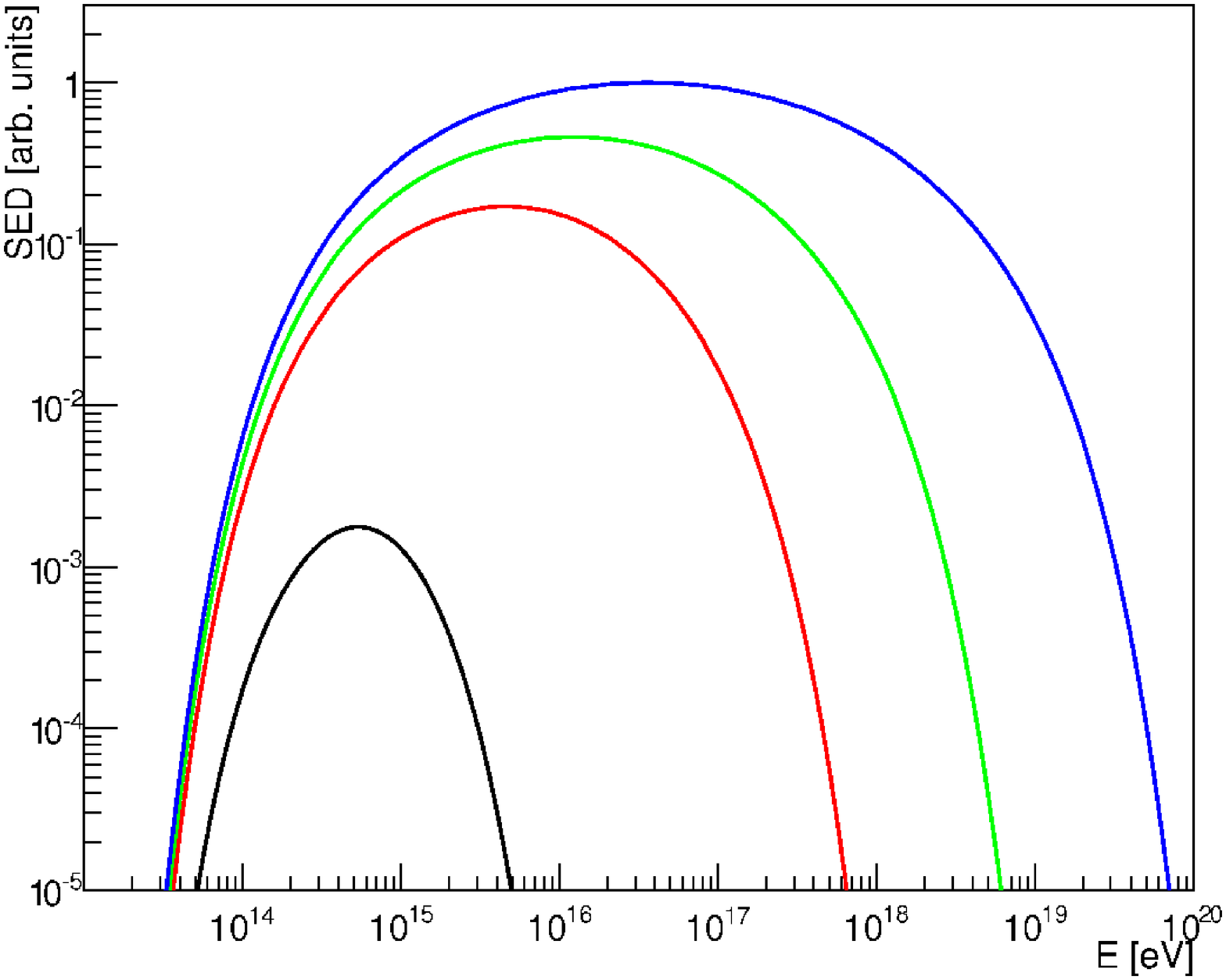}
\caption{Spectra of secondary electrons generated in the pair production process. Black line --- $E_{p0}$= 1 $EeV$, red --- 10 $EeV$, green --- 30 $EeV$, blue --- 100 $EeV$.}
\label{fig2}
\end{figure}

The next step is the calculation of energy spectra of particles produced by primary protons. We follow Kelner \& Aharonian \cite{kelner} to calculate these energy spectra. For the pair production process the spectra of electrons ($dN_{e}/dE_{e}$) and positrons are identical and for the case of pair production on CMB (Kelner \& Aharonian \cite{kelner}, equation (67)):
\begin{eqnarray}
dN_{e}/dE_{e}= -\frac{kT}{2\pi^{2}\gamma_{p}^{3}}\int\limits_{(\gamma_{p}+E_{e})^{2}/2\gamma_{p}E_{e}}^{\infty}{d\omega\omega ln(1-e^{-\omega/(2\gamma_{p}kT)})} \times \nonumber \\ \int\limits_{(\gamma_{p}^{2}+E_{e}^{2})/2\gamma_{p}E_{e}}^{\omega-1}{\frac{dE_{-}}{p_{-}}\Sigma(\omega,E_{-},\xi)},
\end{eqnarray}
where $k$ is the Boltzmann constant; $T=(1+z)T_{0}$ ($T_{0}$= 2.725 $K$ is the CMB temperature at $z$=0); $\gamma_{p}= E_{p}/m_{p}c^{2}$ is proton Lorentz factor; $E_{e}= E_{e}^{cm}/m_{e}c^{2}$, where $E_{e}^{cm}$ is the energy of secondary electron in the comoving frame (i.e. in the same frame where $E_{p}$ is measured); $\omega= (p_{p}\cdot p_{\gamma})/m_{e}^{2}c^{2}$ is the energy of the photon in the rest frame of the proton in units of electron rest energy ($p_{p}$,$p_{\gamma}$ are four-momenta of the proton and photon, respectively); $E_{-}= E_{-}^{rf}/m_{e}c^{2}$, where $E_{-}^{rf}$ and $p_{-}^{rf}= \sqrt{(E_{-}^{rf})^{2}/c^{2}-m_{e}^{2}c^{2}}$ are the energy and the momentum of electron in the rest frame of the proton, respectively ($p_{-}= p_{-}^{rf}/m_{p}c$); $\Sigma(\omega,E_{-},\xi)$ is the double-differential cross section as a function of energy and emission angle of the electron in the rest frame of the proton (Blumenthal \cite{blumenthal70a}, equation (10)) ($\xi= cos(\theta_{-})= (\gamma_{p}E_{-}^{rf}-E_{e}^{cm})/(\gamma_{p}p_{-})$, where the angle between the momenta of the photon and the electron is denoted as $\theta_{-}$). Four examples of calculated electron SED for the case of $z= 0$ are shown in Fig.~\ref{fig2} for different values of primary proton energy $E_{p0}$. The maximum of the SED for the case of $E_{p0}$= 100 $EeV$ was normalized to 1 in this figure.

\begin{figure}
\centering
\includegraphics[width=8cm]{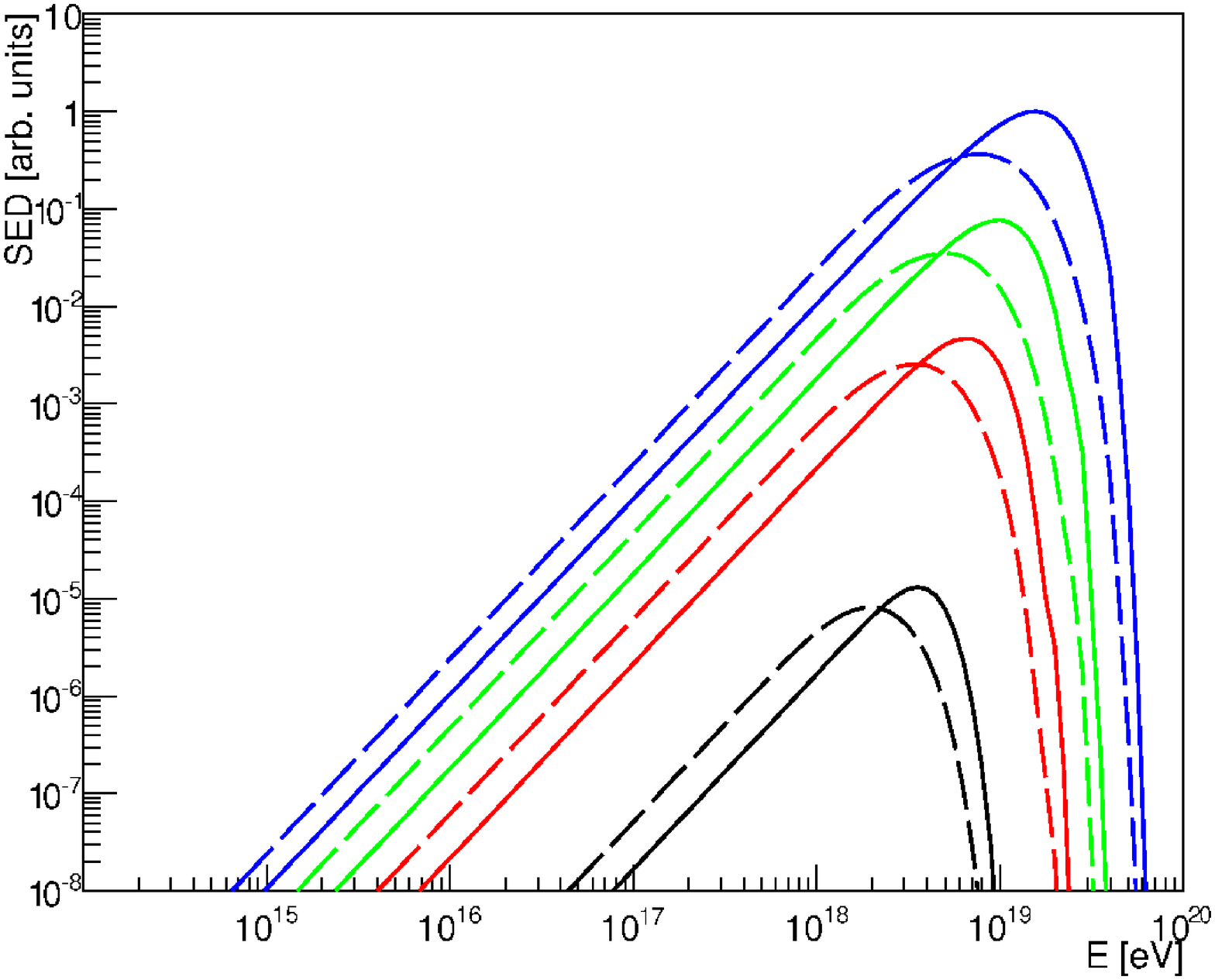}
\caption{Secondary $\gamma$ (solid) and positron (dashed) SEDs for the case of photopion losses. Black lines --- $E_{p0}$= 30 $EeV$, red --- 50 $EeV$, green --- 70 $EeV$ (green), blue --- 100 $EeV$.}
\label{fig3}
\end{figure}

The case of photopion losses was also covered in Kelner \& Aharonian \cite{kelner}; we use equations from Section II of their work with energy of CMB photon $\epsilon=(1+z)\epsilon_{(z=0)}$. As we are interested only in primary energies below 100 $EeV$, we neglect the flux of $e^{-}$ and $\bar{\nu}_{e}$ (which, according to Kelner \& Aharonian \cite{kelner}, Fig. 6 (right panel), contributes only about 0.1 \% of the total flux at $E_{p}$= 100 $EeV$ and $z$=0) and calculate the spectra of ($\gamma$,$e^{+}$,$\nu_{e}$,$\nu_{\mu}$,$\bar{\nu}_{\mu}$). We have checked that both panels of Kelner \& Aharonian \cite{kelner} (Fig. 6) are well reproduced for all types of particles for which we perform our calculations. As an example of this calculation we present Fig.~\ref{fig3} where the SEDs of secondary gamma-rays and positrons are presented for the case of $z$= 0 for different values of primary proton energy. The maximum of the SED for the case of $\gamma$-rays and $E_{p0}$= 100 $EeV$ was normalized to 1 in this figure.

Now we proceed to calculate observable (cascade) $\gamma$-ray spectra. The energies of secondary $\gamma$-rays and electrons in Fig.~\ref{fig2}--\ref{fig3} are mainly above 100 $TeV$, therefore the cascade universality assumption may be applied to reduce computer time requirements; such a calculation is presented in the following Subsubsection 2.2.1. As an alternative, to validate the results obtained in the cascade universality approximation, we performed a full calculation of observable spectra without this assumption (Subsubsection 2.2.2). Synchrotron radiation was neglected in these calculations. In both these Subsubsections we are interested only in the shape of the observable $\gamma$-ray spectrum. The normalization of the spectrum will be discussed in Section 5.

\subsubsection{Observable $\gamma$-ray spectra in the universal spectrum approximation}

\begin{figure}
\centering
\includegraphics[width=8cm]{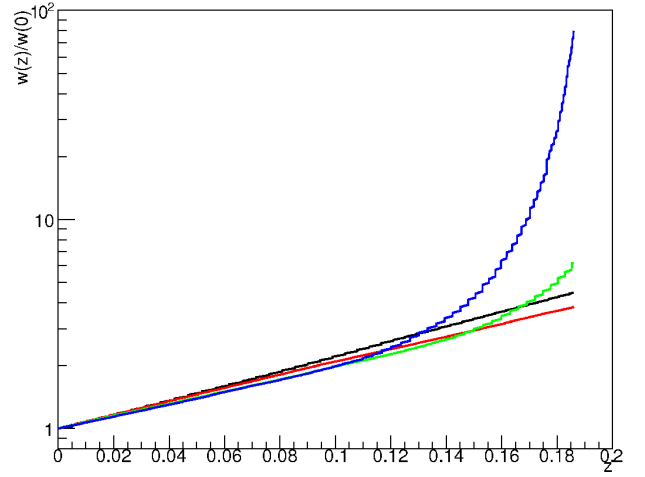}
\caption{$w(z)/w(0)$ dependence. Black line --- $E_{p0}$= 10 $EeV$, red --- 30 $EeV$, green --- 50 $EeV$, blue --- 100 $EeV$.}
\label{fig4}
\end{figure}

Here we assume ``weak universality'', as described above. Using the ELMAG code, we calculated an array of cascade spectra for the case of primary $\gamma$-rays with fixed energy 1 $PeV$ but different $z$, distributed randomly and uniformly from 0 to 0.30. While adiabatic losses do affect the energy of propagating protons, only ``active'' (pair and photopion) losses give rise to new particles and, therefore, eventually produce observable signal in $\gamma$-rays. The energy transferred to these particles on the step $dz$ is:
\begin{eqnarray}
w(z)= \left(\frac{dE_{p}}{dz}\right)_{pair+pion}= -\left(\frac{dE_{p}}{dt}\right)_{pair+pion}\left|\frac{dt}{dz}\right|, \\
\left|\frac{dt}{dz}\right|= \frac{1}{H_{0}\sqrt{\Omega_{m}(1+z)^{3}+\Omega_{\Lambda}}}\frac{1}{1+z}
\end{eqnarray}
Fig.~\ref{fig2}--\ref{fig3} demonstrate that nearly all secondary particles are ultrarelativistic, therefore, nearly all the energy of these particles (except neutrinos) could be transferred to cascade. Several examples of $w(z)/w(0)$ (i.e. $w(z)$ normalized to the value at $z$= 0) are shown in Fig.~\ref{fig4} for different values of primary proton energy. For $E_{p0}$= 50 $EeV$ and especially 100 $EeV$ primary protons quickly lose energy near the source until they reach comparatively low energy so that the energy loss rate becomes much smaller (see Fig. 1 of Berezinsky et al. \cite{berezinsky}). On the other hand, for the case of $E_{p0}$= 10 $EeV$ and 30 $EeV$ protons keep an appreciable fraction of their energy until they reach $z=0$, and, as a result, they keep producing secondaries quite effectively even near the observer.

\begin{figure}
\centering
\includegraphics[width=8cm]{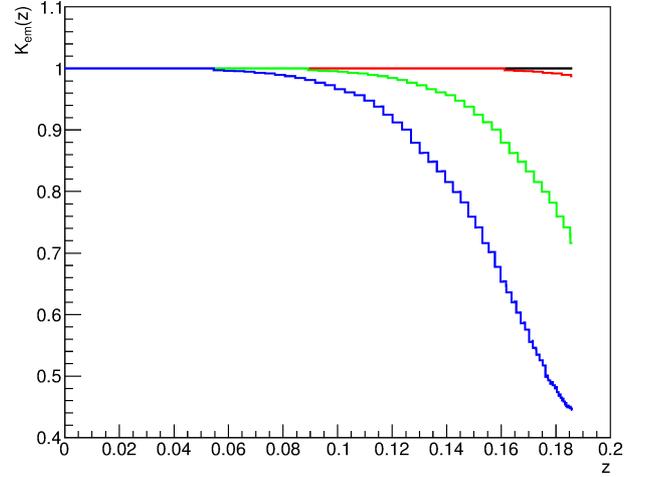}
\caption{$K_{em}(z)$ dependence. Meaning of colors is the same as in Fig.~\ref{fig4}.}
\label{fig5}
\end{figure}

\begin{figure}
\centering
\includegraphics[width=8cm]{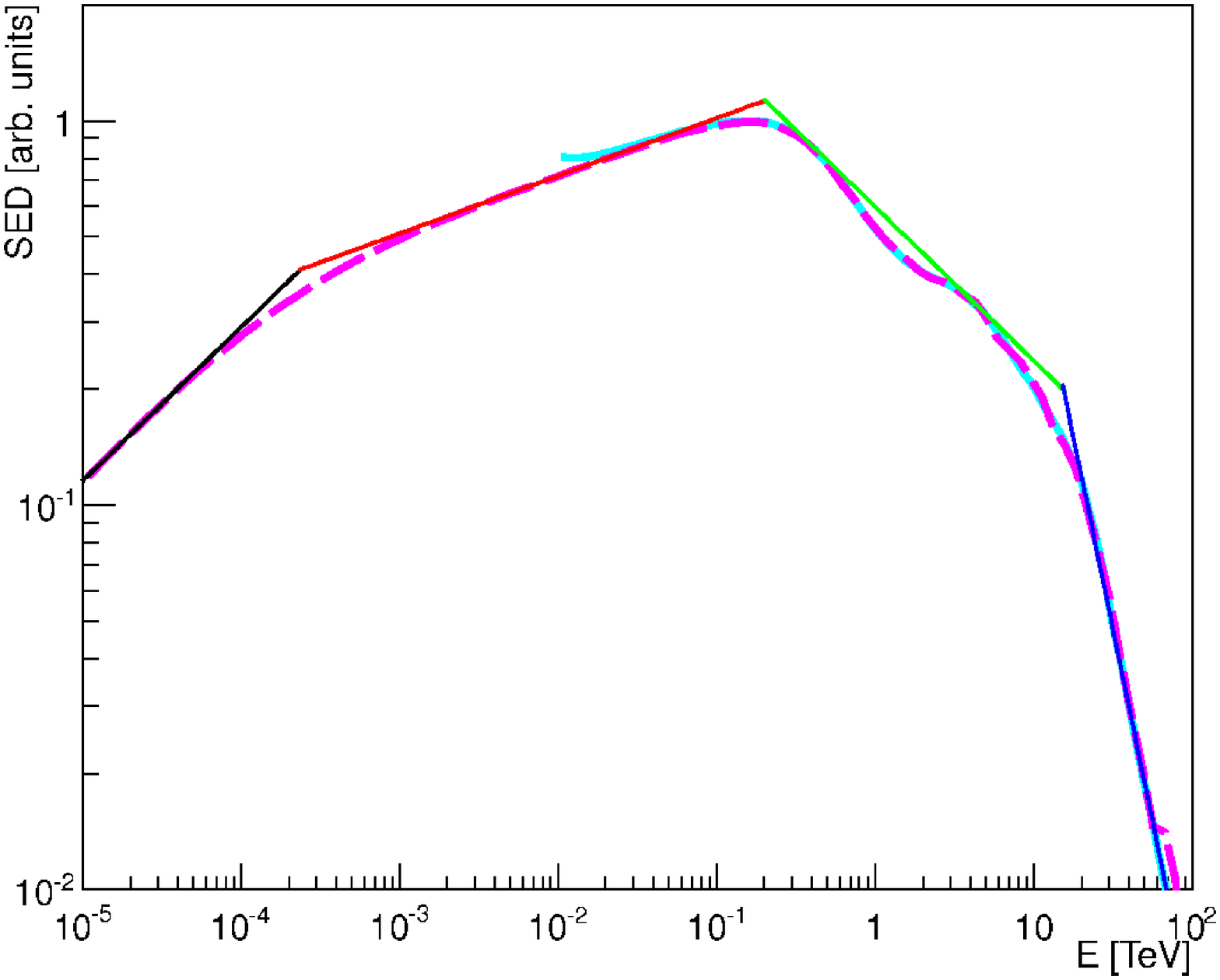}
\caption{Observable $\gamma$-ray SED. Dashed magenta line --- $E_{thr}$= 10 $MeV$, solid cyan --- 10 $GeV$. Black, red, green, and blue lines denote power-law approximations of the spectrum in various energy regions.}
\label{fig6}
\end{figure}

\begin{figure}
\centering
\includegraphics[width=8cm]{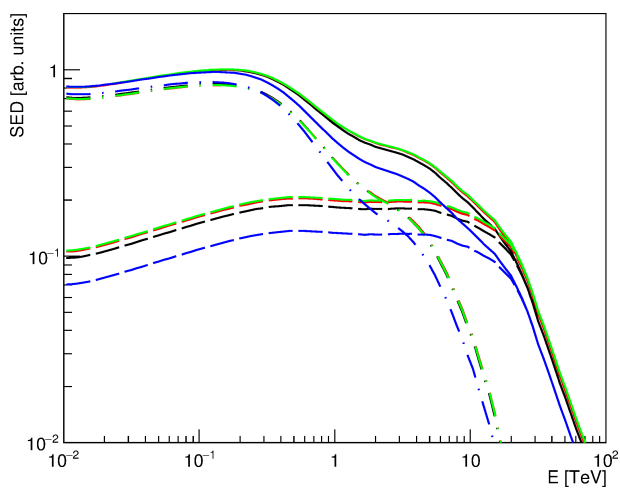}
\caption{Observable SED for various $E_{p0}$ (solid curves): black curve --- $E_{p0}$=10 $EeV$, red --- 30 $EeV$, green --- 50 $EeV$, blue --- 100 $EeV$. SEDs decomposed on the contributions from the primary cascade particles produced at $z<$0.06 (dashed curves) and at $z>$0.06 (dot-dashed curves) are also shown.}
\label{fig7}
\end{figure}

Finally, we calculate the observable spectrum of $\gamma$-rays at $z$=0. For the case of an isotropic source of primary protons:
\begin{equation}
\left(\frac{dN}{dE}\right)_{\gamma-obs}(E)= \int\limits_{0}^{z_{s}}{K_{em}(z)\frac{w(z)}{w(0)}\left(\frac{dN}{dE}\right)_{c}(E,z)dz},
\end{equation}
where $K_{em}(z)$ is the fraction of ``active'' losses transferred to $\gamma$-rays and electrons, and $(dN/dE)_{c}(E,z)$ is the universal spectrum of cascade with starting point at $z$. Several examples of $K_{em}(z)$ are shown in Fig.~\ref{fig5} for different values of primary proton energy. For the case of $E_{p0}\le$ 30 $EeV$, when the dominating source of proton energy loss is the pair-production process, $K_{em}$ is practically equal to 1.

In this paper we study the case of ultrarelativistic primary protons, $E_{p0}>$1 $EeV$, therefore, in the absence of EGMF, the angular distribution of observable $\gamma$-rays is practically coincident with the angular distribution of primary protons. Thus, equation (7) is justified also for the conditions investigated in our work.

The resulting observable SED is shown in Fig.~\ref{fig6} for the case of monoenergetic primary injection with $E_{p0}$= 30 $EeV$ for two options of cascade $\gamma$-ray threshold energy $E_{thr}$ in ELMAG. The signatures of the spectrum are highlighted with additional power-law approximations at various energy regions. The low-energy signatures of the observable spectrum (i.e. the parameters of a ``poly-gonato'' power-law approximation at $E$<200 $GeV$) are practically identical to the case of electromagnetic cascade from primary $\gamma$-ray (see Berezinsky \& Kalashev \cite{berezinsky16} for a detailed discussion of the latter case). Indeed, for the case of our calculations (i.e. primary proton) these parameters are: the shape of the black power-law segment is $\gamma_{1}$= 1.60 (for $dN/dE$ spectrum) below the break at $E_{b1}$= 230 $MeV$, and $\gamma_{2}$= 1.85 below another break at $E_{b2}$= 200 $GeV$ (red segment). The corresponding typical values for electromagnetic cascade from primary $\gamma$-ray are: $\gamma_{1-EM}\approx$ 1.5, $E_{b1-EM}\sim$ 100 $MeV$, $\gamma_{2}\approx$ 1.9, and $E_{b2-EM}\approx$ 200--400 $GeV$ depending on the fitting conventions (see Fig. 3, upper panel, and Fig. 8 of Berezinsky \& Kalashev \cite{berezinsky16}).

However, for the higher energy ($E$>200 $GeV$) the situation is very different: for the case of primary proton the observable spectrum is far less steep than for the case of primary $\gamma$-ray, $\gamma_{3}$= 2.40 between 200 $GeV$ and $E_{b3}$= 15.2 $TeV$ (green segment), and $\gamma_{4}$= 4.00 above 15.2 $TeV$ (blue segment). In this work we are interested only in $E$ below 100 $TeV$; we leave the $E>$100 $TeV$ energy region of $\gamma$-ray observable spectrum for further stidues.

The values of the power-law indices and the break energies were derived in Berezinsky \& Kalashev \cite{berezinsky16} as follows (below we use notations slightly different from this paper). A simplified ``dichromatic'' model of the target photon field (CMB+EBL) was devised, assuming that all CMB photons have energy $\epsilon_{CMB}$= $6.3\cdot10^{-4}$ $eV$, and all EBL photons --- $\epsilon_{EBL}$= $0.68$ $eV$. Then, a quantity $E_{\gamma}^{min}= m_{e}^{2}c^{4}/\epsilon_{EBL} \approx$ $3.84\cdot10^{11}$ $eV\sim E_{b2}$ was introduced meaning the minimum energy of a primary $\gamma$-ray that undergoes effective absorption on EBL photons; the corresponding energy of electrons of the produced pair are $E_{e}^{min} \approx E_{\gamma}^{min}/2$. Secondary $\gamma$-rays produced by electrons with energy $E_{e}^{min}$ typically have energy $E_{X}\approx (4/3)(E_{e}^{min}/(m_{e}c^{2}))^{2}\epsilon_{CMB}$ (Blumenthal \& Gould \cite{blumenthal70b}); putting together the expressions for $E_{X}$, $E_{\gamma}^{min}$, and $E_{e}^{min}$,  $E_{X}\approx (E_{\gamma}^{min}/3)(\epsilon_{CMB}/\epsilon_{EBL}) \approx 1.19\cdot10^{8}$ $eV\approx E_{b1}$.

Denoting the number of cascade electrons that has energy $E_{e}$ during the whole time of cascade development as $q_{e}(E_{e})$ and taking into account that the number of electrons in each cascade generation $N_{e}\approx 2N_{\gamma}$ and the energy of these electrons $\approx E_{\gamma}/2$ has nearly the same distribution as the energy of the $\gamma$-rays of the same generation, in average for the $E_{e}>E_{e}^{min}$ energy range $q_{e}(E_{e})\propto 1/E_{e}$ (as $q_{e}\cdot E_{e} \propto$ the primary energy $E_{0}$). Most electrons are produced above $E_{e}^{min}$, therefore in the low-energy range $E_{e}<E_{e}^{min}$ $q_{e}(E_{e})\approx const$. Finally, the spectrum of cascade photons may be found from the following equation: $dn_{\gamma}(E_{\gamma})= q_{e}(E_{e})dE_{e}/E_{\gamma}$; again using the expression for secondary $\gamma$-ray energy $E_{\gamma}^{sec}\propto E_{e}^{2}$, for $E_{e}<E_{e}^{min}$ (corresponding to $E_{\gamma}<E_{X}$) we obtain for the shape of the observable spectrum $dn_{\gamma}/dE_{\gamma}\propto dn_{\gamma}/(E_{e}dE_{e})\propto 1/E_{e}^{3}= E_{\gamma}^{-3/2}$, which is similar to $E_{\gamma}^{-\gamma_{1}}$. In a like manner, for $E_{\gamma}^{min}>E_{\gamma}>E_{X}$ the shape of the observable spectrum is $dn_{\gamma}/dE_{\gamma}\propto E_{\gamma}^{-2}$; numerical simulations performed in Berezinsky \& Kalashev \cite{berezinsky16} indicate $dn_{\gamma}/dE_{\gamma}\propto E_{\gamma}^{-1.9}$ in the same energy region. The simplified analytic model considered above assumes that all primary $\gamma$-rays with energy $E_{\gamma}>E_{\gamma}^{min}$ are completely absorbed and, consequently, the spectrum has an abrupt cutoff at this energy. By detailed numerical calculations presented here we reproduce the more realistic shape of the high-energy cutoff, both for primary $\gamma$-rays and primary protons.

Fig.~\ref{fig7} shows observable SEDs for different values of $E_{p0}$. For $E_{p0}$= 10, 30, 50 $EeV$ the shape of the spectra are nearly identical, while for the case of $E_{p0}$= 100 $EeV$ the spectrum is somewhat steeper due to a pile-up of the $w(z)$ dependence near the source of primary protons. Indeed, in the latter case more energy is injected near the source, therefore, more electromagnetic cascades start to develop at comparatively high distance from the observer, thus leading to lower intensity at high values of the observable energy.

As well, different components of the observable SEDs are shown --- the one formed by $\gamma$-rays and electrons that were produced by primary protons at $z<$0.06, as well as the other contribution for $z>$0.06. The first component dominates at $E>$5 $TeV$ and has almost the same shape of the spectrum for all $E_{p0}$ considered. This effect may be qualitatively understood as follows. The spectrum of the first component is defined by an equation similar to equation (7), but with limits of integration on redshift equal to 0 and 0.06. Primary protons with energy $>$50 $EeV$ quickly lose energy to pion production (see Berezinsky et al. \cite{berezinsky}), until they reach the energy range where pair production losses dominate. Therefore, $K_{em}(z)\approx 1$ at $z<$0.06 even for $E_{p0}$=100 $EeV$ (see Fig. 5). The $w(z)$ dependence is also similar for all considered $E_{p0}$ at $z<$0.06 (see Fig. 4). Again, this is due to the fact that for $E_{p0}>$50 $EeV$ protons quickly lose energy so that their energy at $z<$0.06 is similar irrespectively of the $E_{p0}$ value. Thus, the shape of the observable SEDs at $E>$20 $TeV$, where the first component is strongly dominant, is almost the same. In the 200 $GeV$--10 $TeV$ energy region and $E_{p0}=$100 $EeV$ the first component has lower relative normalization (with respect to the second component), once again due to the fact that in this case a larger fraction of primary energy is lost at $z>$0.06 than for $E_{p0}<$50 $EeV$. The second component for $E_{p0}=$100 $EeV$ is also somewhat steeper for $E_{p0}=$100 $EeV$ due to the same effect. Therefore, the overall observable SED in the 200 $GeV$--10 $TeV$ energy region is steeper for $E_{p0}=$100 $EeV$ than for the other considered values of $E_{p0}$. The difference in the slope between the observable SEDs in this energy range may be estimated as follows. The total normalization of the second component is $I_{2}= \int\limits_{0.06}^{z_{s}}{K_{em}(z)w(z)dz}$, and of the first component --- $I_{1}= \int\limits_{0}^{0.06}{K_{em}(z)w(z)dz}$; $K_{nz}= C_{nz}I_{1}/I_{2}$ is the ratio of these quantities; the normalization factor $C_{nz}$ is chosen so that $K_{nz}=1$ at $E_{p0}=$30 $EeV$. Direct numerical calculation yields the values $K_{nz}= 0.91$ at $E_{p0}=$10 $EeV$, $K_{nz}= 1$ at $E_{p0}=$30 $EeV$, $K_{nz}= 1.01$ at $E_{p0}=$50 $EeV$, and $K_{nz}= 0.63$ corresponding to $E_{p0}=$100 $EeV$; this translates to the difference in the power-law slope $\delta \gamma=0.07-0.09$ between the case of $E_{p0}=$100 $EeV$ and other cases.

\subsubsection{Observable $\gamma$-ray spectra in the full hybrid approach \label{ssec:hybrid}}

To obtain the observable spectrum without the assumption of cascade universality, we calculated an array of cascade spectra from primary $\gamma$-rays and electrons of energies 10 $TeV$--100 $EeV$. From 10 $TeV$ to 10 $EeV$ the primary spectrum of these cascades ($dN/dE_{0}$) had a power-law slope 1, and the slope from 10 $EeV$ to 100 $EeV$ was set to 2. As in the universal spectrum approach, we propagated primary protons, but now, instead of the universal spectrum, we utilized our new database of cascade spectra. We calculated a vast two-dimensional array of secondary particle spectra for the case of the pair production process, as described above, as well as for the photopion process for ($\gamma$,$e^{+}$,$\nu_{e}$,$\nu_{\mu}$,$\bar{\nu}_{\mu}$), both for the case of 101 logarithmically spaced values of $E_{p}$ from 1 $EeV$ to 100 $EeV$, and for 30 linearly spaced values of $z$ from 0 to 0.30. These arrays were used to calculate the observable gamma-ray spectrum, that is:
\begin{eqnarray}
\left(\frac{dN}{dE}\right)_{\gamma-obs}(E)= \int\limits_{0}^{z_{s}}{dz(F_{e}(E,z)+F_{\gamma}(E,z))} \\
F_{e}(E,z)=  \int\limits_{0}^{\infty}{dE_{s}\left(\frac{dN}{dE_{s}}\right)_{e-p}(E_{s},z)\left(\frac{dN}{dE}\right)_{e-c}(E_{s},E,z)} \\
F_{\gamma}(E,z)= \int\limits_{0}^{\infty}{dE_{s}\left(\frac{dN}{dE_{s}}\right)_{\gamma-p}(E_{s},z)\left(\frac{dN}{dE}\right)_{\gamma-c}(E_{s},E,z)},
\end{eqnarray}
where $(dN/dE)_{e-c}(E_{s},E,z)$ and $(dN/dE)_{\gamma-c}(E_{s},E,z)$ are, as before, the spectra of cascades, but now the shape of these spectra may depend not only on $z$, but also on the primary energy $E_{s}$ and the type of the primary particle. $(dN/dE_{s})_{e-p}(E_{s},z)$ and $(dN/dE_{s})_{\gamma-p}(E_{s},z)$ are the spectra of secondary electrons and $\gamma$-rays, respectively, that were produced by primary protons.

\begin{figure}
\centering
\includegraphics[width=8cm]{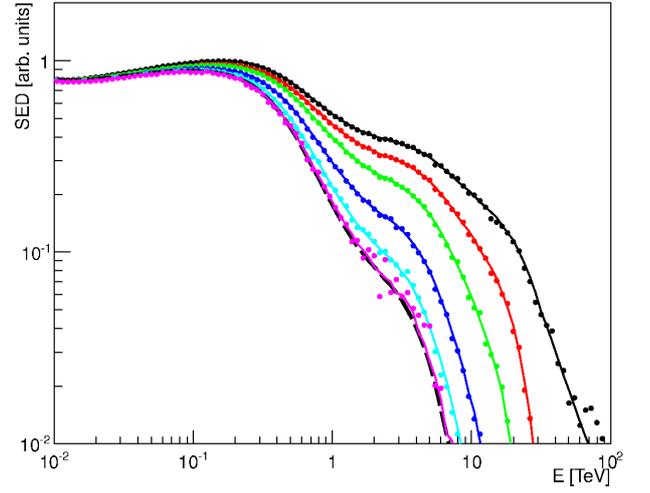}
\caption{The comparison between observable SED obtained in the full hybrid approach (circles) and the universal hybrid method approach (solid curves) for different values of the $z_{c}$ parameter: black --- $z_{c}$=0, red --- $z_{c}$=0.02, green --- $z_{c}$= 0.05, blue --- $z_{c}$= 0.10, cyan --- $z_{c}$= 0.15, magenta --- $z_{c}$= 0.18. Primary proton energy $E_{p0}$= 30 $EeV$. The universal spectrum for the case of primary $\gamma$-rays with energy 1 $PeV$ is shown for comparison (dashed black line). Relative normalization of all spectra is done at E= 10 $GeV$.}
\label{fig8}
\end{figure}

\begin{figure}
\centering
\includegraphics[width=8cm]{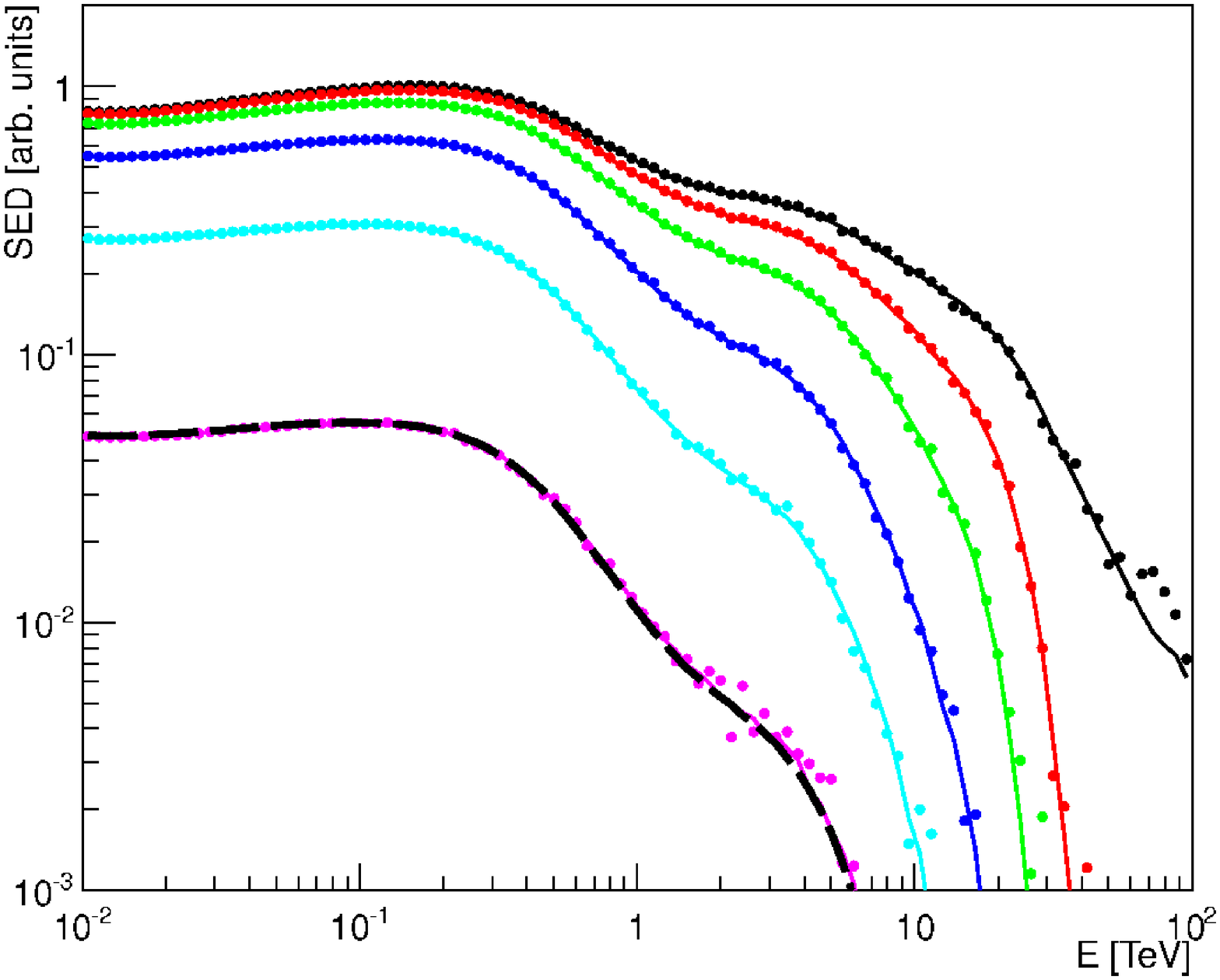}
\caption{Same as in Fig.~\ref{fig8}, but without relative normalization of spectra with different $z_{c}$.}
\label{fig9}
\end{figure}

\begin{figure}
\centering
\includegraphics[width=8cm]{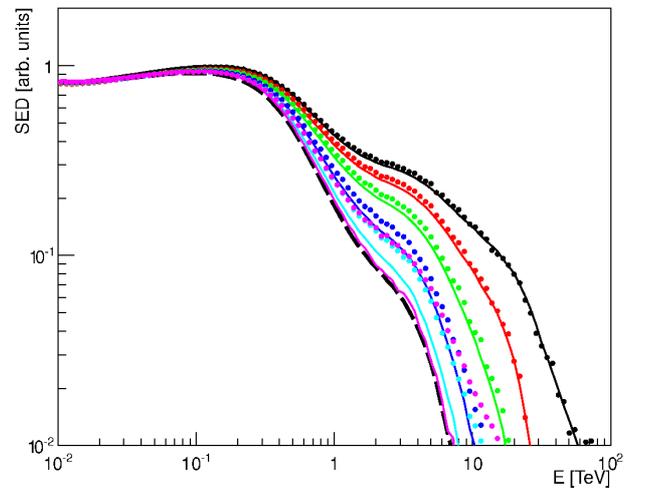}
\caption{Same as in Fig.~\ref{fig8}, but for $E_{p0}$= 100 $EeV$.}
\label{fig10}
\end{figure}

\begin{figure}
\centering
\includegraphics[width=8cm]{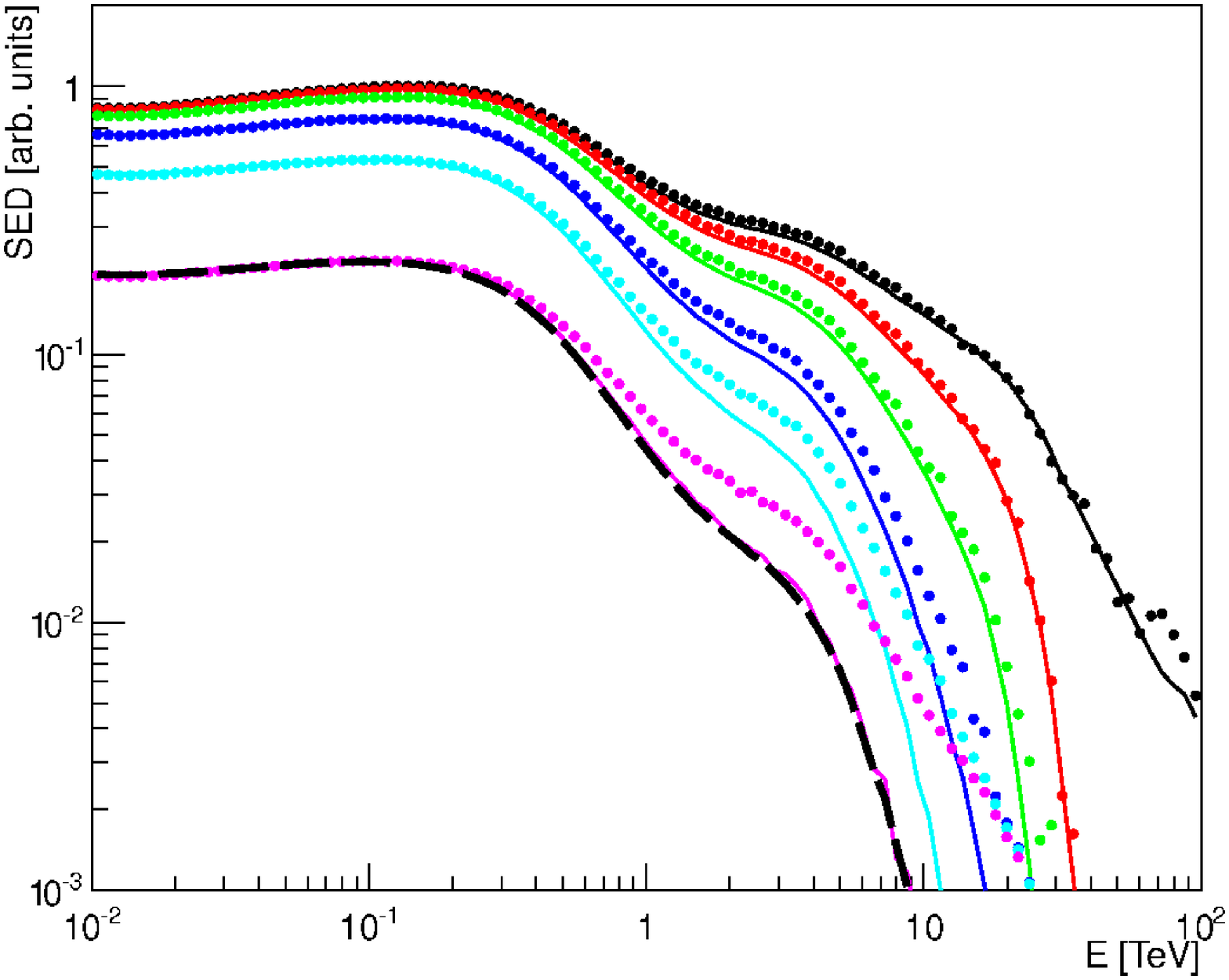}
\caption{Same as in Fig.~\ref{fig10}, but without relative normalization of spectra with different $z_{c}$.}
\label{fig11}
\end{figure}

The results of our calculations of observable SEDs are presented in Figs.~\ref{fig8}--\ref{fig11}. The comparison graph of spectra calculated with two different approaches for the case of $E_{p0}$= 30 $EeV$ (black solid line and circles in Fig.~\ref{fig8}) shows very good agreement. We also considered a possibility of a highly structured extragalactic magnetic field (for instance, a galaxy cluster) that is situated at some redshift $z_{c}$ between the source and the observer. We assume that magnetic field strength in this cluster is sufficiently high so that the proton beam traversing the cluster is practically dissolved. Several cases of observable spectra for different $z_{c}$ are also shown in Fig.~\ref{fig8}. In all these cases the agreement of results obtained with two different approaches is good. Finally, Fig.~\ref{fig8} shows that in the case when $z_{c}\approx z_{s}$ the observable spectrum is very near to a universal cascade spectrum from primary $\gamma$-rays. The same SEDs, but without relative normalization, are shown in Fig.~\ref{fig9}.

\begin{table*}
\caption{Observations of extreme $TeV$ blazars used in this paper}
\label{table:1}
\centering
\begin{tabular}{c c c c c} 
\hline\hline
  N & Source                & $z$   & Observational period    &       Reference                           \\
\hline
 1  & H 1426+428            & 0.129 &      1999-2000          & Aharonian et al. \cite{aharonian03}       \\
 2  & H 1426+428            & 0.129 &      1998-2000          & Djannati-Atai et al. \cite{djannati-atai} \\
 3  & H 1426+428            & 0.129 &      2001               & Horan et al. \cite{horan}                 \\
 4  & 1ES 0229+200          & 0.140 &      2005-2006          & Aharonian et al. \cite{aharonian07a}      \\
 5  & 1ES 0229+200          & 0.140 &      2010-2012          & Aliu et al. \cite{aliu}                   \\
 6  & 1ES 1218+304          & 0.182 &      2012-2013          & Madhavan et al. \cite{madhavan}           \\
 7  & 1ES 1101-232          & 0.186 &      2004-2005          & Aharonian et al. \cite{aharonian07b}      \\
 8  & 1ES 1101-232          & 0.186 &      2004-2005          & Aharonian et al. \cite{aharonian06}       \\
 9  & 1ES 0347-121          & 0.188 &      Aug.-Dec. 2006     & Aharonian et al. \cite{aharonian07c}      \\
10  & 1ES 0414+009          & 0.287 &      2005-2009          & Abramowski A. et al. \cite{abramowski12}  \\
\hline
\end{tabular}
\end{table*}

The same results as in Fig.~\ref{fig8}, but for $E_{p0}$= 100 $EeV$, are shown in Fig.~\ref{fig10}. For $z_{c}\le$0.05 the agreement between the results obtained with our two methods is rather good, while for $z_{c}\ge$0.10 SEDs resulting from the universal hybrid approach are more steep at comparatively high values of observable energy, $E$>500 $GeV$. This property of observable SEDs is due to a possible hardening of electromagnetic cascade spectra shape for the case of high primary energy, $E_{\gamma0}\ge$1 $EeV$, see Fig.~\ref{figA3}. This hardening is the result of the ``extreme high-energy cascade regime'' that is realized when $E_{\gamma0}\epsilon>>m_{e}c^{2}$, where $\epsilon$ is the energy of EBL/CMB photon, and $m_{e}$ is the mass of electron. In this regime, for the case of the pair production process, one electron of every produced pair gets almost all of the energy of the primary photon, and, likewise, the secondary photon in the IC process gets almost all of the energy of the primary electron (for comparatively recent discussions see Khangulyan et al. \cite{khangulyan}, Aharonian et al. \cite{aharonian12}). As a result, electromagnetic cascade develops much more slowly than in the universal regime.

We note, however, that the ELMAG code does not account neither for a possible impact of the so-called universal radio background (URB) to the cascade development (Protheroe \& Biermann \cite{protheroe96}, Settimo \& De Domenico \cite{settimo}), nor for the higher-order quantum electrodynamics (QED) processes such as double pair production (e.g. Brown et al. \cite{brown}, Demidov \& Kalashev \cite{demidov}) or ``triplet production'' (Mastichiadis et al., \cite{mastichiadis}). The results presented in Settimo \& De Domenico \cite{settimo} (Fig. 2, bottom) show that the mean free path for the IC process is likely to be affected by these effects even for the primary electron energy of 10 $EeV$, thus making the ``extreme high-energy cascade regime'' less pronounced. In what follows we mainly use results obtained in the universal cascade regime. Finally, the same SEDs as in Fig.~\ref{fig10}, but without relative normalization, are shown in Fig.~\ref{fig11}.

\section{The sample of blazar spectra}

In this study we focus on a subsample of AGN, the so-called extreme $TeV$ blazars (Bonnoli et al. \cite{bonnoli}). Blazars are defined as a class of AGN that has peculiar multi-wavelength properties from radio to $\gamma$-ray bands of the spectrum. From the point of view of a gamma-ray astronomer, a blazar may be defined simply as an active galactic nucleus that is a bright source of $\gamma$-ray emission. Extreme $TeV$ blazars are believed to have particularly high peak energy in their SED, $E_{peak}>$1 $TeV$ (Bonnoli et al. \cite{bonnoli}).

During the last decade, these sources became very important for the studies of $\gamma\gamma$ absorption on EBL (Aharonian et al. (H.E.S.S. Collaboration) \cite{aharonian06}). Indeed, these sources allow to study this effect for the highest values of $\tau_{\gamma\gamma}$ available now. Additionally, they may have a very hard primary spectrum of $\gamma$-rays, thus making the contribution of the cascade component significant even in the VHE energy range.

The sample of observations of extreme $TeV$ blazars used in this paper is presented in Table 1. It contains 9 independent observations of 6 sources, performed with imaging Cherenkov telescopes HEGRA, CAT, Whipple, H.E.S.S., and VERITAS. For the whole history of observations of these sources in the VHE energy region up to 2016-01-01 (Wakely \& Horan \cite{wakely}), 5 out of 6 of them display a rather slow (if any) variability (with characteristic period of months or even years). The only exception is 1ES 1812+304; for this source a rapid flare with full width on half magnitude about 2-3 days was observed in 2008 by the VERITAS telescope \cite{veritas}. 

\section{Fitting the spectra of extreme $TeV$ blazars}

In this section we compare the predictions of different models with observations. Before we proceed to present the fits for various sources, some remarks are in order. The collaborations operating Cherenkov telescopes usually reconstruct the spectrum of primary $\gamma$-rays, and not just the histogram of measured energy values. Therefore, while presenting our results, we will not perform any convolution of model spectra with instrumental energy resolution templates. Of course, the procedure of the primary spectrum deconvolution always results in some additional systematic uncertainty of the reconstructed spectrum. This systematics, however, for the case of observations listed in Table 1 is usually subdominant with respect to statistical uncertainties, especially in the optically thick region, to which we pay most attention in this work.

Another remark concerns the selection of EBL model for the case of absorption-only fits. Meyer et al. \cite{meyer12} calculated the significance $Z_{a}$ of the VHE anomaly for different EBL models (Franceschini et al. \cite{franceschini} (hereafter F08), Kneiske \& Dole \cite{kneiske10} (already denoted as KD10), and Dominguez et al. \cite{dominguez} (D11)) and found that $Z_{a}$ may depend non-trivially on the total normalization of the EBL intensity in a certain wavelength band. The lowest $Z_{a}$ was obtained for the case of the F08 model. Therefore, in what follows we present absorption-only fits using the F08 model; for comparison, we also include the fits for G12 model, as well as for the model directly derived from the ELMAG code (see next subsection for more details).

\subsection{The case of 1ES 0347-121}

Let us start our discussion with the case of blazar 1ES 0347-121 and the absorption-only model. The redshift of this source $z$= 0.188 is very near to 0.186, for which all calculations presented in Section 2 are applicable. The shape of the primary spectrum was chosen as:
\begin{equation}
dN/dE_{0} \propto E_{0}^{-\gamma}exp(-E_{0}/E_{0c}), 
\label{eqn11}
\end{equation}
where $E_{0c}$ is the cutoff energy. For this model we neglect adiabatic losses, because they do not change the shape of the spectrum.

The ROOT analysis framework with the integrated minimization system MINUIT (James \& Roos \cite{james}) (namely, the gradient optimization routine MIGRAD) was used to obtain this fit. During the optimization procedure, every experimental bin was divided to 20 small parts; the flux extinction factor $\propto exp(-\tau)$ was calculated for each of these sub-bins in order to ensure realistic implementation of the $\gamma$-ray absorption process. After that, a histogram of model SED was evaluated, and a $\chi^{2}$ minimization was performed with the MIGRAD routine. The output of the routine includes the optimized values of primary spectrum parameters.

\begin{figure}[t]
\centerline{\includegraphics[width=0.50\textwidth]{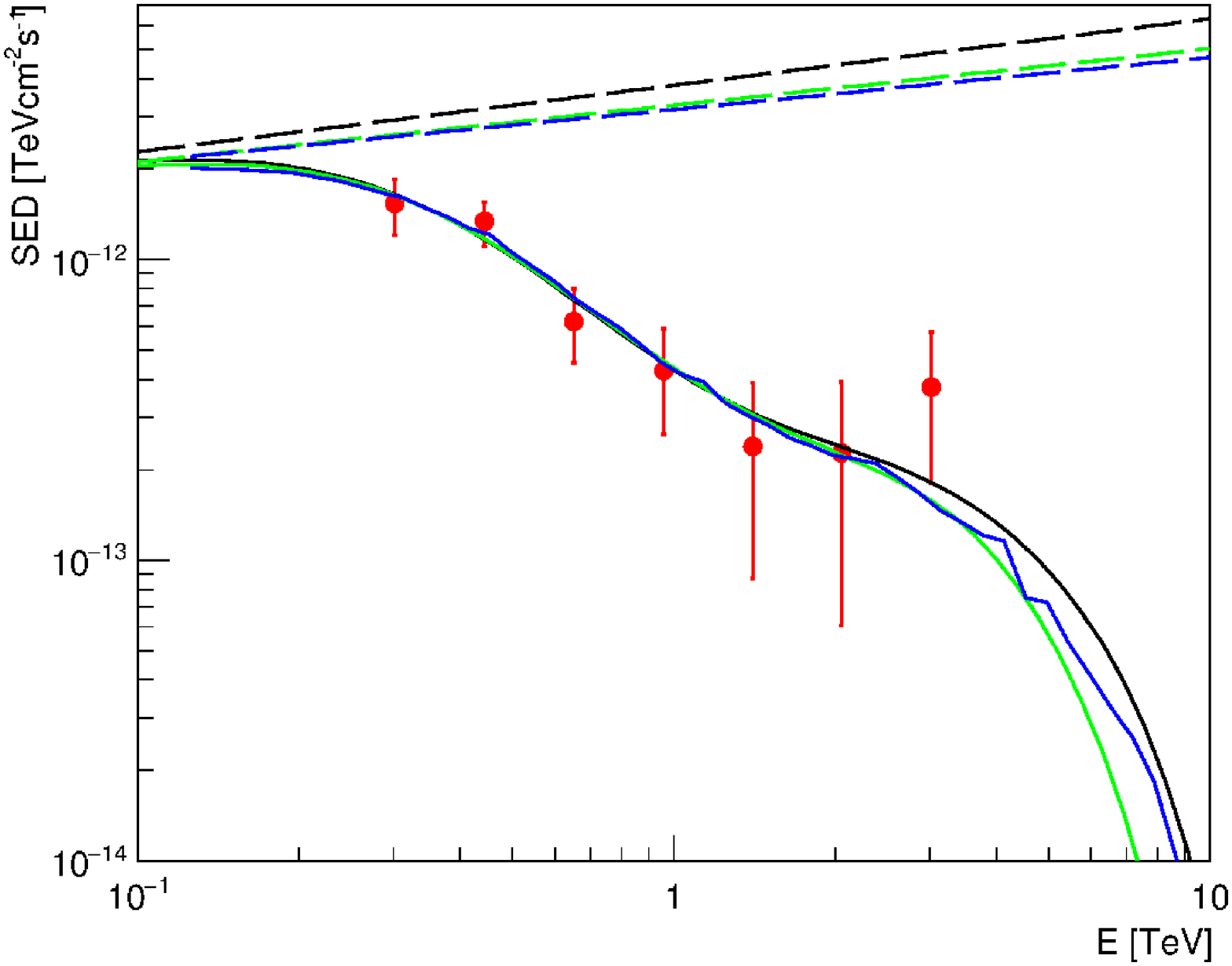}\includegraphics[width=0.50\textwidth]{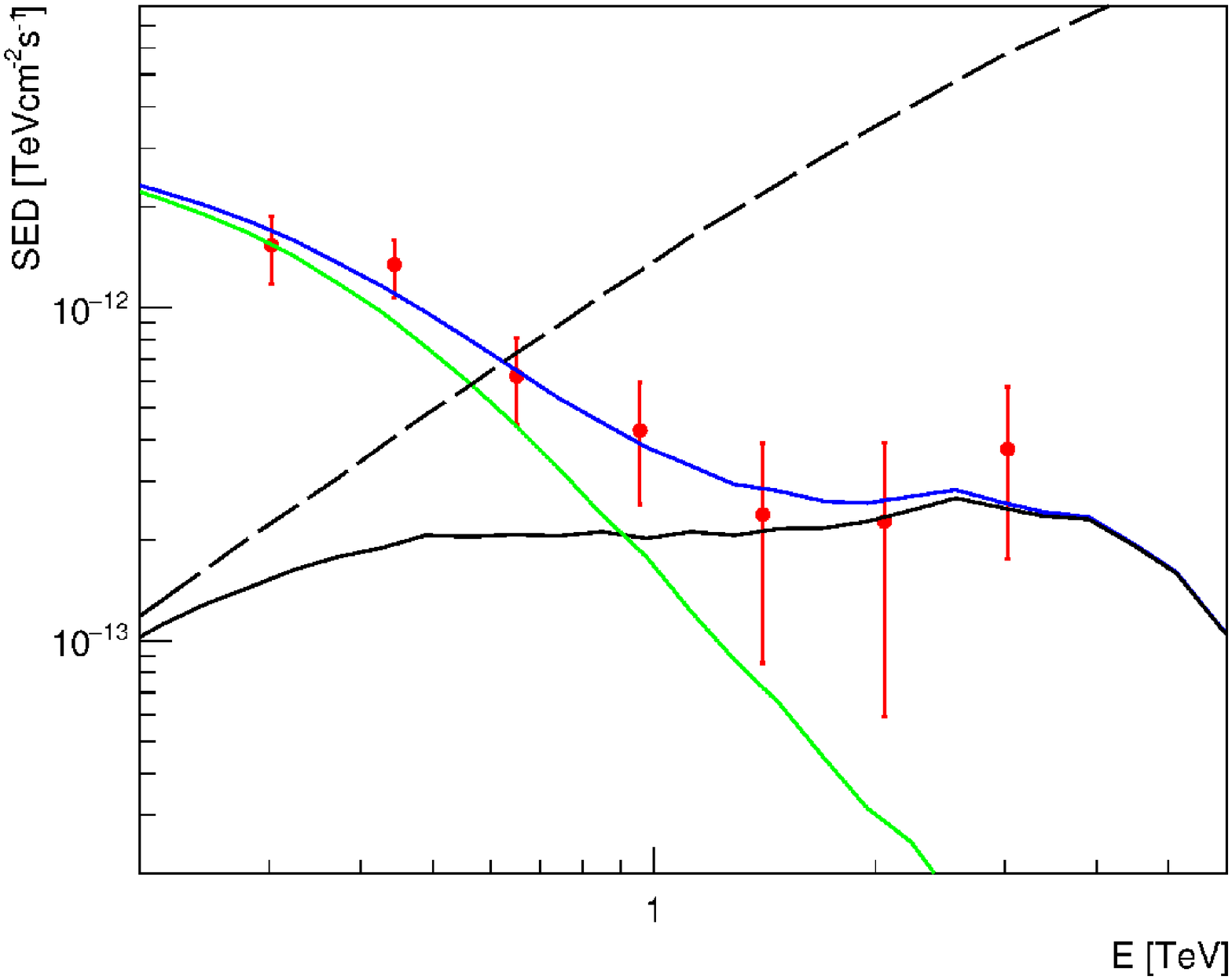}}
\centerline{\includegraphics[width=0.50\textwidth]{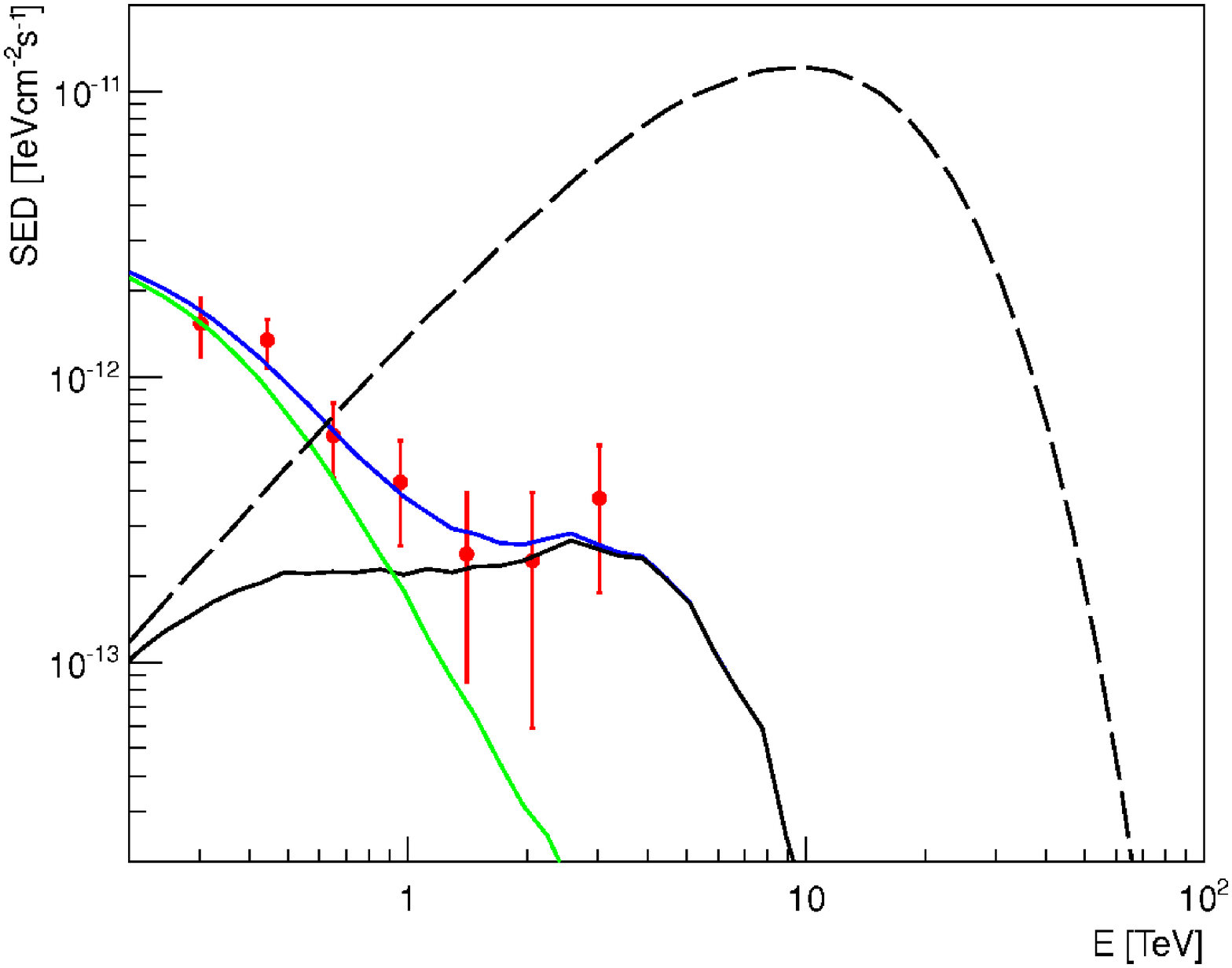}\includegraphics[width=0.50\textwidth]{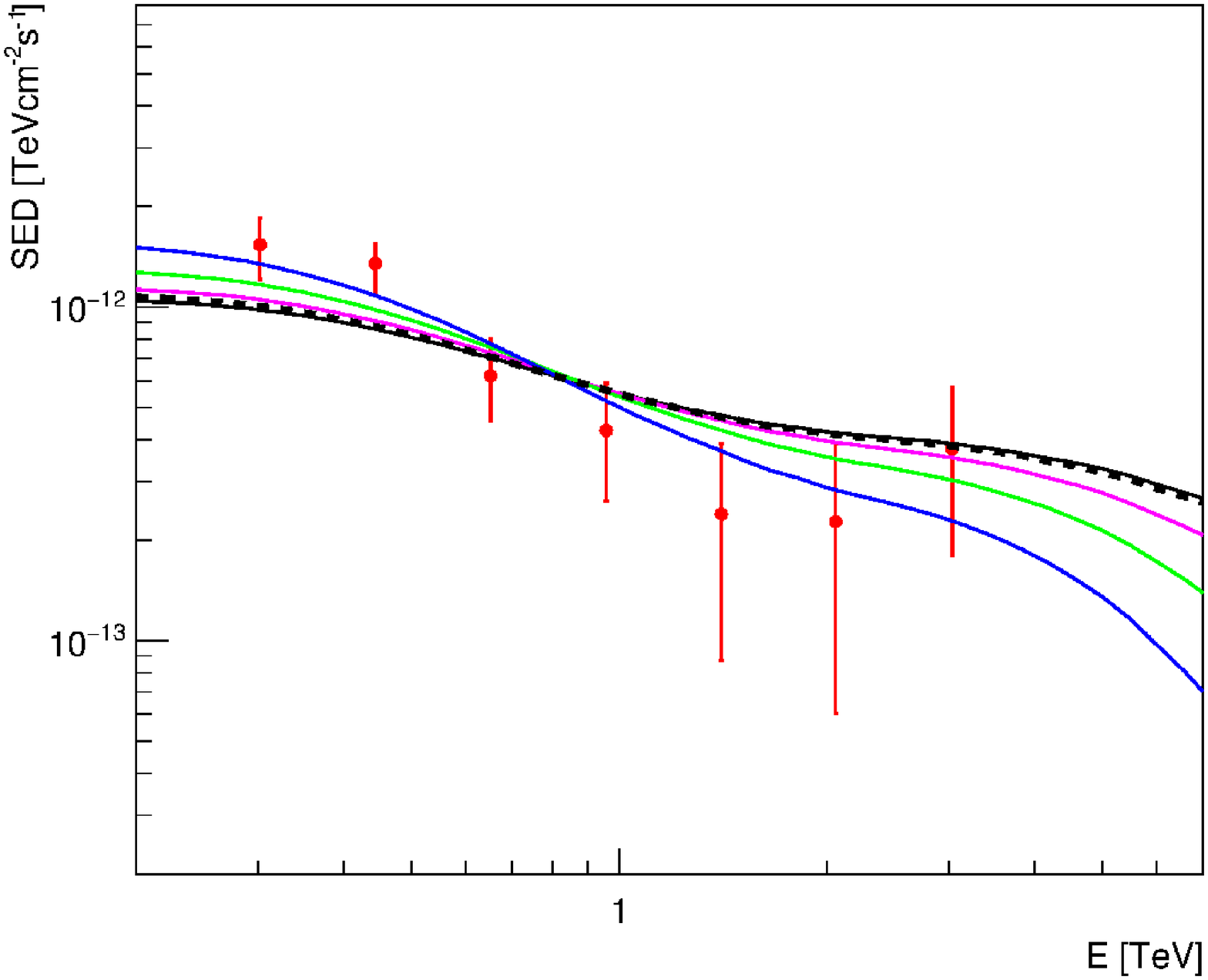}}
\centerline{\includegraphics[width=0.50\textwidth]{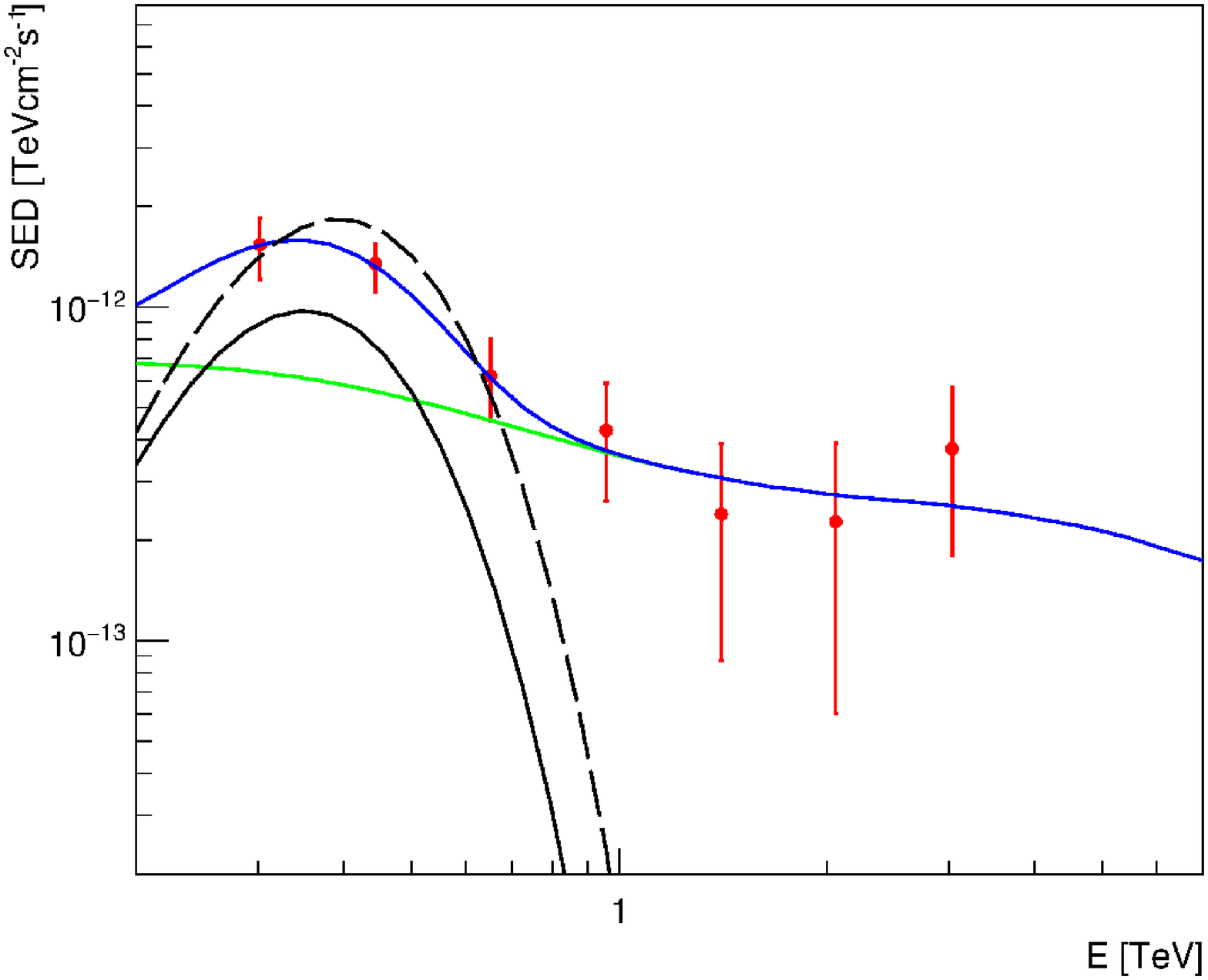}\includegraphics[width=0.50\textwidth]{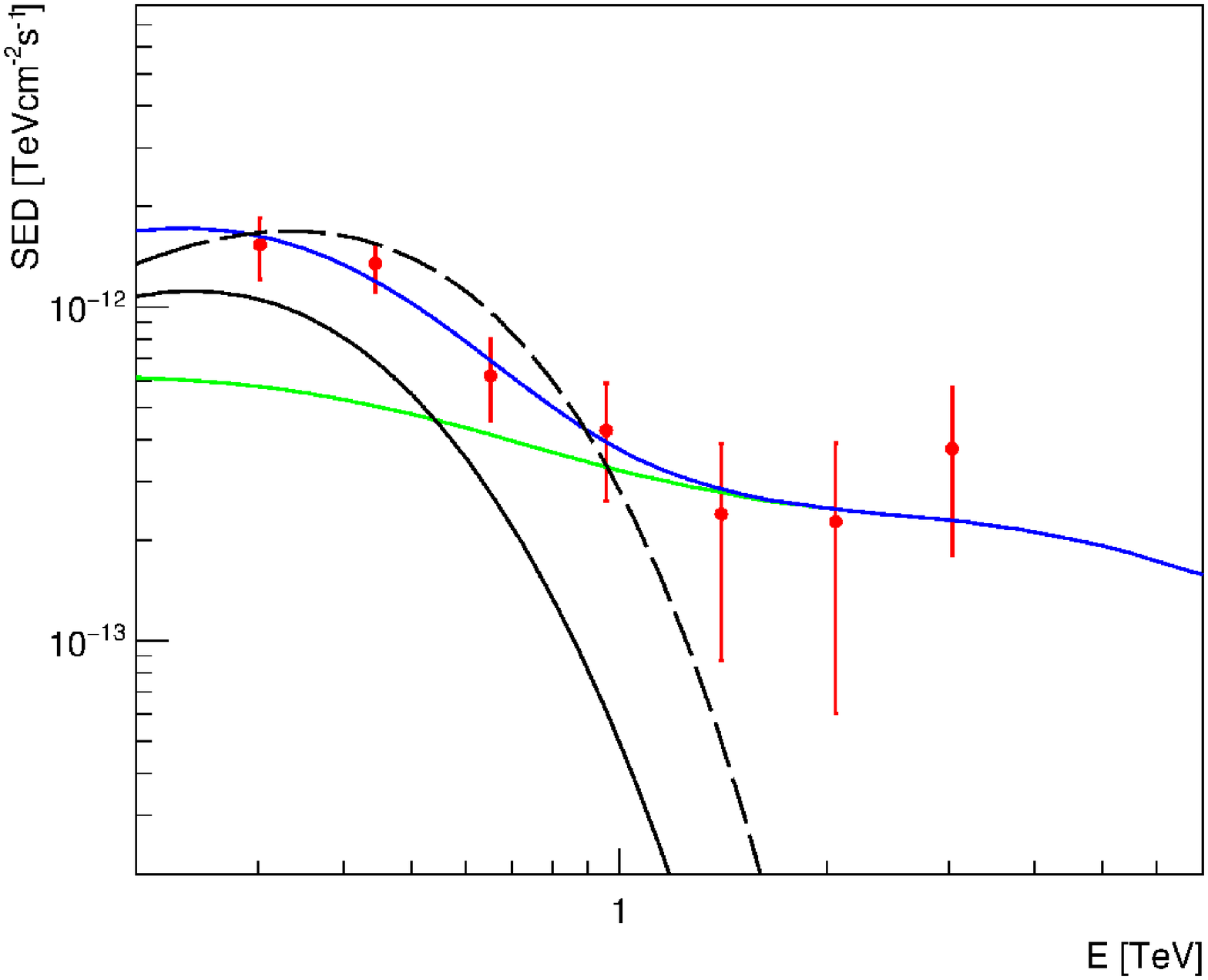}}
\caption{Model fits for the case of 1ES 0347-121. Red circles denote measurements, bars --- their uncertainties; long-dashed lines --- primary $\gamma$-ray spectra. Top-left --- absorption-only model: black --- G12 EBL model, green --- F08, blue --- KD10 model as implemented in ELMAG; solid lines --- observable spectra. Top-right --- electromagnetic cascade model: blue solid line denotes observable spectrum, black --- absorbed component, green --- cascade component. Middle-left: the same as for top-right, but showing the high-energy cutoff of the primary spectrum. Middle-right --- basic hadronic model: black solid line denotes $z_{c}$= 0, magenta --- $z_{c}$= 0.02, green --- $z_{c}$= 0.05, blue --- $z_{c}$= 0.10; short-dashed line --- power-law spectrum of primary protons with $z_{c}$= 0. Bottom-left --- modified hadronic model, the same meaning of colors as in the middle-left panel. Bottom-right --- another fit for the case of modified hadronic model.}\label{fig12}
\end{figure}

The result of fitting for the absorption-only model is shown in Fig.~\ref{fig12}, top-left panel. We note that this and the following spectra are presented in the form of histograms with dense energy sampling using narrow model bins rather than somewhat sparse experimental histograms that were in fact computed during the optimization procedure. Together with fits for the case of G12 and F08 EBL models, we present a similar result for the case of EBL model as implemented in ELMAG. To do this, we evaluated optical depth vs. energy for this model as:
\begin{equation}
\tau_{i}= ln\left(\frac{N_{Prim-i}}{N_{Abs-i}}\right),
\label{eqn12}
\end{equation}
where $i$ is the number of bin in the energy histogram; $N_{Prim}$ is the histogram of primary photons, but redshifted in order to account for adiabatic losses; $N_{Abs}$ is the histogram of detected photons, that were not absorbed (i.e. $i$ corresponds to certain narrow interval of observable energies for both histograms). Appendix C demonstrates that for the case of KD10 model the ELMAG code slightly (about 10-20 \%) overestimates $\tau$ in the 1-10 $TeV$ energy range. However, as the actual EBL level is still uncertain, this EBL model that differs from the original KD10 model is still perfectly acceptable for our study.

Top-right panel of Fig.~\ref{fig12} presents a fit in the framework of the electromagnetic cascade model of Dzhatdoev \cite{dzhatdoev15a}, Dzhatdoev \cite{dzhatdoev15b}. To obtain this fit, we generated a large array of cascades from primary gamma-rays using the ELMAG code with power-law spectrum $dN_{base}/dE_{0} \propto E_{0}^{-1}$ and primary energies from 100 $GeV$ to 100 $TeV$. Observable spectra for any other primary spectrum of $\gamma$-rays $dN_{prim}/dE_{0}$ may be calculated using a re-weighting procedure with a weight defined as:
\begin{equation}
W_{s}(E_{0})=\frac{dN_{prim}/dE_{0}}{dN_{base}/dE_{0}}.
\label{eqn13}
\end{equation}

To test this spectrum synthesis procedure we present a comparison graph (see Appendix D, Fig.~\ref{figD1}) of spectra calculated by Vovk et al. \cite{vovk} using the F08 EBL model and our results obtained with the primary spectrum parameters from Vovk et al. \cite{vovk}. Notwitstanding slightly different EBL models, the agreement between the total observable model spectra is rather good.

As for the case of the absorption-only model, we generated histograms of observable spectra for a set of parameters $(\gamma,E_{c})$ and performed optimization over these parameters, this time evaluating $\chi^{2}$ on a grid of 10$\times$10 cells with exhaustive calculation. The values of $(\gamma,E_{c})$ that ensure the minimal value of $\chi^{2}$ on this grid were found, and the optimization run was repeated on a similar grid, but with a smaller step $(\delta \gamma,\delta E_{c})$. This procedure was repeated several times until $\delta \gamma<10^{-2}$ and $\delta E_{c}< 0.1$ $TeV$. The step decreasing factor was set to 3. We have put considerable efforts into ensuring that the global minimum of $\chi^{2}$ is captured by our optimization method. 

The best-fit spectrum consists of two distinct components --- the primary (absorbed) that dominates at high energy, and the cascade one, that has rather steep spectrum and contributes mainly in the low-energy region of the total spectrum. These two components, when put together, produce an ankle-like spectral feature at an energy around 1 $TeV$. The same fit, but with different energy and intensity scale, is shown in Fig.~\ref{fig12}, middle-left panel. It is clearly seen that the cutoff in the primary spectrum is situated well below 100 $TeV$.

Fig.~\ref{fig12}, lower-left panel was calculated for the case of the ``basic hadronic cascade model'' of Essey et al. \cite{essey10b}, where all observable $\gamma$-rays are of secondary nature, i.e. were produced by protons on the way from the source to the observer. Apart from the option without any considerable EGMF on the line-of-sight, we also include several options with a possible magnetic field structure at $z_{c}$, as was described in Subsubsection 2.2.2. Such calculations presented in this section were performed in the weak universality approximation with $E_{p0}$= 30 $EeV$. The case of power-law spectrum of primary protons from 1 $EeV$ to 100 $EeV$ with $dN_{p0}/dE_{p0} \propto E_{p0}^{-2}$ is also included to this Figure.

The case of the ``modified hadronic cascade model'' (Essey \& Kusenko \cite{essey14}) that includes both primary and secondary components is shown in two lower panels of Fig.~\ref{fig12}. Optimization was performed on 4 parameters: the normalization factors of the primary $\gamma$-ray and cascade components, and the parameters of the shape of the primary component $(\gamma,E_{c})$ (see equation (11)).

\onecolumn
\twocolumn

Exhaustive search is hardly possible on such a four-dimensional grid; therefore, we have developed a code in the ROOT framework using the Minuit package to perform gradient optimization. The formal best fit obtained with this procedure is presented in Fig.~\ref{fig12}, bottom-left. For the case of the primary $\gamma$-ray component we neglected weak cascade emission produced by this component. In this case the spectrum of the primary component appears to be very narrow, with sharp lower- and higher-energy cutoffs. In Fig.~\ref{fig12}, bottom-right we show, however, that even in the case of a more realistic primary component a good fit to the observed SED is obtainable.

Now let us introduce a quantity called the modification factor (following Berezinsky et al. (2006)) also known as the flux boost factor (Sanchez-Conde et al. (2009)) defined as the ratio between the spectra in the electromagnetic cascade model (ECM) and the absorption-only model (AOM):
\begin{equation}
K_{B}(E)=\frac{\left(dN/dE\right)_{ECM}}{\left(dN/dE\right)_{AOM}}.
\label{eqn14}
\end{equation}

\begin{figure}
\centering
\includegraphics[width=8cm]{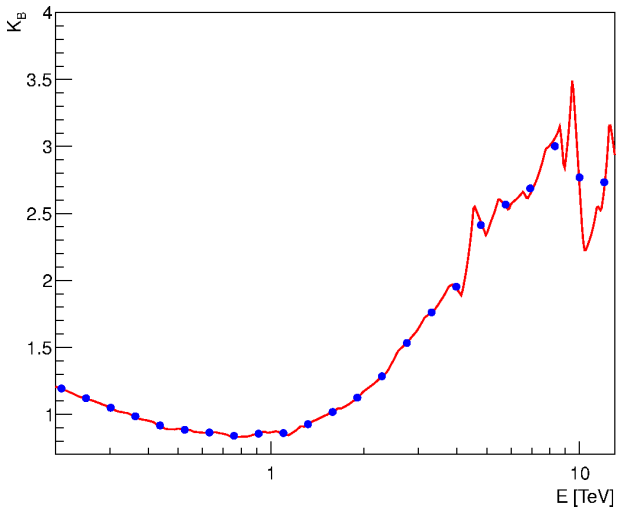}
\caption{Flux boost factor $K_{B}(E)$ for 1ES 0347-121 (red curve). Blue circles --- $K_{B}(E)$ averaged over energy intervals to suppress statistical fluctuations.}
\label{fig13}
\end{figure}

The graph of $K_{B}(E)$ for blazar 1ES 0347-121 and KD10 EBL model, as implemented in the ELMAG code, is shown in Fig.~\ref{fig13}. For calculations of the boost factor, $\left(dN/dE\right)_{ECM}$ and $\left(dN/dE\right)_{AOM}$ were interpolated to 1000 bins. $K_{B}(E)$ is clearly greater than unity for $E$>2 $TeV$; the ratio of the maximal to the minimal values of $K_{B}(E)$ is about 3.5. Therefore, electromagnetic cascade model predicts that the intensity in the optically thick region may significantly exceed the one deduced from the fitting of the optically-thin part of experimental SED in the framework of the absorption-only model. Indeed, the cascade component, dominating at low energies, ensures good quality of fit at these energies notwithstanding a very hard primary spectrum. The cascade component, in effect, conceals, or ``masks'', the primary component in the optically-thin energy region (hence the name of the present paper). On the other hand, at high energies where $\tau$>2, a very hard primary spectrum provides enough photons to explain observations even after extinction caused by the EBL.

\subsection{The case of 1ES 0229+200}

\begin{figure}[t]
\centerline{\includegraphics[width=0.50\textwidth]{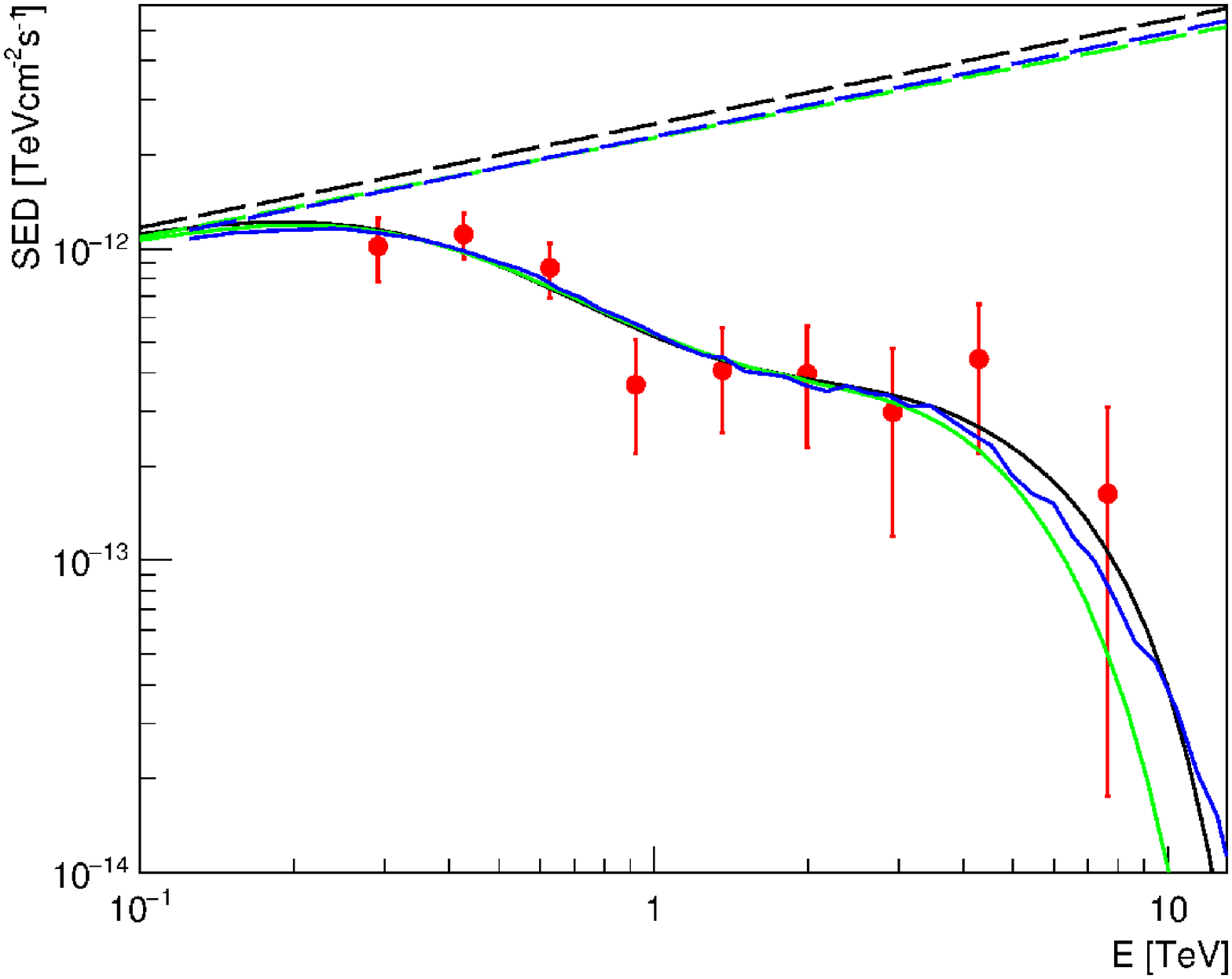}\includegraphics[width=0.50\textwidth]{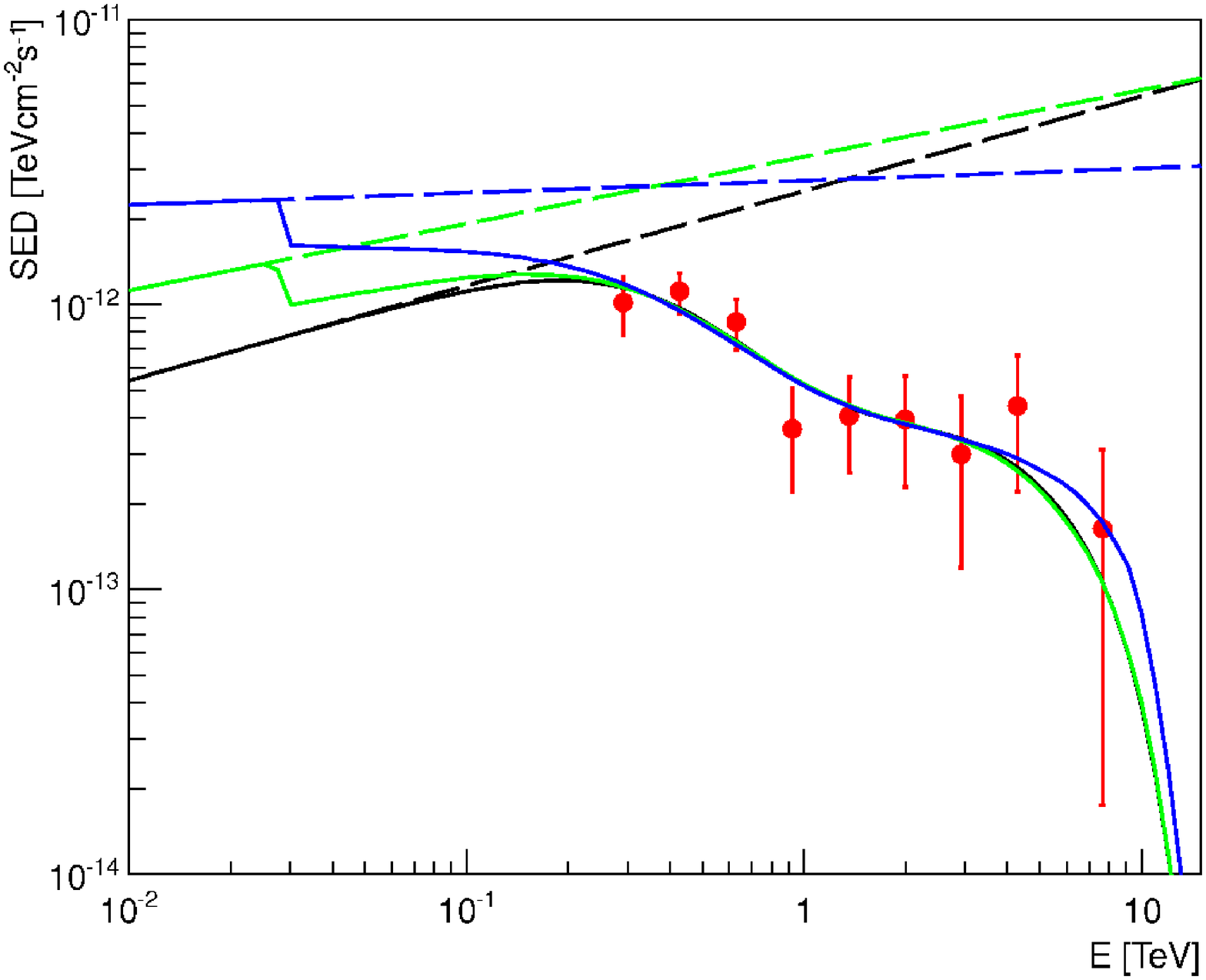}}
\centerline{\includegraphics[width=0.50\textwidth]{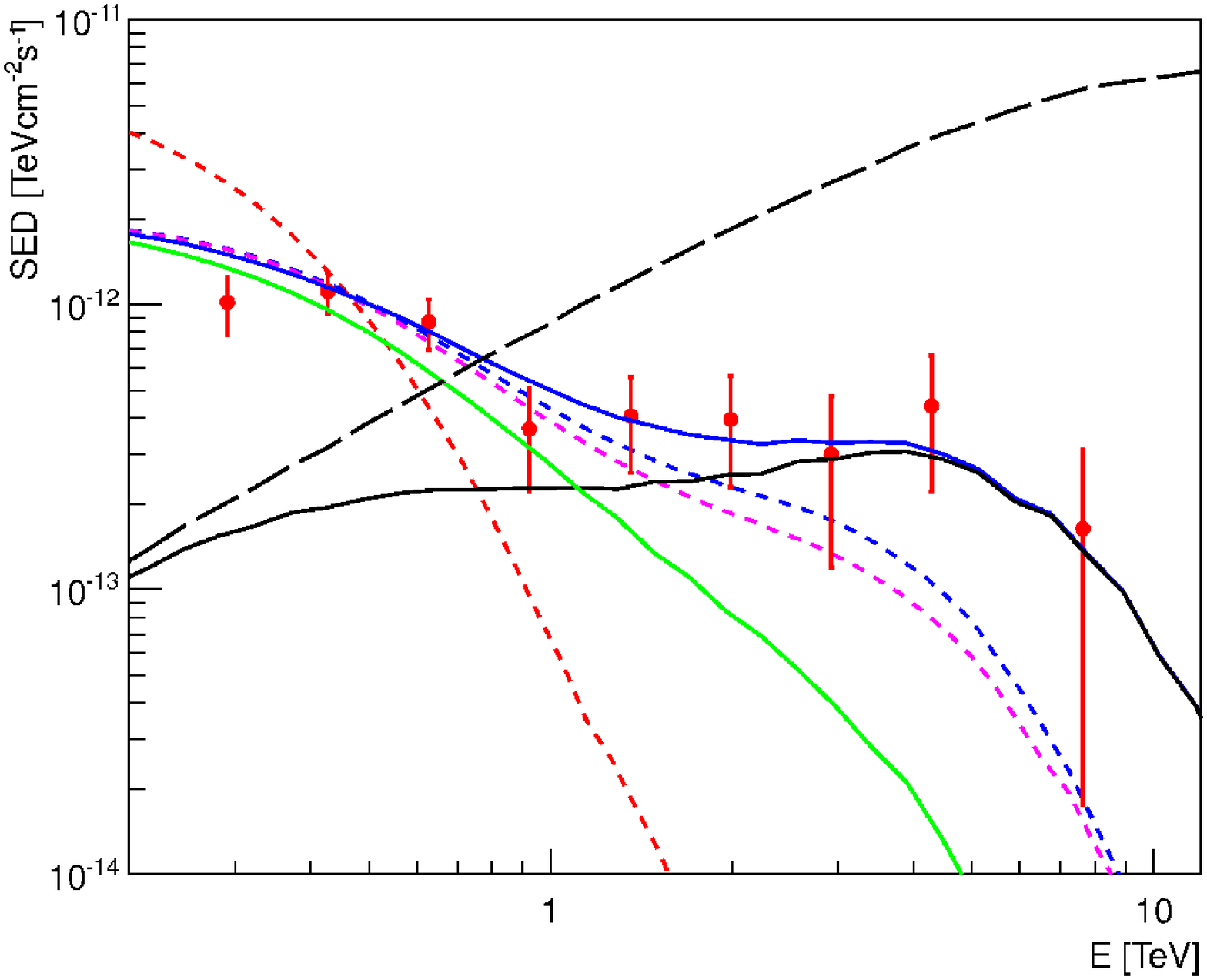}\includegraphics[width=0.50\textwidth]{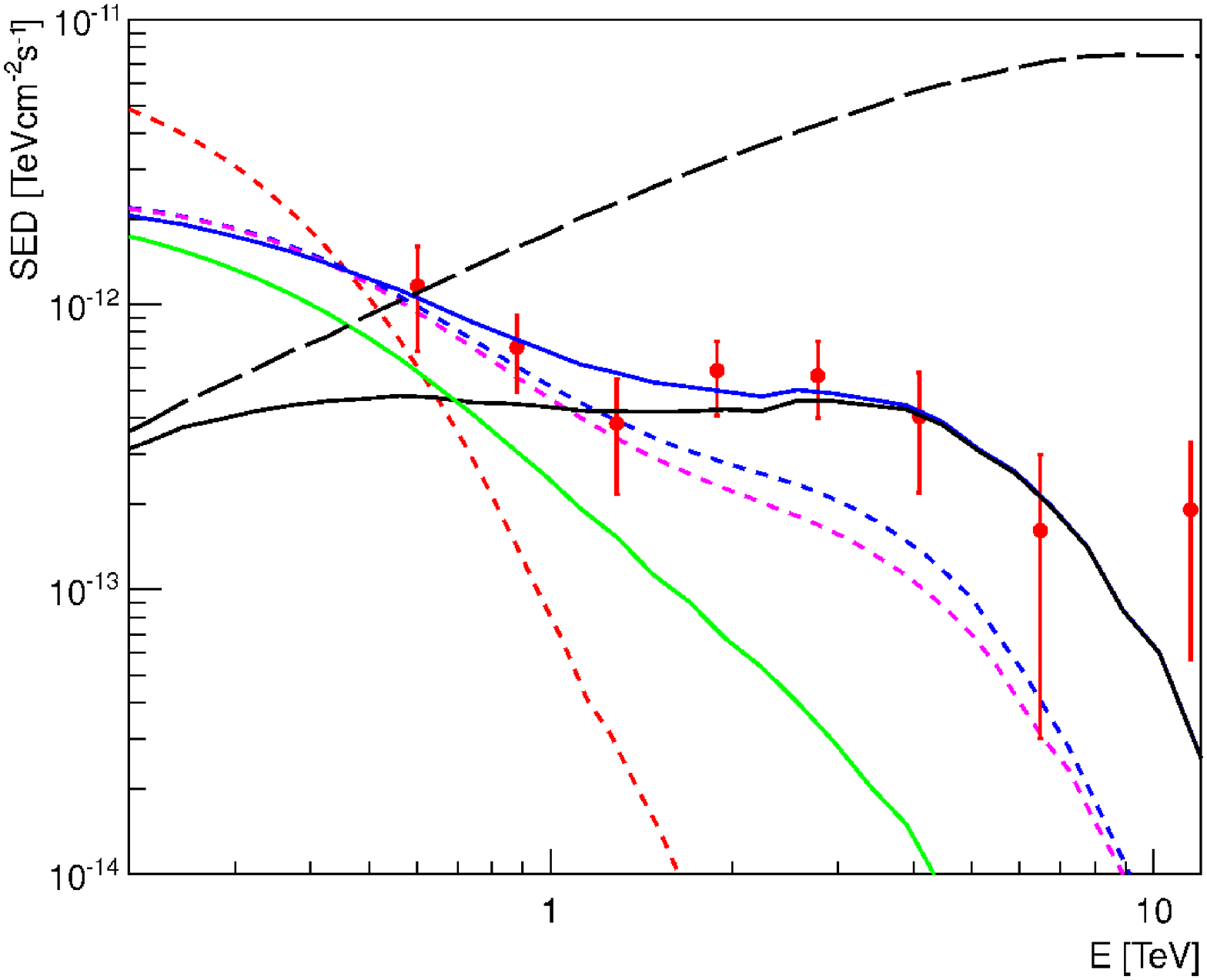}}
\centerline{\includegraphics[width=0.50\textwidth]{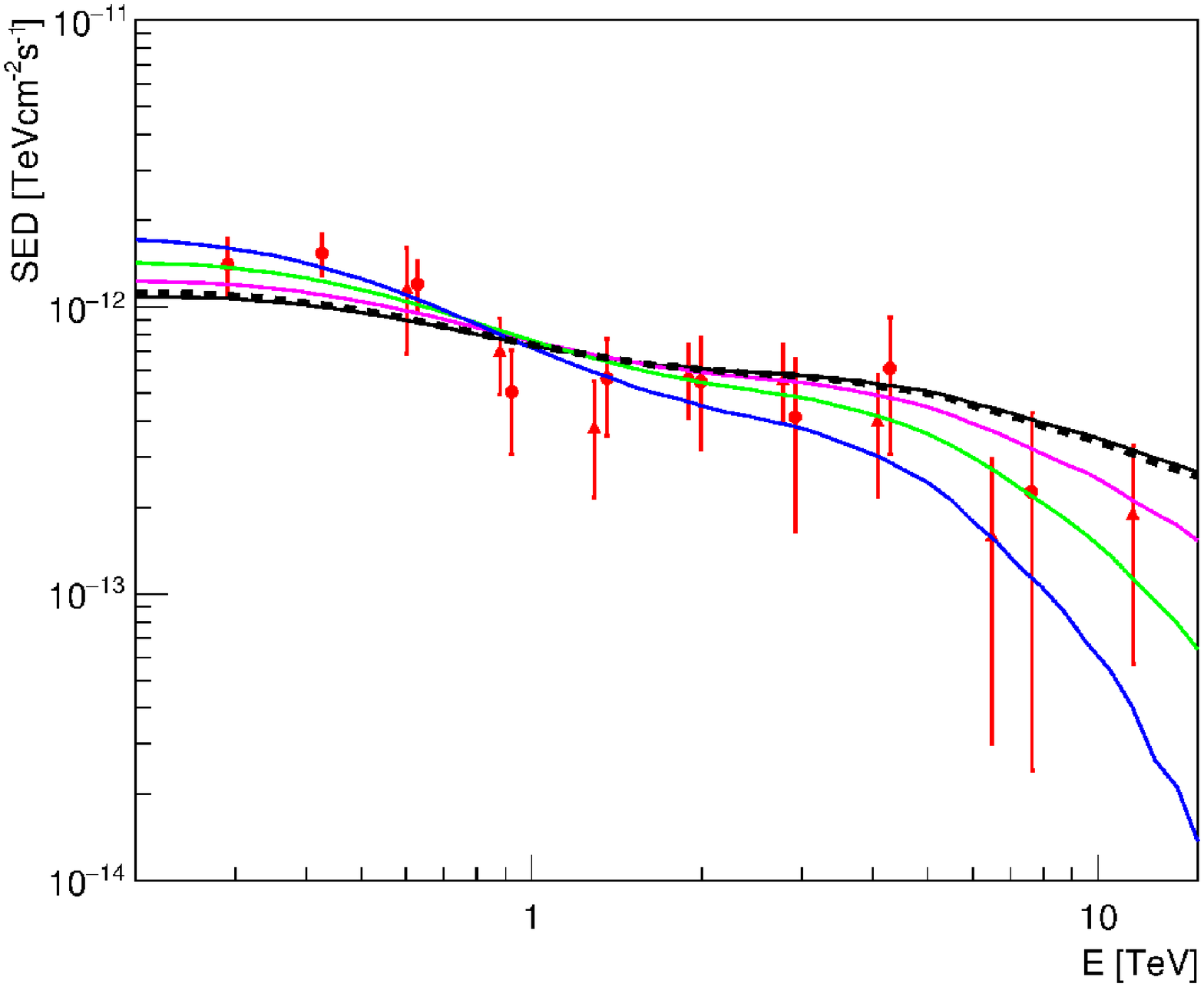}\includegraphics[width=0.50\textwidth]{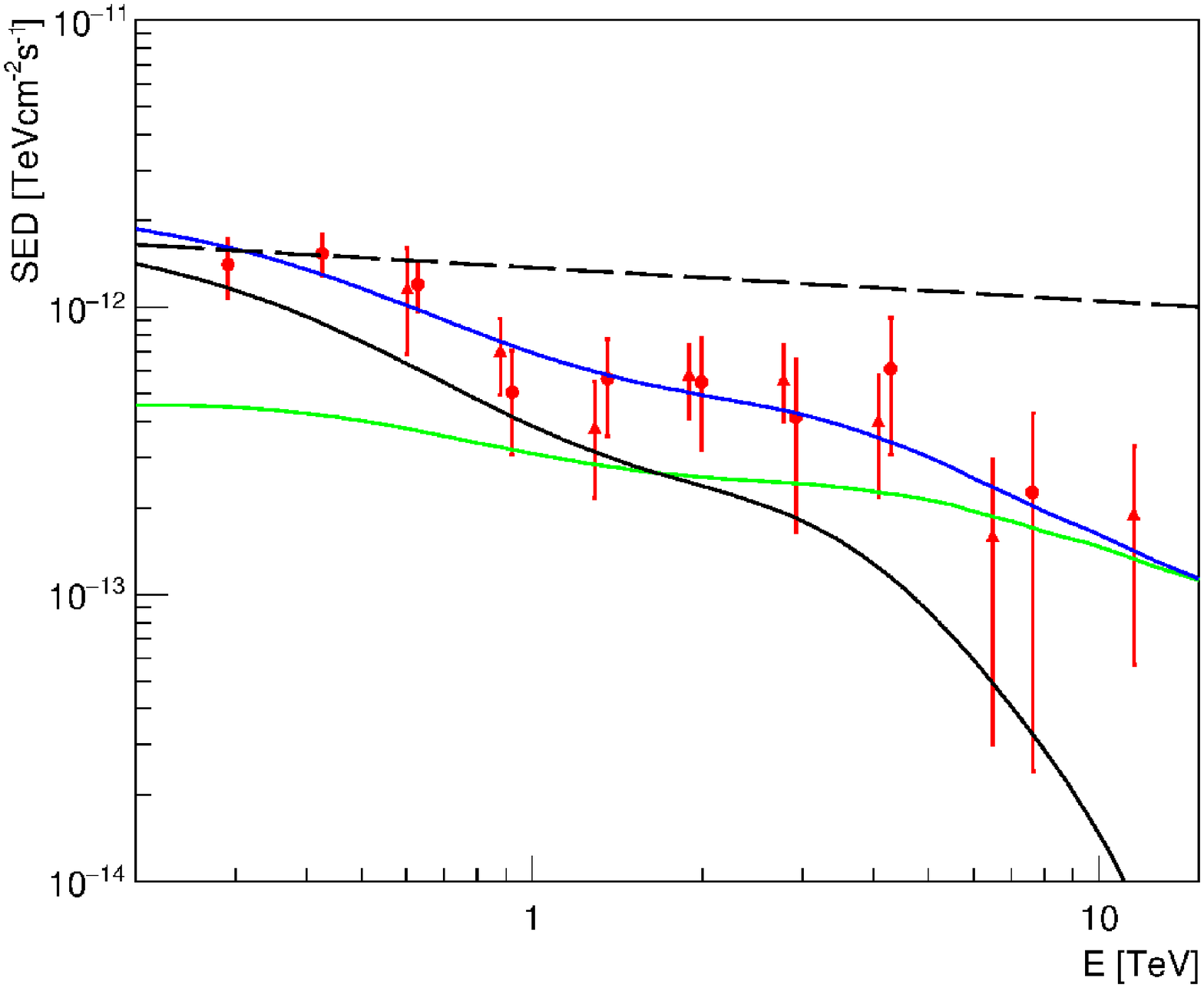}}
\caption{Model fits for the case of 1ES 0229+200. Top-left --- absorption-only model, all notions are the same as in Fig.~\ref{fig11}, top-left. Top-right --- comparison of the absorption-only and $\gamma$-ALP mixing models: black line denotes absorption-only model, green --- weak $\gamma$-ALP flux enhancement, blue --- strong $\gamma$-ALP flux enhancement. Middle panels --- electromagnetic cascade model for the case of the VERITAS (left) and HESS (right, red triangles) observations together with cascade spectra for monoenergetic primary $\gamma$-ray injection (short-dashed lines): red --- $E_{\gamma0}$= 10 $TeV$, blue --- $E_{\gamma0}$= 100 $TeV$, magenta --- $E_{\gamma0}$= 1 $PeV$. Bottom-left --- basic hadronic cascade model ($z_{c}$ are the same as in Fig.~\ref{fig12} middle-right); bottom-right --- modified CR beam model.}\label{fig14}
\end{figure}

Blazar 1ES 0229+200 is a frequent source in many discussions of extragalactic $\gamma$-ray propagation due to its hard spectrum, slow variability, and comparatively high total flux. The fits for the case of the absorption-only model, which are similar in every respect to those presented in Fig.~\ref{fig12} for 1ES 0347-121, are shown in top-left panel of Fig.~\ref{fig14}.

For this source we also present two fits for a model that includes the $\gamma \rightarrow ALP$ oscillation process (top-right panel of Fig.~\ref{fig14}), but without account of any secondary (cascade) emission. As was discussed in Introduction, extra photons with respect to the case of the absorption-only model in the optically-thick region of the spectrum may be ascribed to this process. To evaluate the effect of $\gamma$-ALP mixing in the spectrum of 1ES0229+200, we used the results of Sanchez-Conde et al. (2009) that were calculated for the case of similar $z$= 0.116. Sanchez-Conde et al. (2009) presented graphs for boost factor (Fig. 7 of their work) for the case of Primack et al., \cite{primack} EBL model, as well as for the Kneiske et al. \cite{kneiske} EBL model. In the former case, the values of $K_{Boost}$ in the optically-thick region are typically smaller than in the latter case. These two cases (weak and strong flux enhancement, respectively) will serve to qualitatively demonstrate more or less strong $\gamma$-ALP mixing effects. However, in Fig.~\ref{fig12} we actually use fixed EBL model G12.

More detailed study of $\gamma$-ALP mixing is underway and will be reported elsewhere. Given huge uncertainties of EGMF strength and a vast room for the $\gamma$-ALP mixing parameter values, we find it appropriate to use these estimates, even though they were performed for the case of a slightly different redshift and different EBL models.

It is interesting that the case of weak mixing is practically indistinguishable from the absorption-only model fit. The primary spectrum in the case of the exotic model is indeed less hard than without it, but that does not induce any significant observable effect in the 300 $GeV$ --- 10 $TeV$ energy range. Moreover, the difference between the spectra in these two cases is clearly insufficient to choose between the models on purely astrophysical grounds. On the other hand, in the low-energy range, together with a spectral irregularity at $E$=20-30 $GeV$ that was qualitatively discussed in Introduction, another feature is present: namely, the spectral slope for the case of exotic model is steeper than the one for the absorption-only fit, reflecting the shape of the primary spectrum. Therefore, it appears that for the case of weak mixing the observable spectrum is the same or even steeper at all energies from the above-mentioned irregularity up to 10 $TeV$, and never harder. Up to our knowledge, this highly unexpected result was never reported before.

For the case of strong mixing the situation is different --- the intensity of the observable model spectrum at $E>$3 $TeV$ is greater than the ones in two other cases, while all other features of the spectrum are qualitatively similar. However, according to Ajello et al. (2016), this strong mixing regime was strongly constrained considering the non-detection of the low-energy spectral irregularity, as was discussed in Introduction.

The fits for the case of electromagnetic cascade model are presented in the middle panels of Fig.~\ref{fig12} for the case of VERITAS (left) and HESS (right) observations. In both cases the ankle feature, which is formed by the intersection of the primary (absorbed) and secondary (cascade) components, is clearly seen. The position of the ankle is similar for both cases, again indicating that this feature, however faint, is not just a statistical fluctuation. For comparison, cascade spectra for the case of monoenergetic primary injection with various energies are also shown. Finally, low panels of Fig.~\ref{fig14} were computed for the case of the basic (left) and modified (right) hadronic cascade models. Both VERITAS and HESS data are presented in the picture, with normalization to the HESS data.

\onecolumn
\twocolumn

These fits were obtained using exactly the same procedure as those presented in Fig.~\ref{fig12}, middle-right and bottom-left. In Appendix D, Fig.~\ref{figD2} we present a comparison graph of spectra calculated for basic hadronic model by Essey et al. (2011), Murase et al. (2012), and by us for the case of $z_{s}$=0.14.

For this source we also present several fits for the case of a simple model of structured EGMF. Namely, following Furniss et al. \cite{furniss}, we use the voidiness parameter $V_{LoS}$ defined as the fraction of the (comoving) line-of-sight distance covered by underdense regions of space (voids) to the total line-of-sight distance from the source to the observer. In fact, a spatial region that is comparatively close to the source (typically 10-100 $Mpc$ from the source) is the most important for the cascade development, for it contains most of cascade electrons. EGMF strength in voids may be sufficiently low to allow cascade electrons radiate secondary photons before these electrons are strongly deflected from the line-of-sight. On the contrary, denser regions of space, as compared to voids, typically contain comparatively strong magnetic fields $B\sim 10^{-6}-10^{-11}$ $G$; cascade electrons produced in these regions are strongly deflected and delayed, and secondary photons do not contribute to the observable spectrum. In effect, to account for possible structures with strong magnetic field, we multiply the cascade component to the suppression factor $0<K_{V}<1$.

\begin{figure}
\centering
\includegraphics[width=8cm]{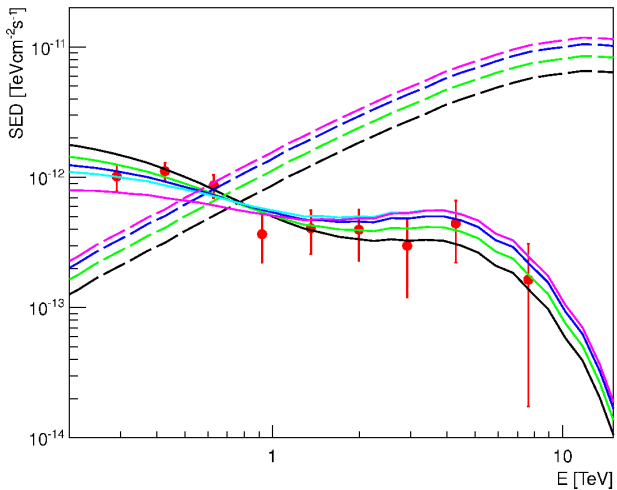}
\caption{Fits to the spectrum of 1ES 0229+200 for electromagnetic cascade model and different values of $K_{V}$. Dashed lines denote primary spectrum near the source, solid lines --- total model spectra; black --- $K_{V}$= 1, green --- 0.6, blue --- 0.4, cyan --- 0.3, magenta --- 0.2.}
\label{fig15}
\end{figure}

\begin{figure}
\centering
\includegraphics[width=8cm]{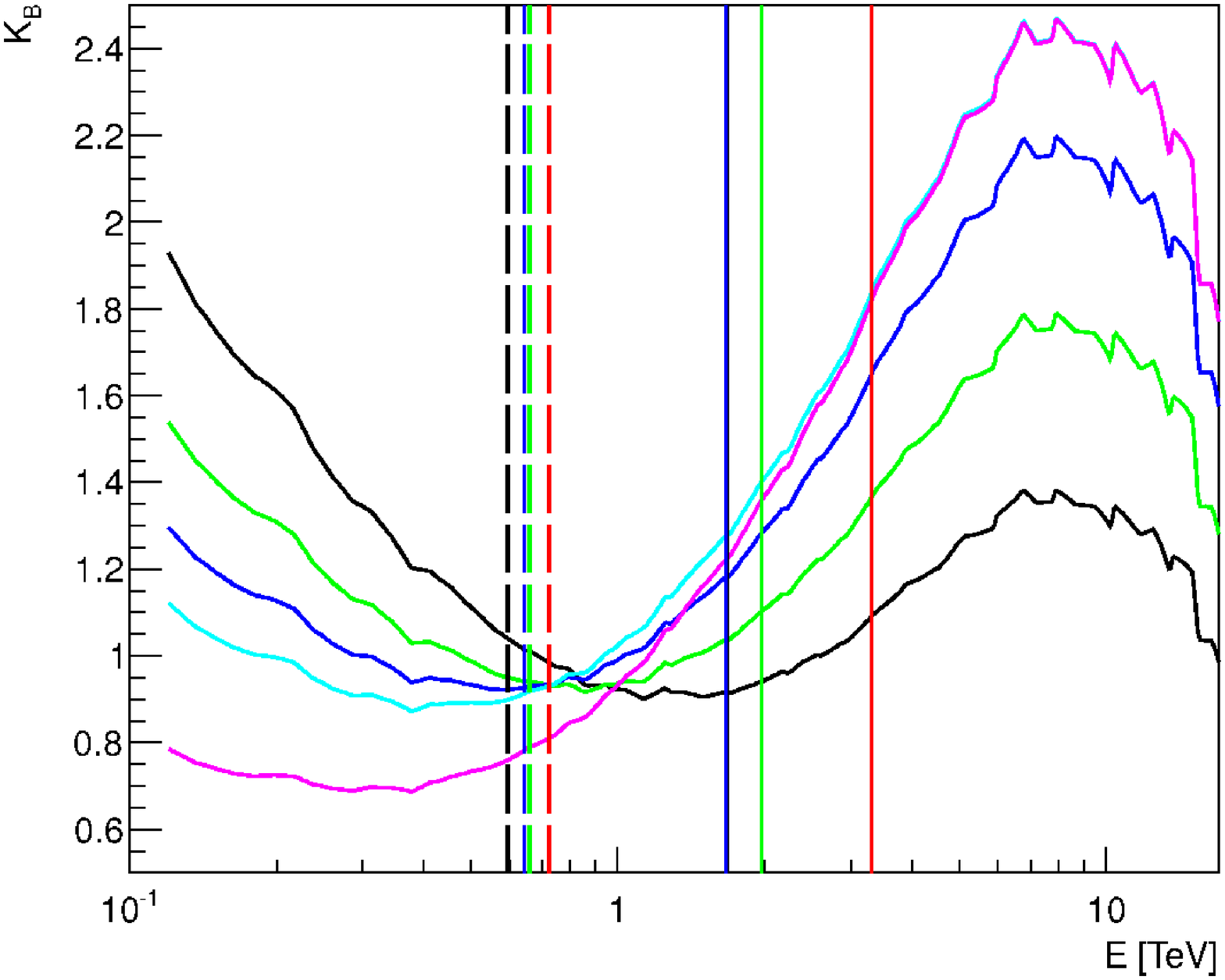}
\caption{Boost factor $K_{B}$ vs. energy for the fits presented in Fig.~\ref{fig13}. Solid lines denote $K_{B}(E)$ for different values of $K_{V}$: black --- $K_{V}$= 1, green --- 0.6, blue --- 0.4, cyan --- 0.3, magenta --- 0.2.}
\label{fig16}
\end{figure}

Several fits to the spectrum of 1ES 0229+200 for different values of $K_{V}$ from 0.2 to 1.0 are presented in Fig.~\ref{fig15}. The corresponding graphs of boost factor for these fits are shown in Fig.~\ref{fig16}. This figure demonstrates that the ankle signature in the total model spectrum does not disappear for $K_{V}<1$, but, on the contrary, only gets more prominent as $K_{V}$ falls in the range of the considered $K_{V}$ values.

\subsection{The case of 1ES 1101-232, 1ES 1812+304, 1ES 0414+009, and H 1426+428}

1ES 1101-232 was observed by the HESS Collaboration in 2004-2005, and the result of spectral measurements was published in Aharonian et al. \cite{aharonian06}. One year later, a reanalysis of these observations was performed, and the new result on the spectrum was presented in Aharonian et al. \cite{aharonian07b}.

\begin{figure}[t]
\centerline{\includegraphics[width=0.50\textwidth]{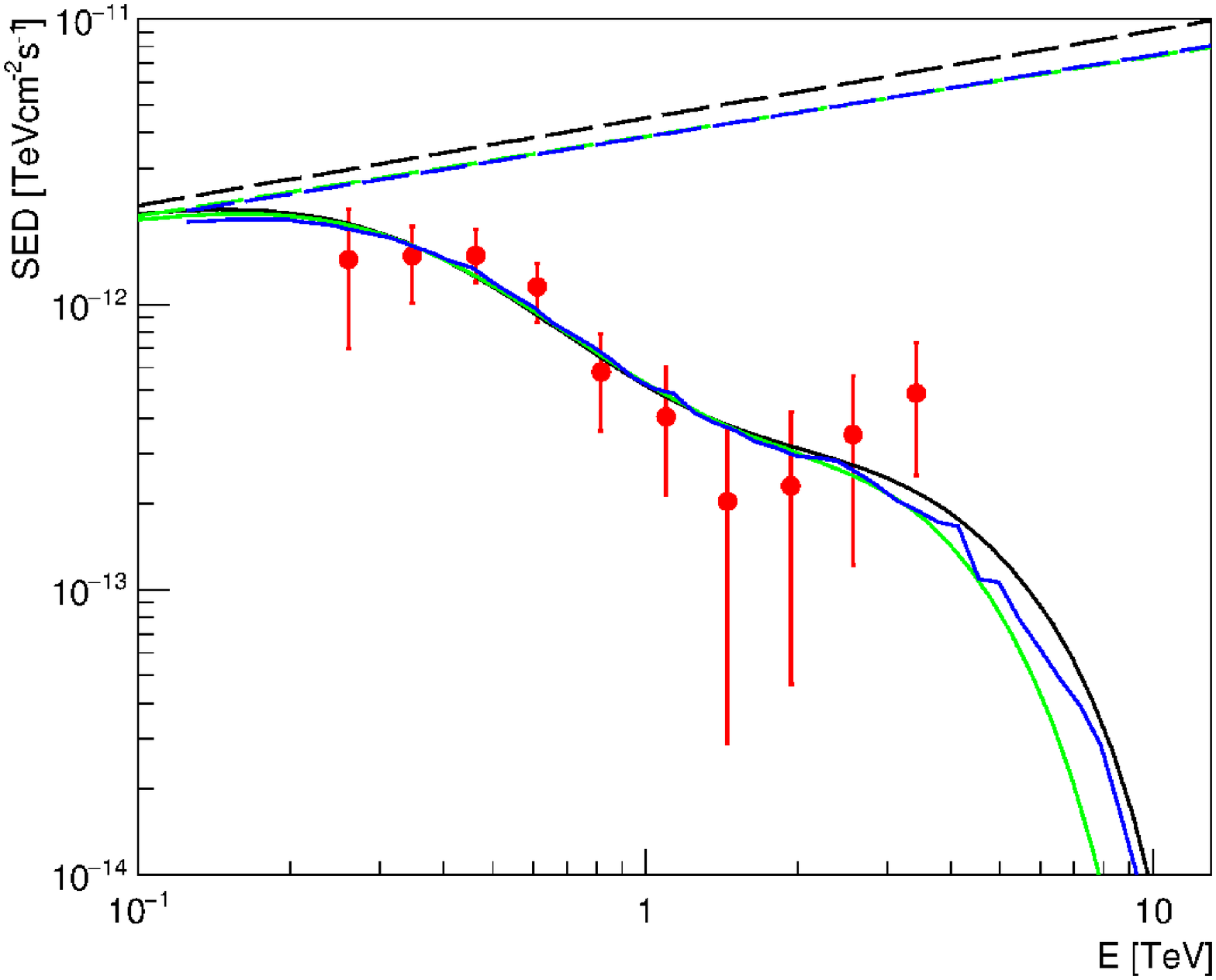}\includegraphics[width=0.50\textwidth]{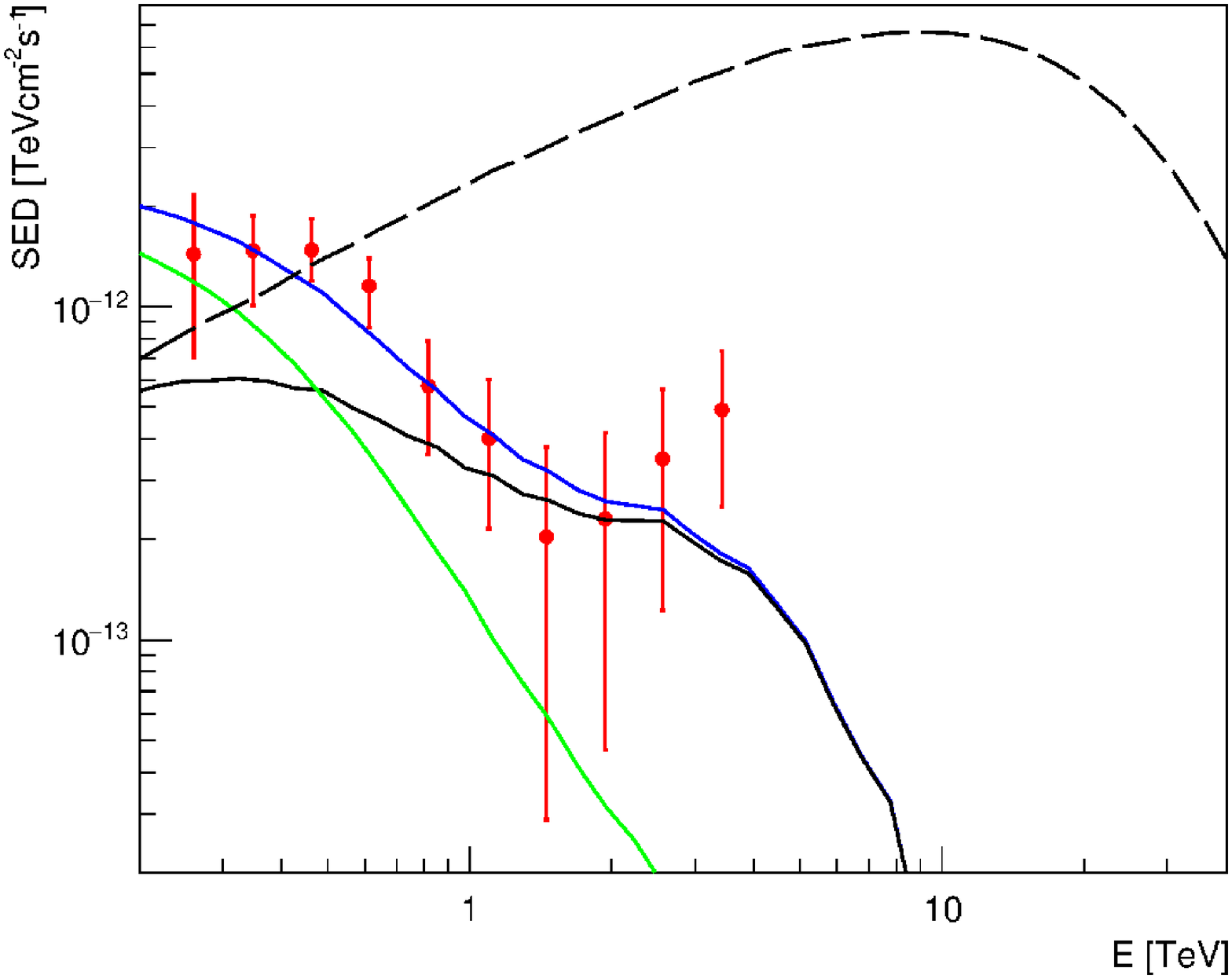}}
\centerline{\includegraphics[width=0.50\textwidth]{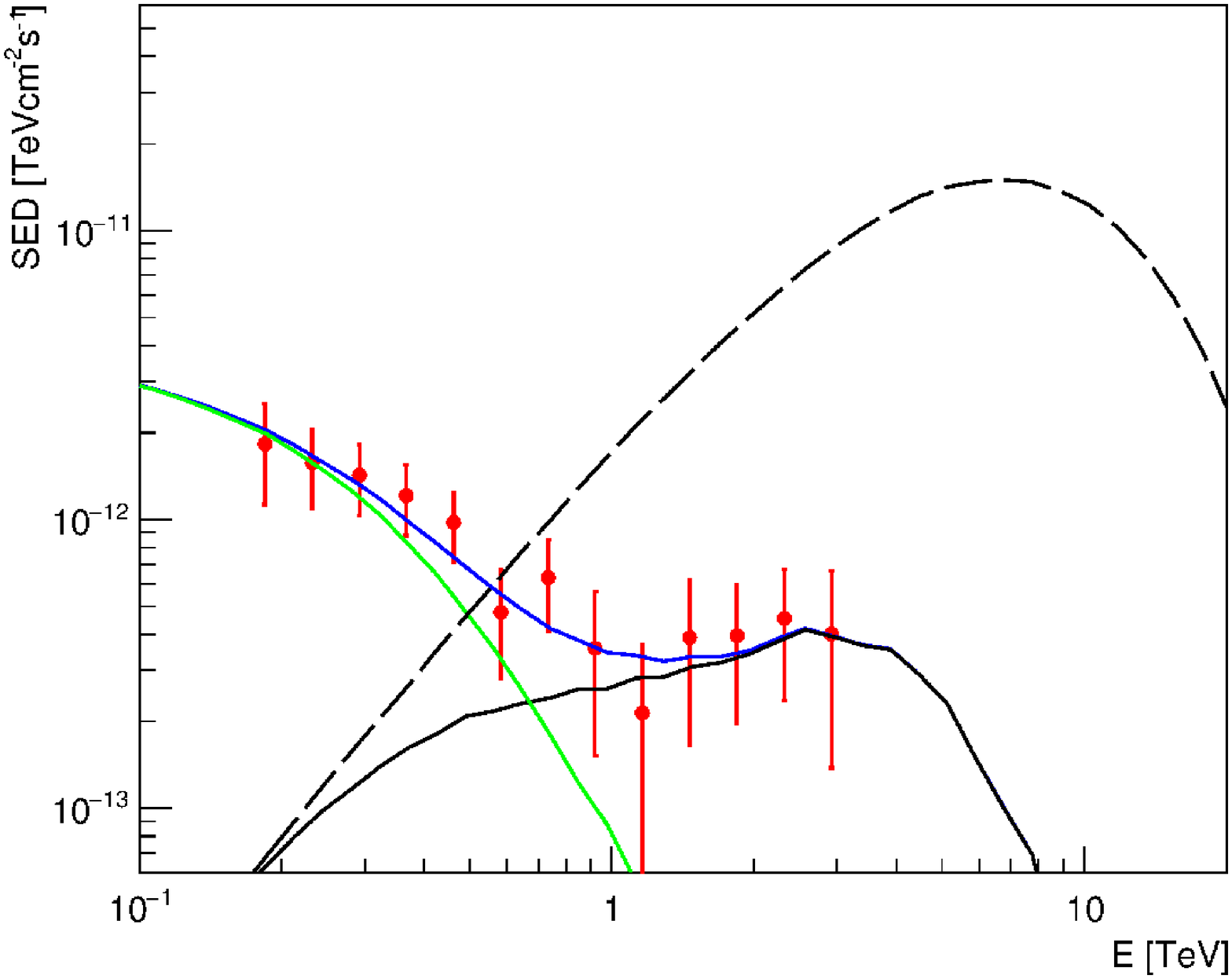}\includegraphics[width=0.50\textwidth]{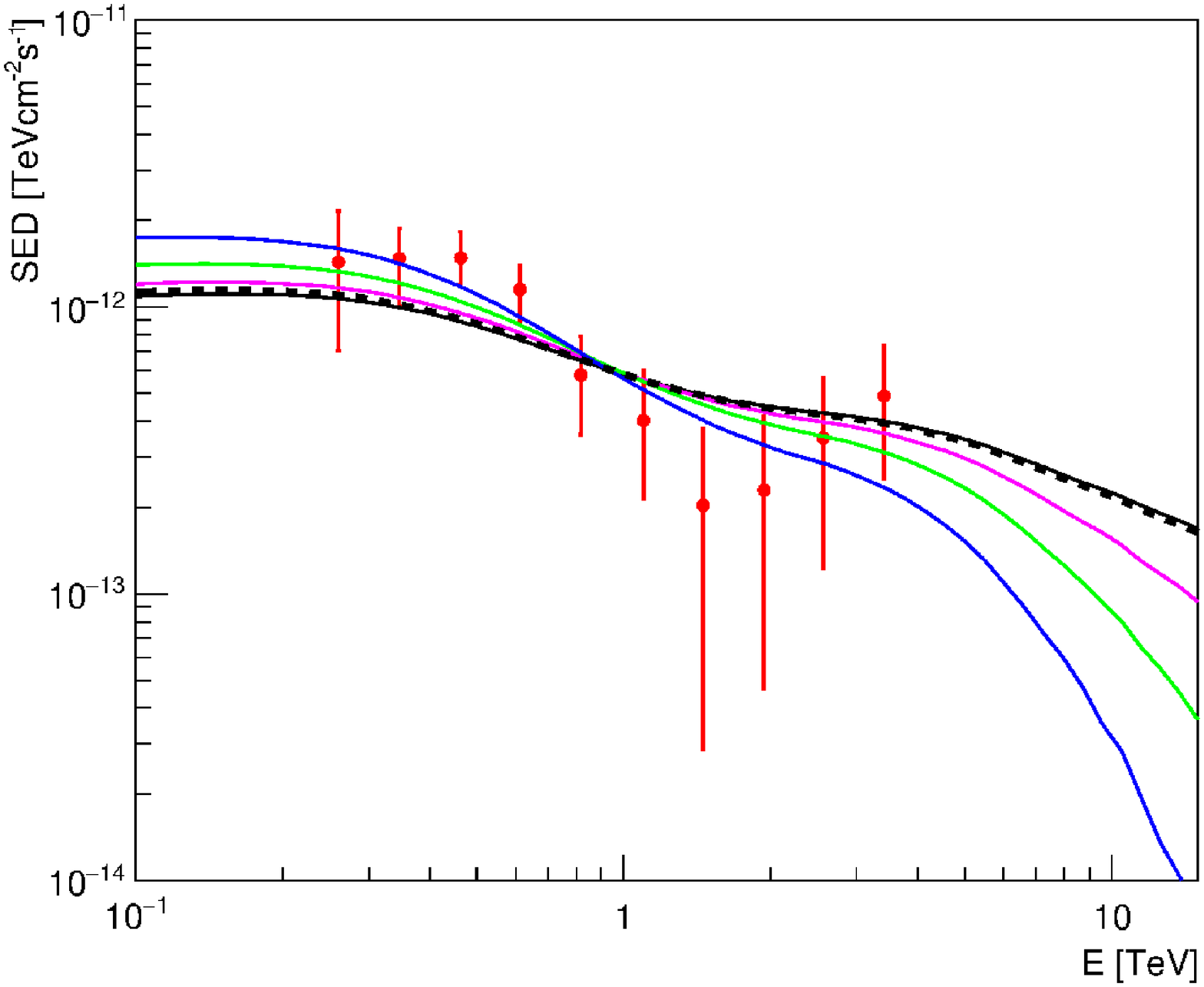}}
\centerline{\includegraphics[width=0.50\textwidth]{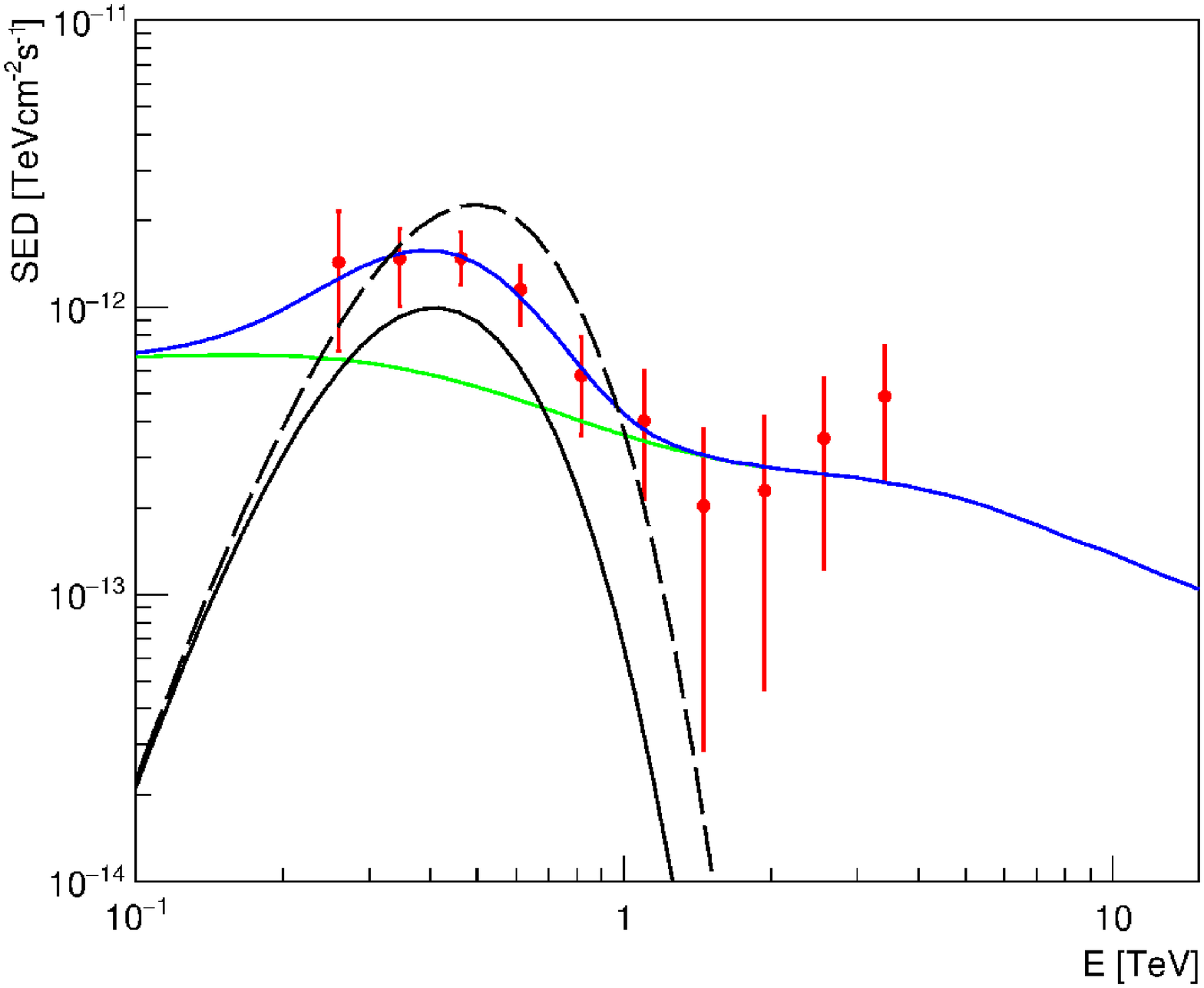}\includegraphics[width=0.50\textwidth]{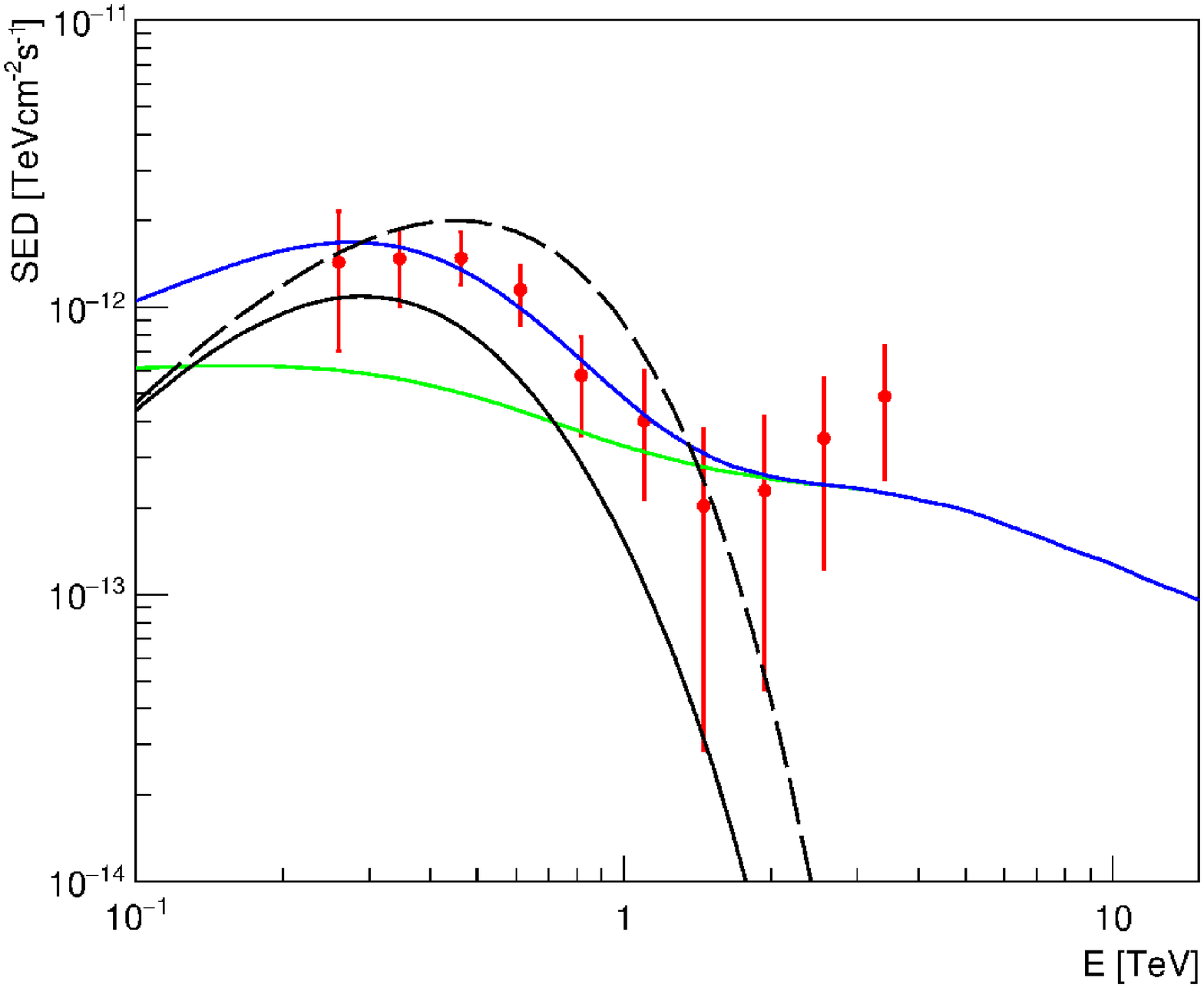}}
\caption{Model fits for the case of 1ES 1101-232. Top --- absorption-only model (left) and electromagnetic cascade model for the case of 2007 reanalysis (right); all notions are as in corresponding panels of Fig.~\ref{fig10}. Middle-left --- electromagnetic cascade model for the case of 2006 analysis. Middle-right --- basic hadronic cascade model ($z_{c}$ are the same as in Fig.~\ref{fig10} middle-right). Bottom-left --- modifed hadronic model (formal best fit). Bottom-right --- another fit for the case of the modifed hadronic model.}\label{fig17}
\end{figure}

In our work we mainly use the latter results, but we also present a fit for the case of the electromagnetic cascade model for the former analysis. Fig.~\ref{fig17}, top-left panel contains a fit to the SED of 1ES 1101-232 in the absorption-only model; top-right and middle-left panels of this figure contain the fits for the case of the electromagnetic cascade model for reanalysis and 2006 analysis, respectively.

A basic hadronic model fit is presented in Fig.~\ref{fig17}, middle-right panel. The formal best fit for the case of the modified hadronic model is shown in Fig.~\ref{fig17}, bottom-left. As for the case of 1ES 0347-121, primary component in this figure is very narrow, therefore, we also present another, more physically plausible fit for the same model in bottom-right panel of Fig.~\ref{fig17}.

\begin{figure}[t]
\centerline{\includegraphics[width=0.35\textwidth]{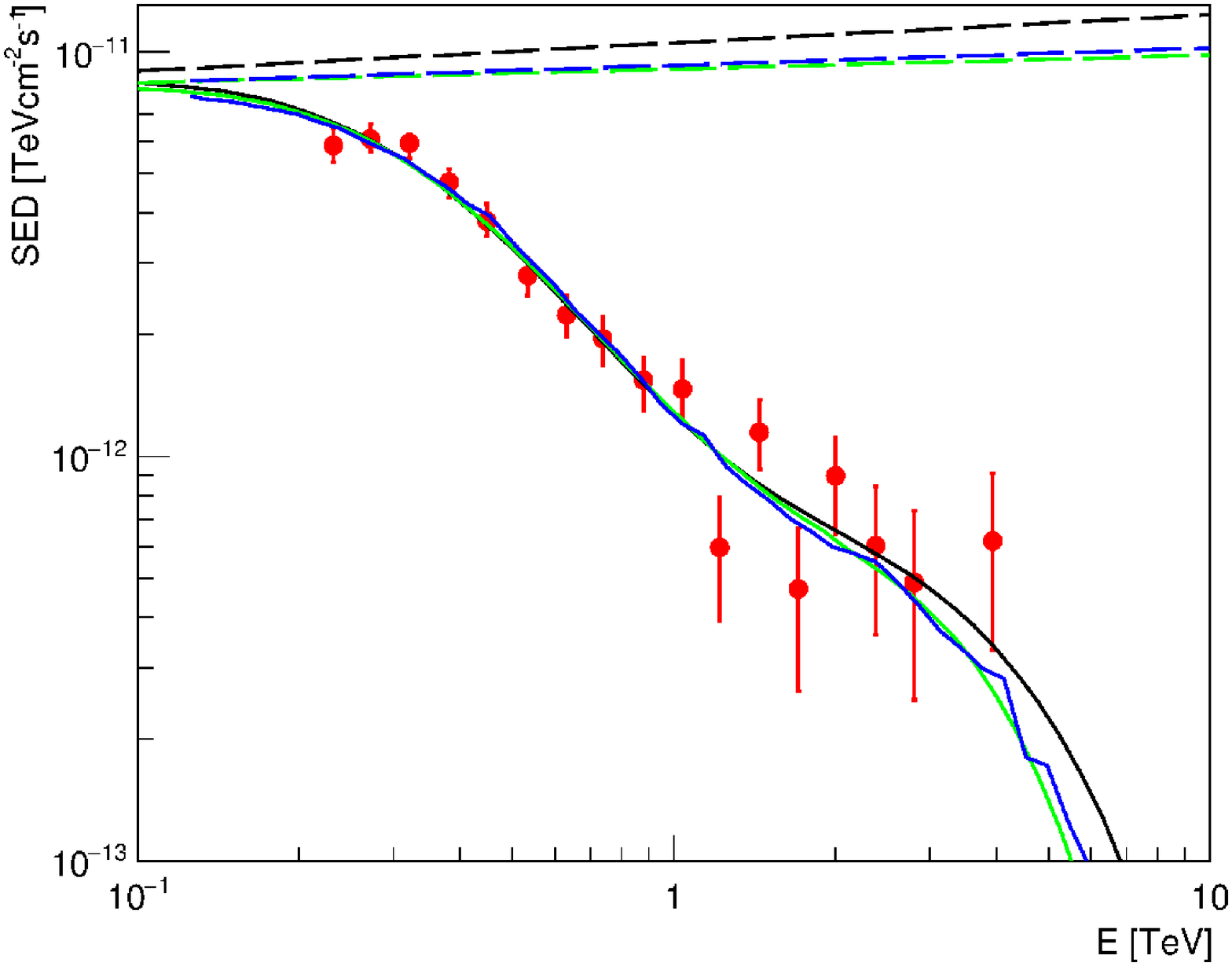}\includegraphics[width=0.35\textwidth]{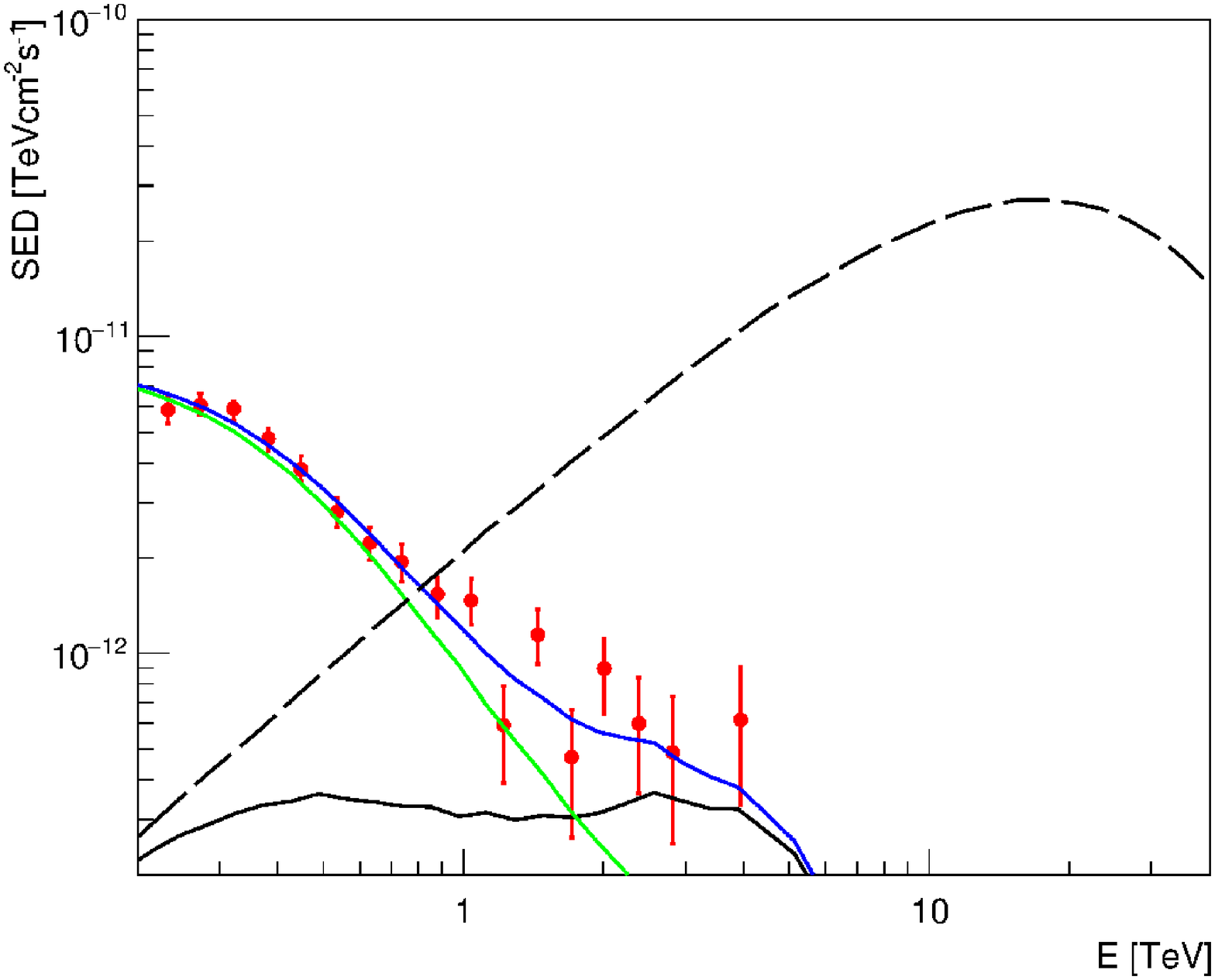}}
\centerline{\includegraphics[width=0.35\textwidth]{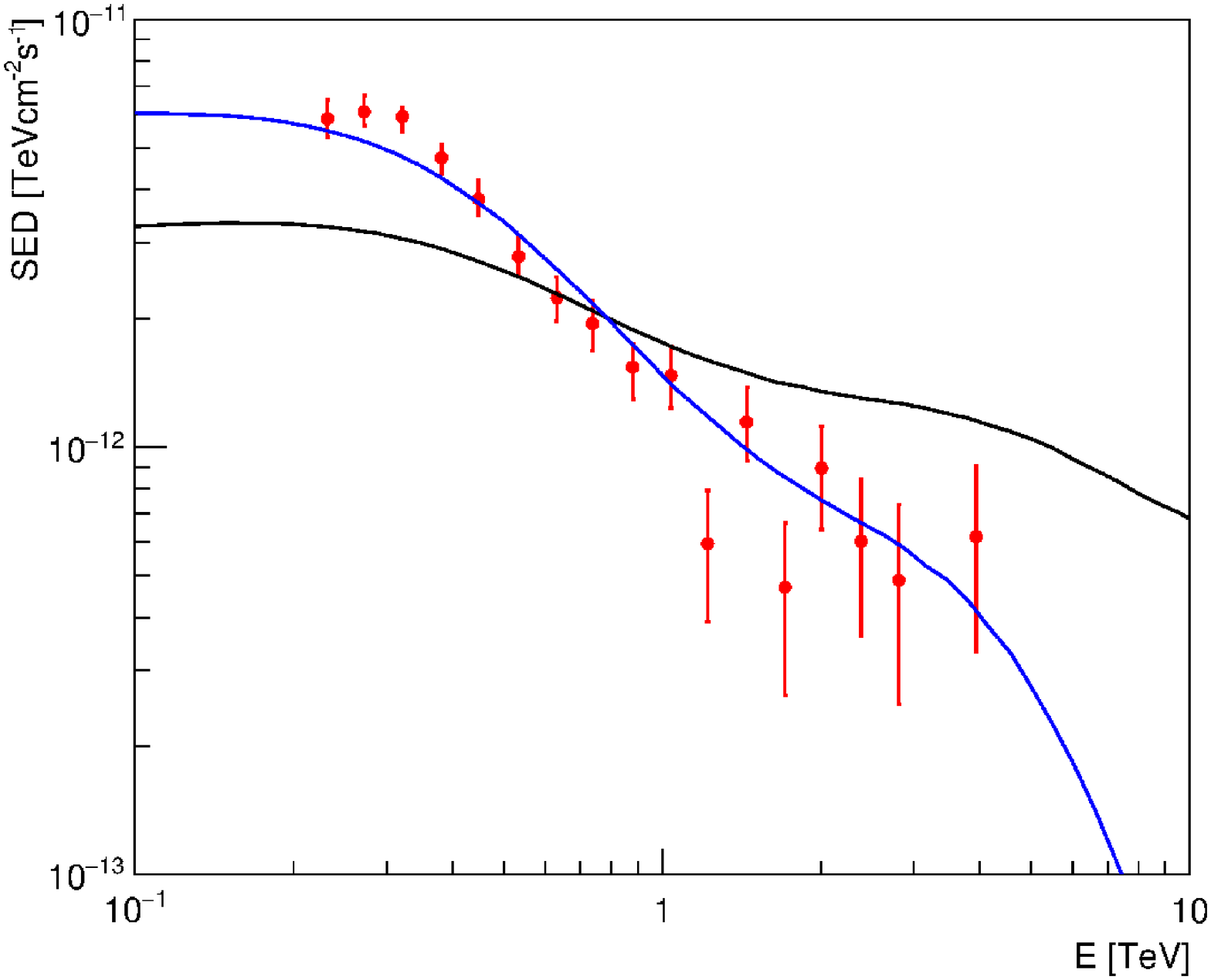}\includegraphics[width=0.35\textwidth]{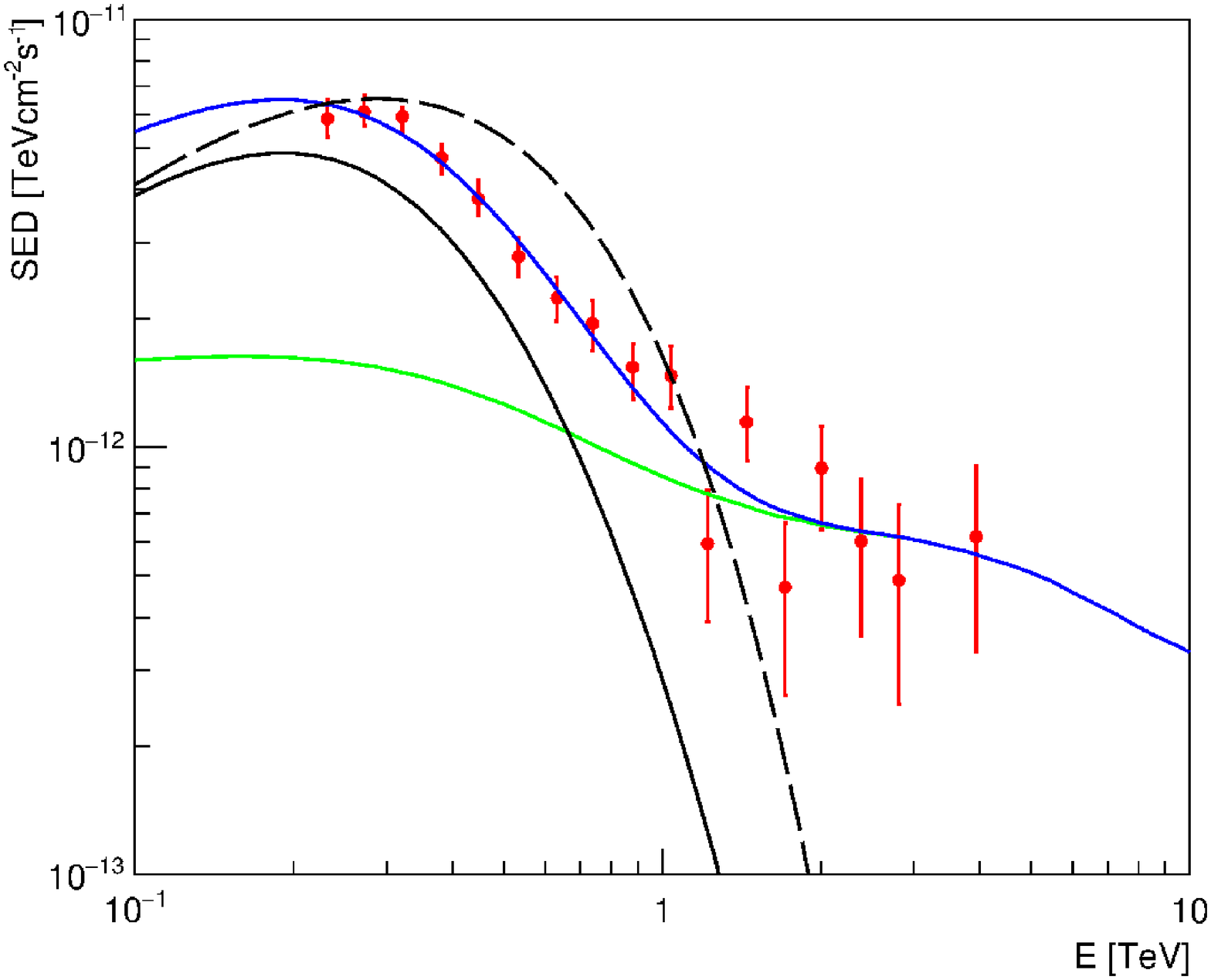}}
\centerline{\includegraphics[width=0.35\textwidth]{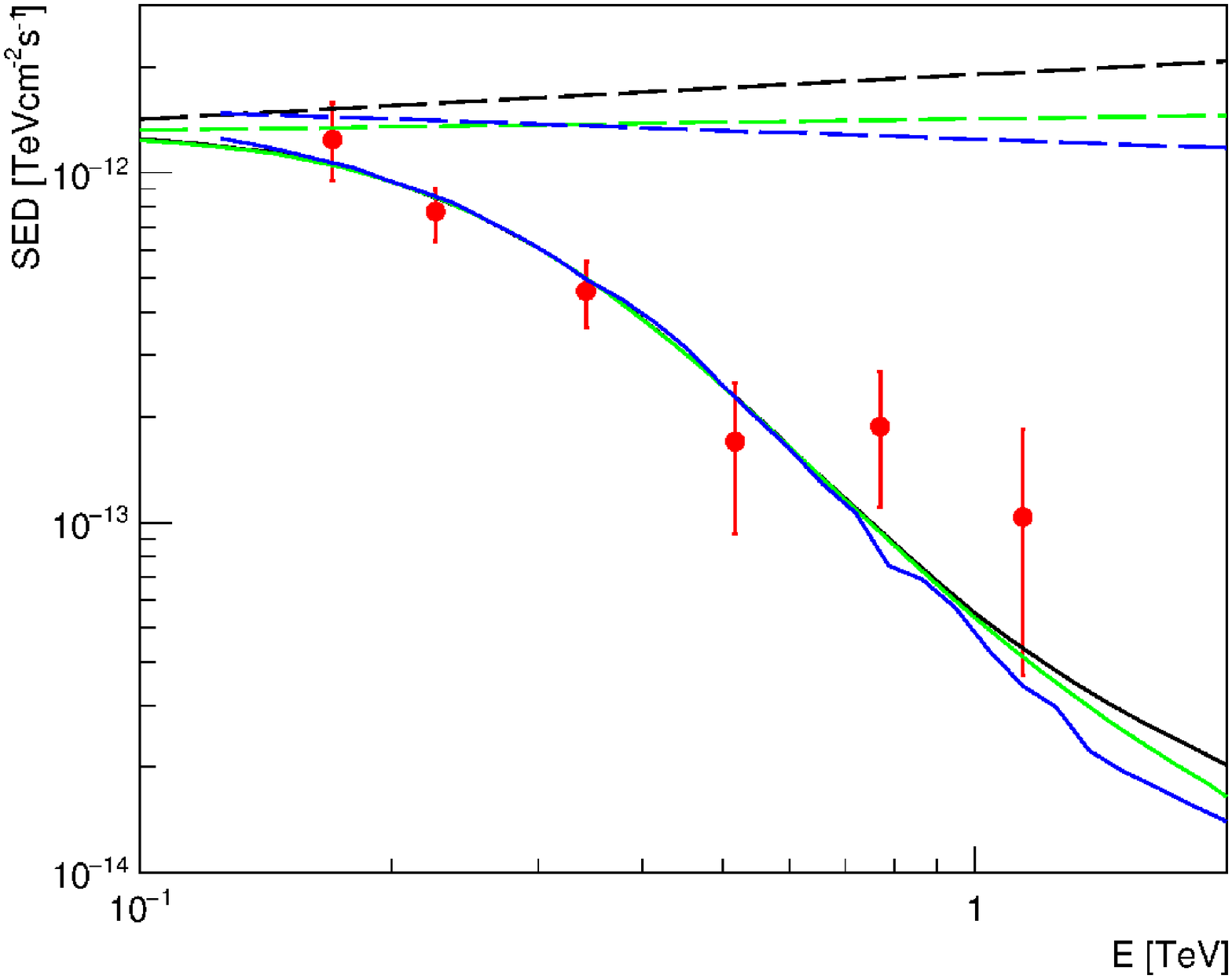}\includegraphics[width=0.35\textwidth]{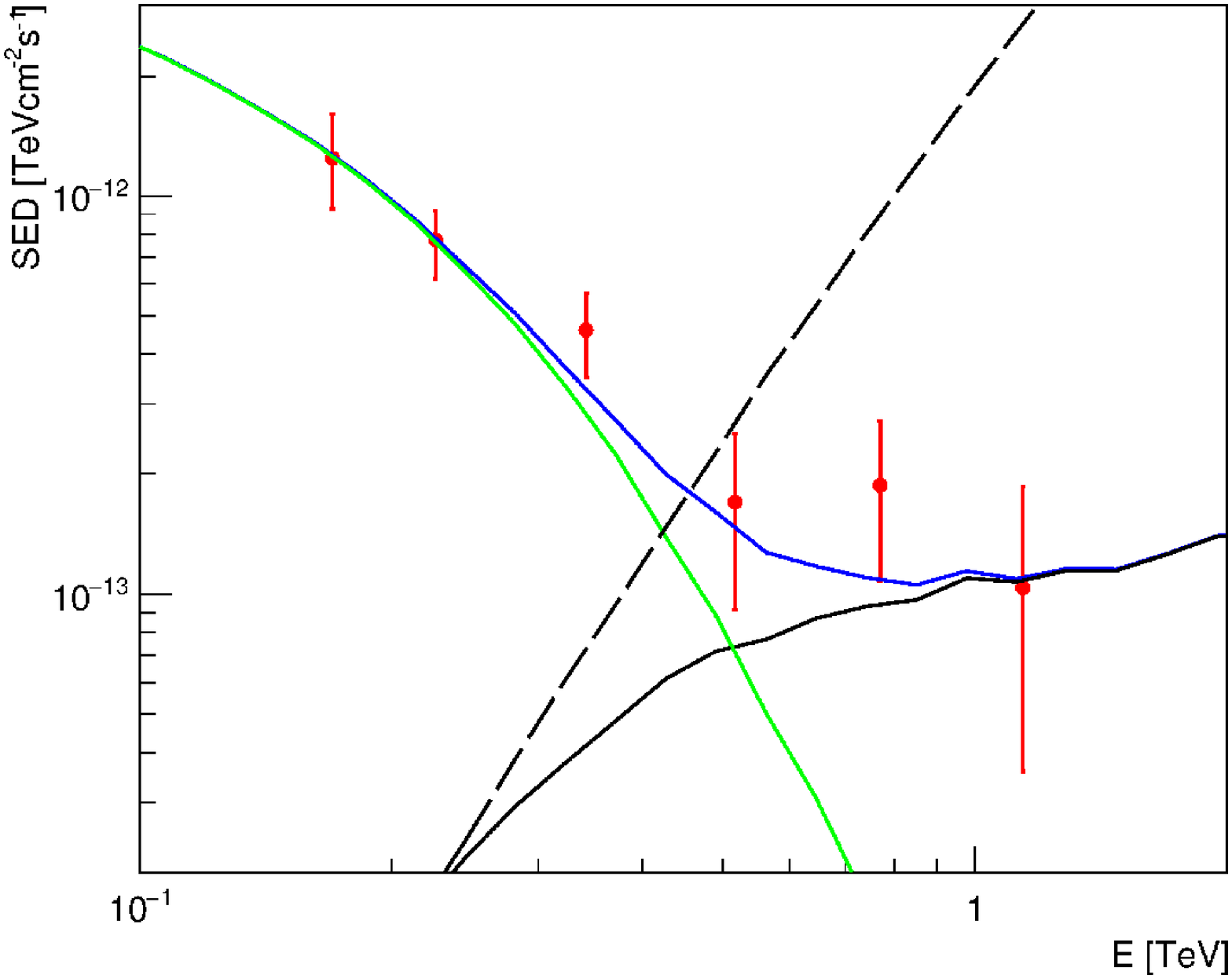}}
\centerline{\includegraphics[width=0.35\textwidth]{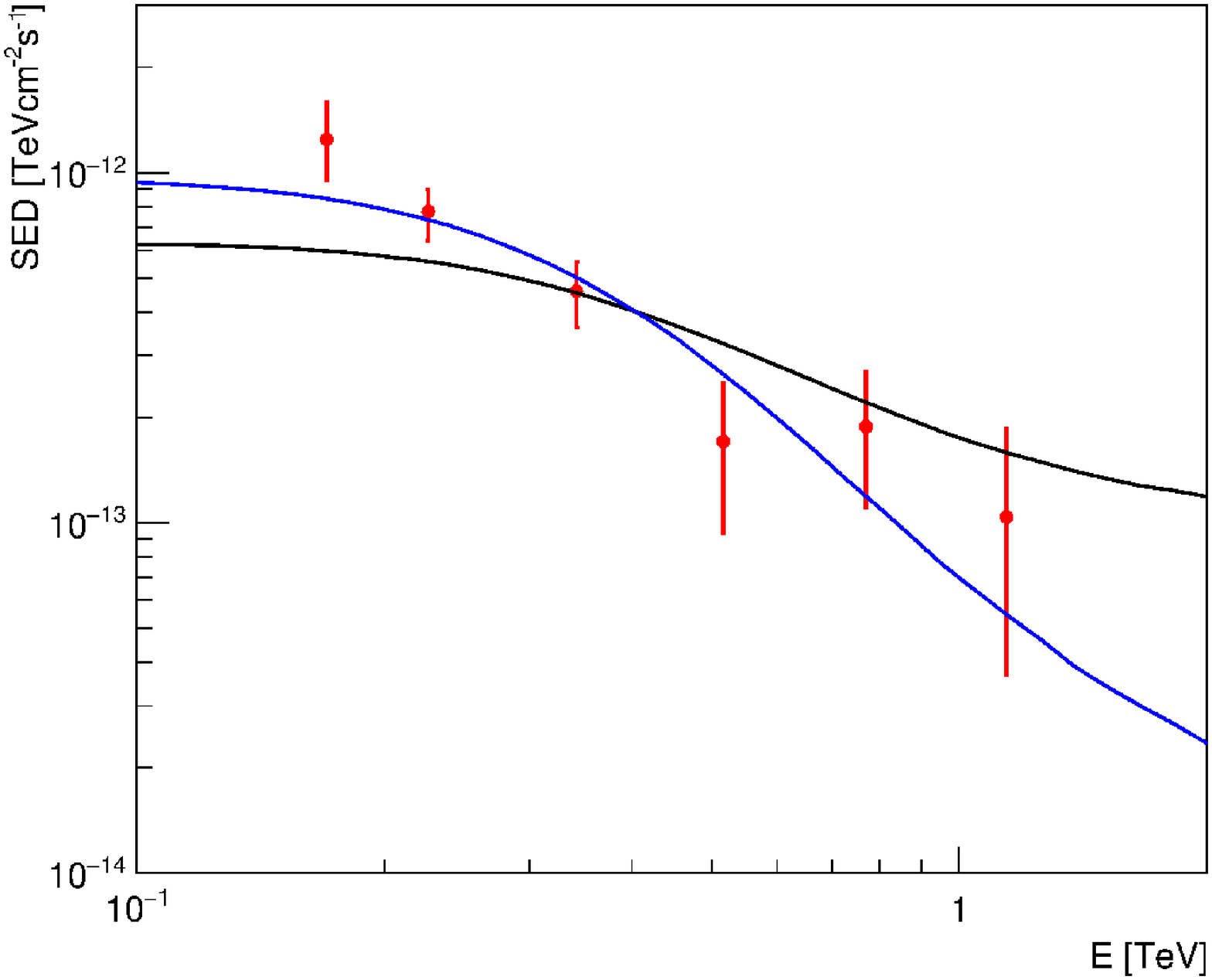}\includegraphics[width=0.35\textwidth]{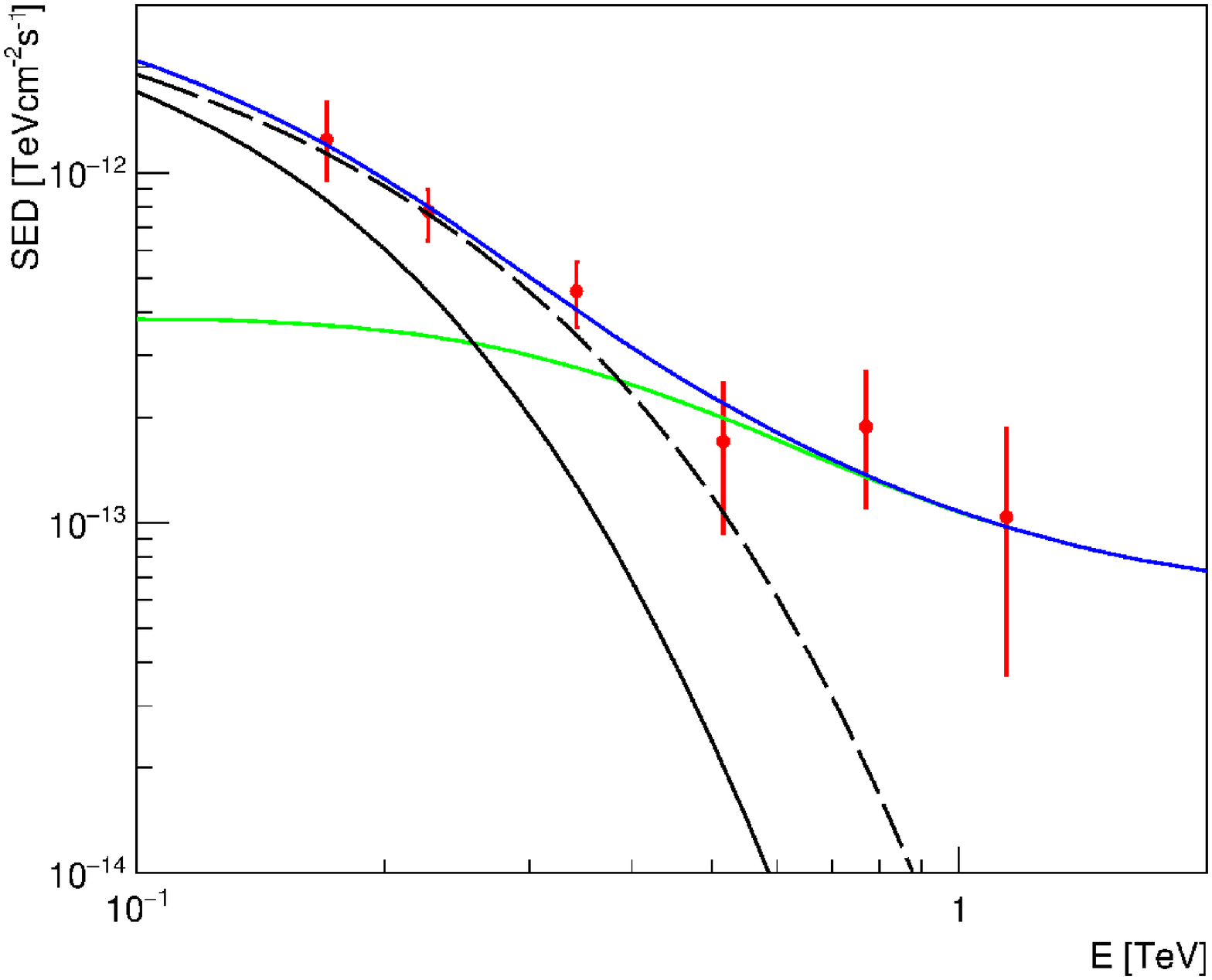}}
\caption{Model fits for the case of 1ES1218+304 (first row, left --- absorption-only model; first row, right --- electromagnetic cascade model; second row, left --- basic hadronic model (black curve --- $z_{c}$= 0, blue curve --- $c_{c}$= 0.15); second row, right --- modified hadronic model) and 1ES 0414+009 (third row, left --- absorption-only model; third row, right --- electromagnetic cascade model; bottom-left --- basic hadronic model  (black curve --- $z_{c}$= 0, blue curve --- $c_{c}$= 0.25); bottom-right --- modified hadronic model).}\label{fig18}
\end{figure}

Similar fits for 1ES 1812+304 and 1ES 0414+009 are shown in Fig.~\ref{fig18}. The spectrum of 1ES 1812+304 is not far from the case of universal spectrum of gamma-rays. In the framework of the basic hadronic model this corresponds to the case of $z_{c}$ only slightly less than $z$ of the source. 

\begin{figure}[t]
\centerline{\includegraphics[width=0.50\textwidth]{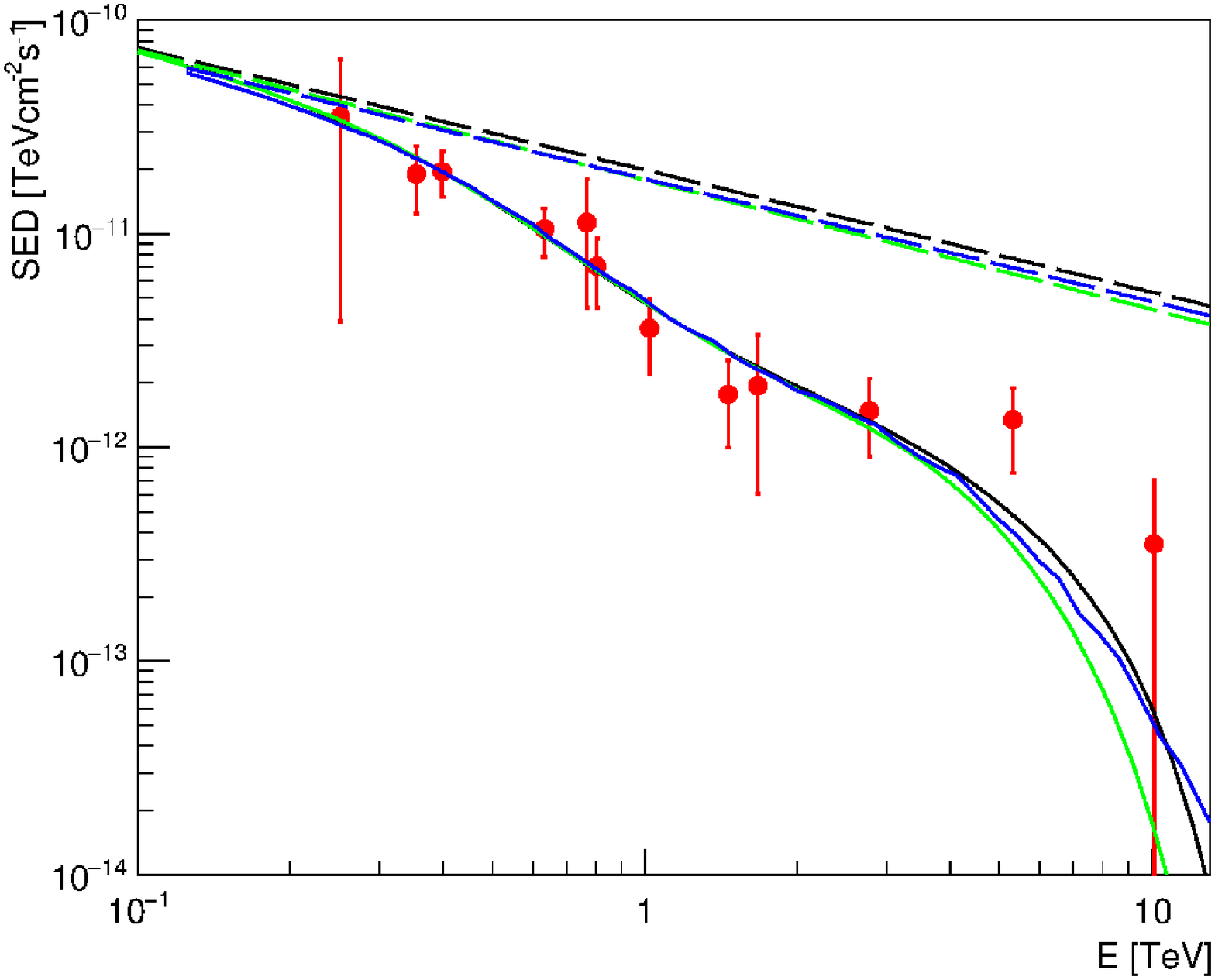}\includegraphics[width=0.50\textwidth]{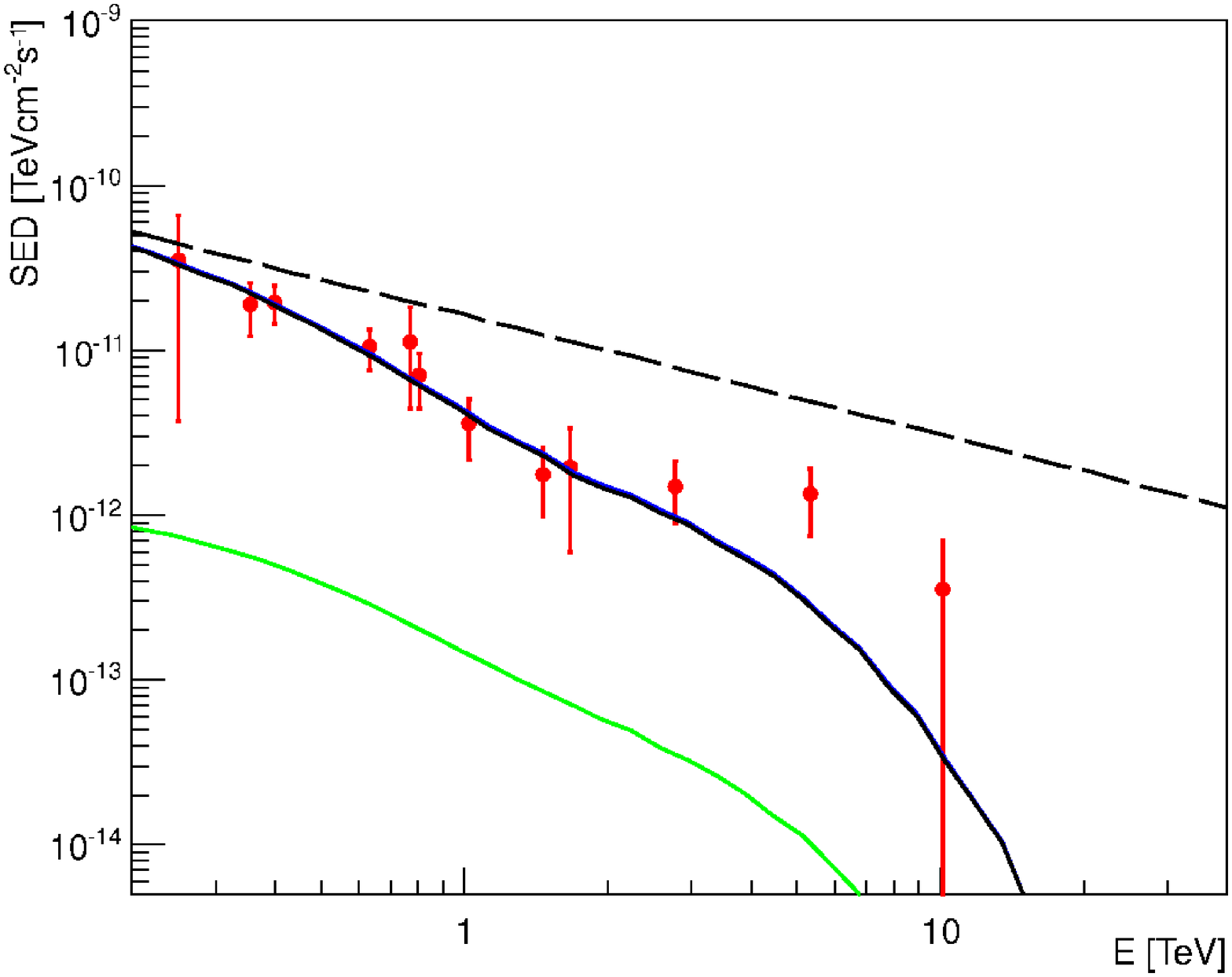}}
\centerline{\includegraphics[width=0.50\textwidth]{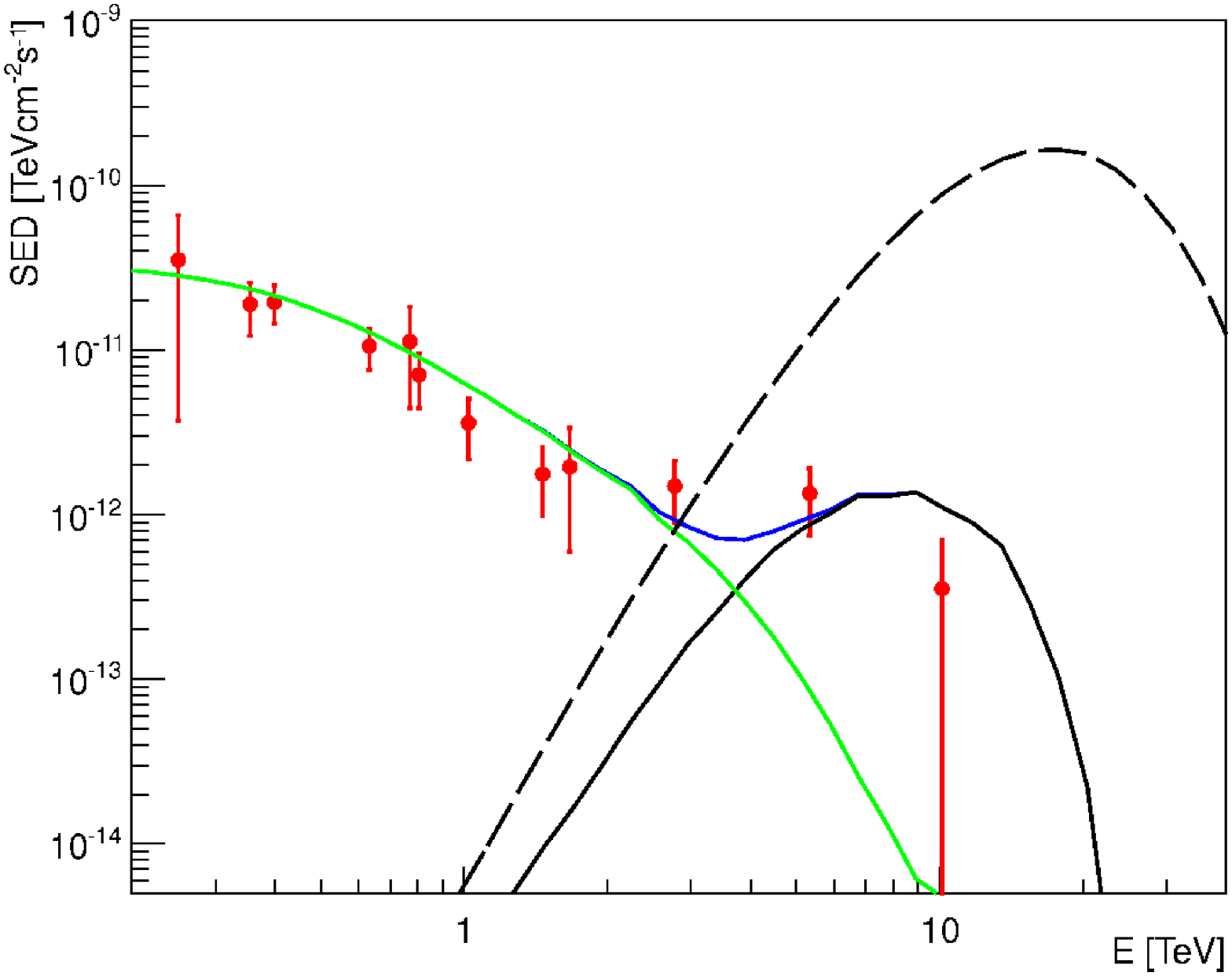}\includegraphics[width=0.50\textwidth]{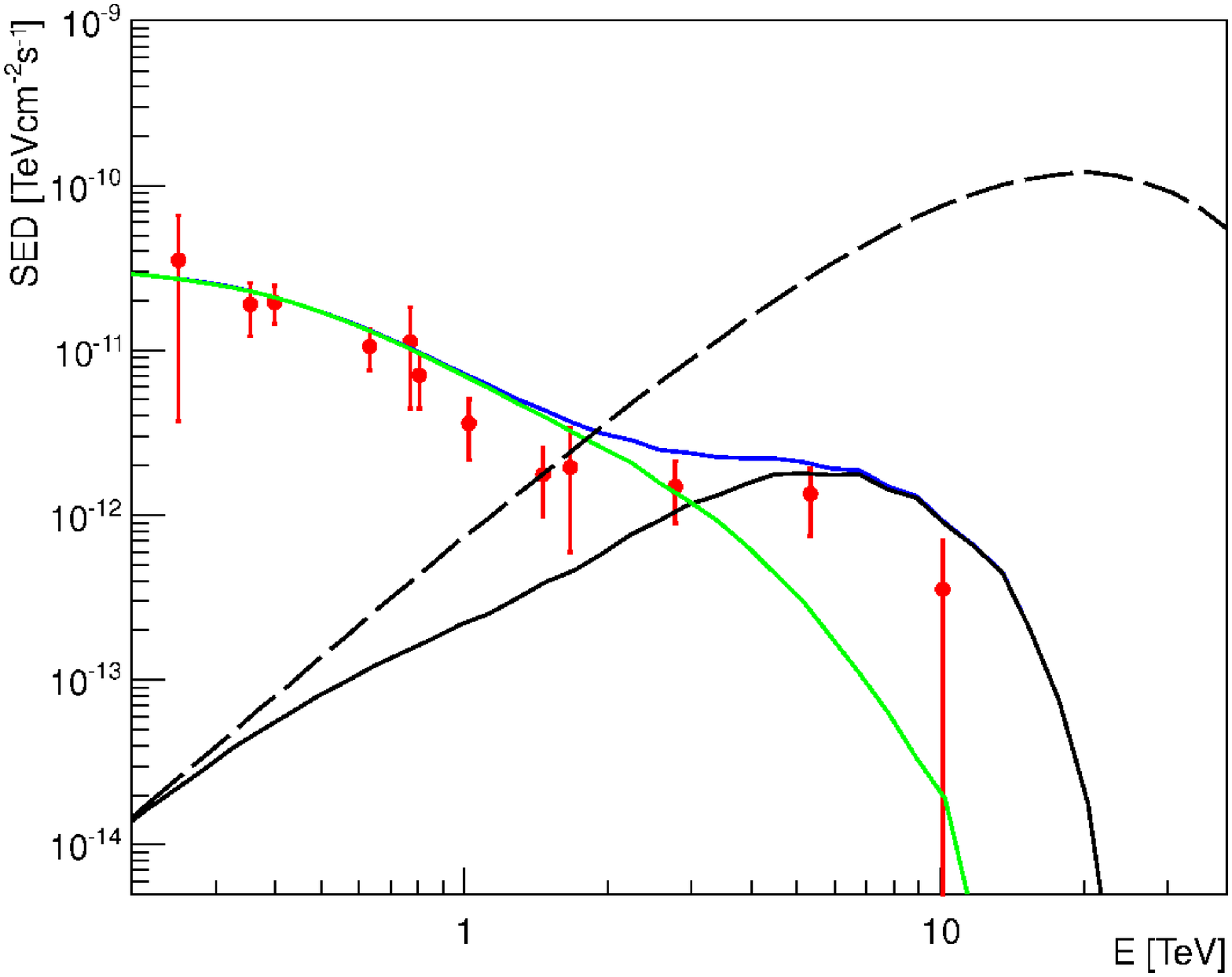}}
\centerline{\includegraphics[width=0.50\textwidth]{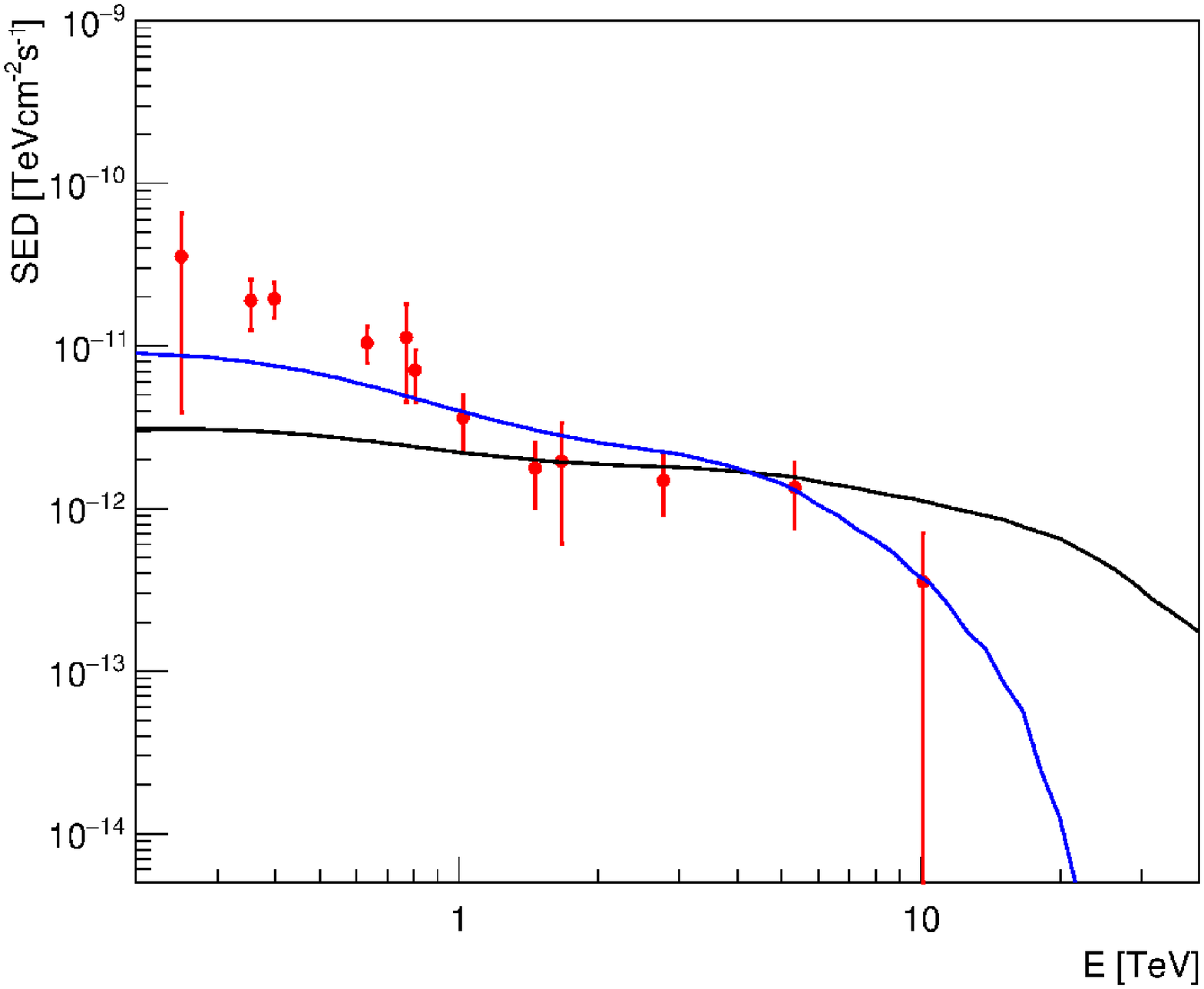}\includegraphics[width=0.50\textwidth]{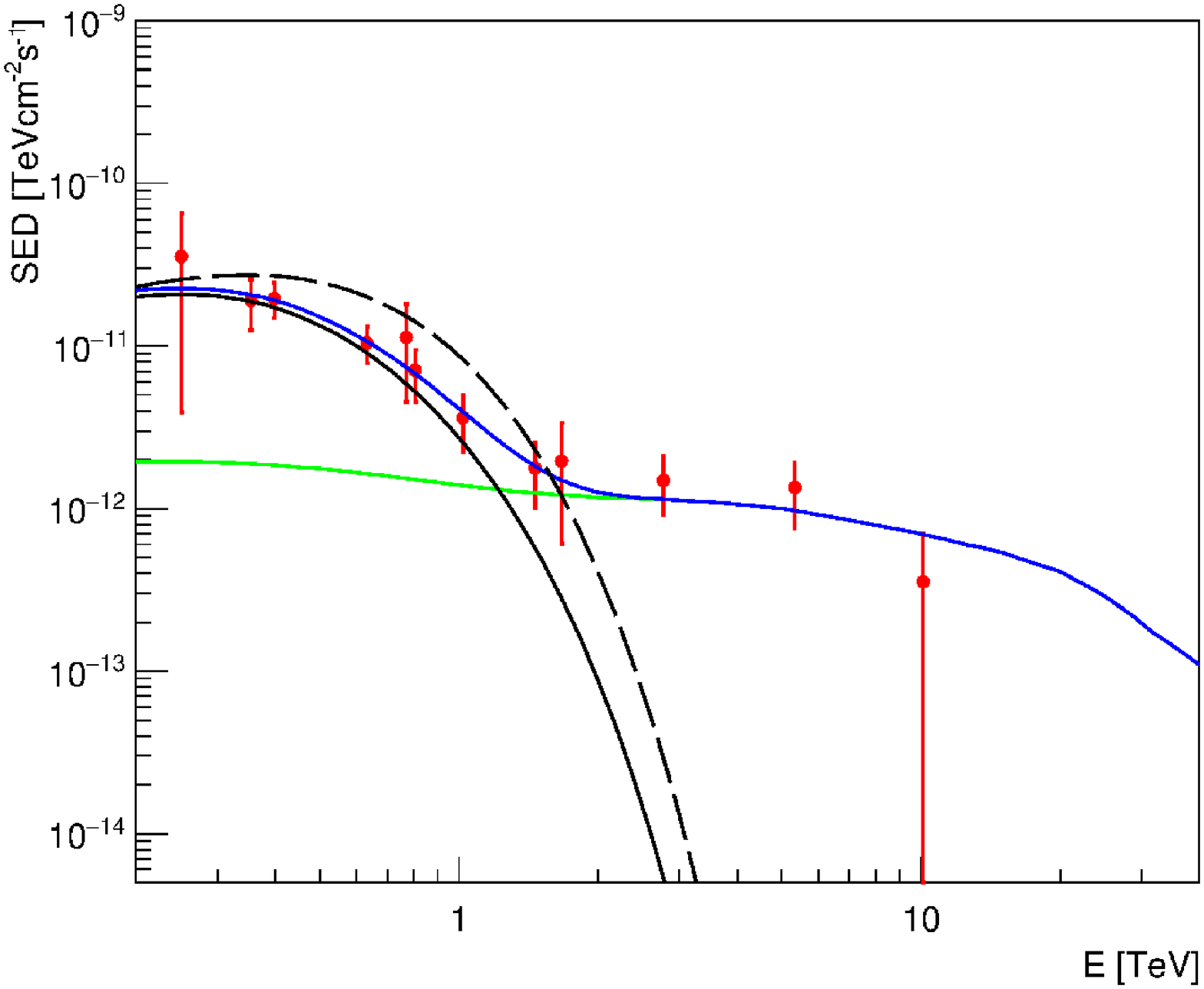}}
\caption{Model fits for the case of H 1426+428. Top-left --- absorption-only model, top-right --- electromagnetic cascade model (formal best fit). Middle panels --- two fits for the electromagnetic cascade model with large contribution of cascade component at low energy. Bottom-left --- basic hadronic model (black curve denote $z_{c}$= 0, blue curve --- $c_{c}$= 0.1), bottom-right --- modified hadronic model.}\label{fig19}
\end{figure}

Finally, fits for the source H 1426+428 are shown in Fig.~\ref{fig19}. For this object, the formal best fit for the electromagnetic cascade model is practically coincident with the case of the absorption-only model. The contribution of the cascade component to the total flux in this case is vastly subdominant, only about several percent. In this respect, it is interesting that Furniss et al. \cite{furniss} estimated $V_{LOS}\approx$ 0 for this source, i.e. that there is small room on the line of sight where electromagnetic cascade could develop. Nevertheless, we also present two fits for the case of dominant cascade contribution at low energy ($E$< 1 $TeV$) for completeness. The fits for the case of basic and modified hadronic model are presented as well.

\onecolumn
\twocolumn

\onecolumn
\twocolumn

\onecolumn
\twocolumn

\section{Constraints on hadronic cascade models}

As we have seen in the previous section, both electromagnetic and hadronic cascade models provide reasonable fits to the observed shape of the spectra of extreme $TeV$ blazars considered in this paper. Some additional information is required in order to favour one model over another. One of the criteria upon which it could be done is the plausibility of the source emission model in context of total normalization of a certain emission component. Tavecchio \cite{tavecchio} developed such a model (hereafter the T14 model) specifically for the hadronic cascade scenario. In what follows we use the parameters of blazar emission presented in Tavecchio \cite{tavecchio} (Table 2).

In fact, central emitting object of a blazar is circumvented by magnetic fields of an appreciable strength and spatial scale. In what follows we assume the model of the galaxy cluster turbulent magnetic field of Meyer et al. \cite{meyer} (equations (14),(15)) with parameters $r_{core}$= 200 $kpc$, $\beta$= 2/3, $\eta$= 1/2, coherence scale 10 $kpc$ (the spatial dimensions of all magnetic field domains are equal) and variable parameter $B_{ICMF}^{0}$ (hereafter denoted simply as $B_{0}$).

These magnetic fields, in fact, modify the angular pattern of the proton beam, thus effectively dimming the observable emission. Initial angular pattern of the hadronic beam was set the same as the angular pattern of the leptonic component. The T14 model provided a following constraint to the total power of proton emission: it cannot be greater than the total ``magnetic luminosity'' $P_{B}$. Indeed, in the context of the T14 model, protons during the acceleration are confined to the region of magnetic field, therefore, the energy density of these protons can not be higher than the energy density of magnetic field, otherwise they escape from the region, thus terminating the acceleration process. T14 introduced a parameter of the maximal proton acceleration energy, $E_{max}$. We assume the accelerated proton spectrum shape $dN_{p}/dE\propto E^{-\gamma_{p}}exp(-E_{p}/E_{max})$.

It is convenient for us to count the total power of accelerated protons $L_{p}$ with respect to the power-law spectrum, introducing an additional factor
\begin{equation}
K_{cut}= \frac{\int\limits_{E_{min}}^{E_{max}}{E^{-\gamma_{p}}dE}}{\int\limits_{E_{min}}^{E_{max}}{E^{-\gamma_{p}}exp(-E_{p}/E_{max})dE}},
\label{eqn15}
\end{equation}
so that $L_{p}<K_{cut}P_{B}$. In what follows we introduce additional modification factors of the proton intensity with respect to the power-law spectrum of protons $dN_{p}/dE\propto E^{-\gamma_{p}}$. We set $E_{min}$= 1 $EeV$ and $E_{max}$= 100 $EeV$.

We estimate the flux modification factor $K_{M}$ as:
\begin{equation}
K_{M}= \frac{1-cos(\theta_{j})}{1-cos(\theta_{d}+\theta_{j})}exp(-E_{p}/E_{Max})(1+\theta_{d}/\pi),
\label{eqn16}
\end{equation}
where $\theta_{j}= 1/\Gamma$, $\Gamma$ is the bulk Lorentz factor of the blazar jet. The deflection angle of protons in the cluster magnetic field is estimated as:
\begin{equation}
\theta_{d}\approx\sqrt{\sum_{i=1}^{N}{\theta_{i}^{2}}},
\label{eqn17}
\end{equation}
where $N$=200 is the number of magnetic field domains and $\theta_{i}$ are partial deflections of proton in these domains:
\begin{equation}
\theta_{i}\approx\left(\frac{B}{1 mkG}\right)\left(\frac{10 EeV}{E_{p}}\right)
\label{eqn18}
\end{equation}
This approximation works reasonably well in the small deflection angle limit, $\theta_{d}<<$1 $rad$. As we will see, the small deflection angle assumption is also well justified.

\begin{figure}
\centering
\includegraphics[width=8cm]{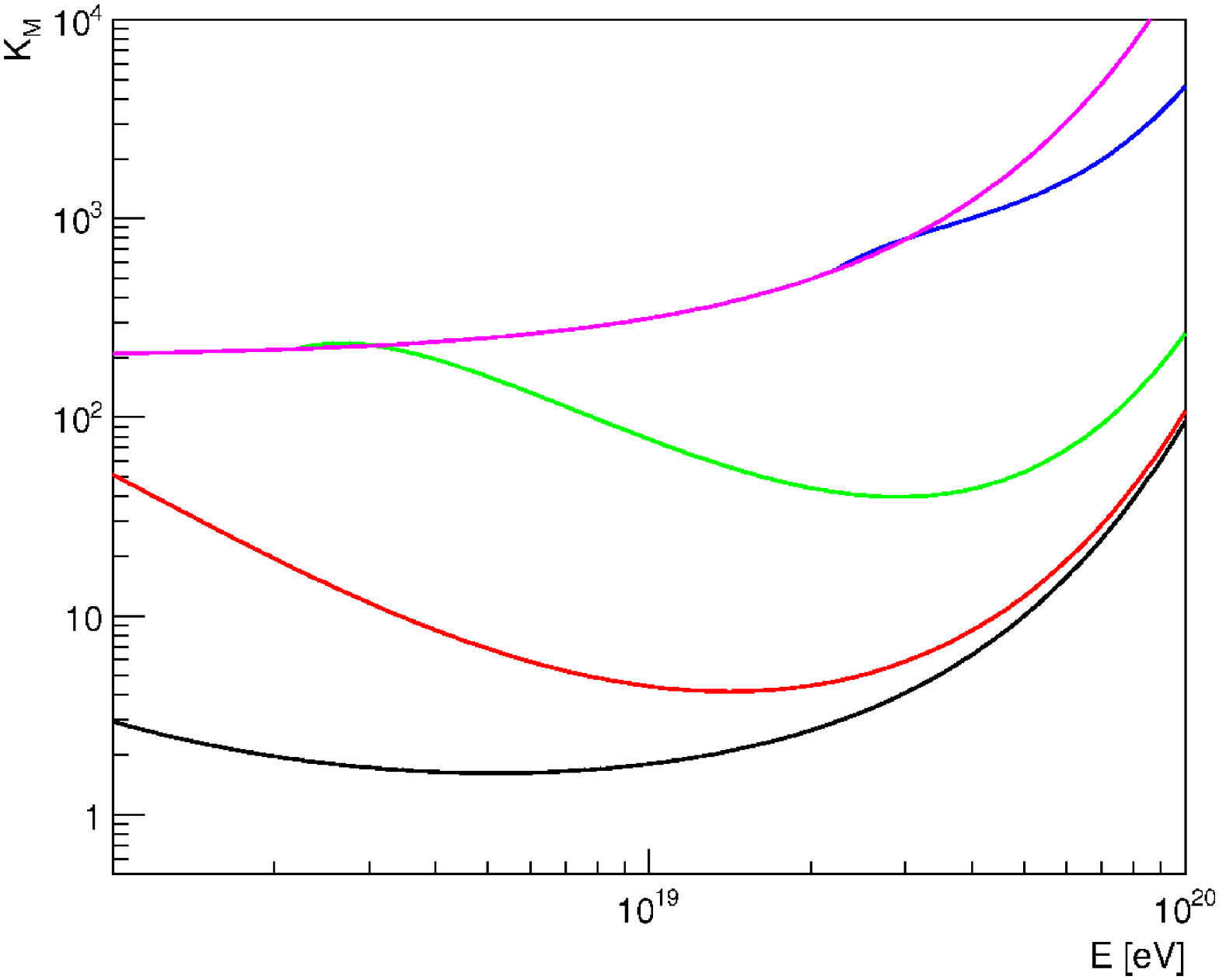}
\caption{Modification factor $K_{M}$ for different values of $B_{0}$. Black curve denotes $B_{0}$= 1 $nG$, red curve --- 10 $nG$, green --- 100 $nG$, blue --- 1 $mkG$, magenta --- 10 $mkG$.}
\label{fig20}
\end{figure}

The interpretation of the equation (16) is as follows: the jet angular pattern is broadened by the deflection of protons in the cluster magnetic field, high energy protons at $E_{p}>E_{max}$ effectively leave the region of acceleration and only a small number of particles is accelerated far above $E_{max}$; finally, the factor $(1+\theta_{d}/\pi)$ roughly accounts for the effect of the visibility of the second jet (counter-jet) when the deflection angle is large. The modification factor for the case of several values of the $B_{0}$ parameter is shown in Fig.~\ref{fig20}.

After that, we calculate the spectrum of observable $\gamma$-rays for different model parameters. We assume that all energy of secondary electrons and photons was converted to the energy of cascade $\gamma$-rays. As we are interested only in setting an upper limit to the intensity of the observable $\gamma$-rays produced by protons on the way from the source to the observer, this assumption is justified.

\begin{figure}
\centering
\includegraphics[width=8cm]{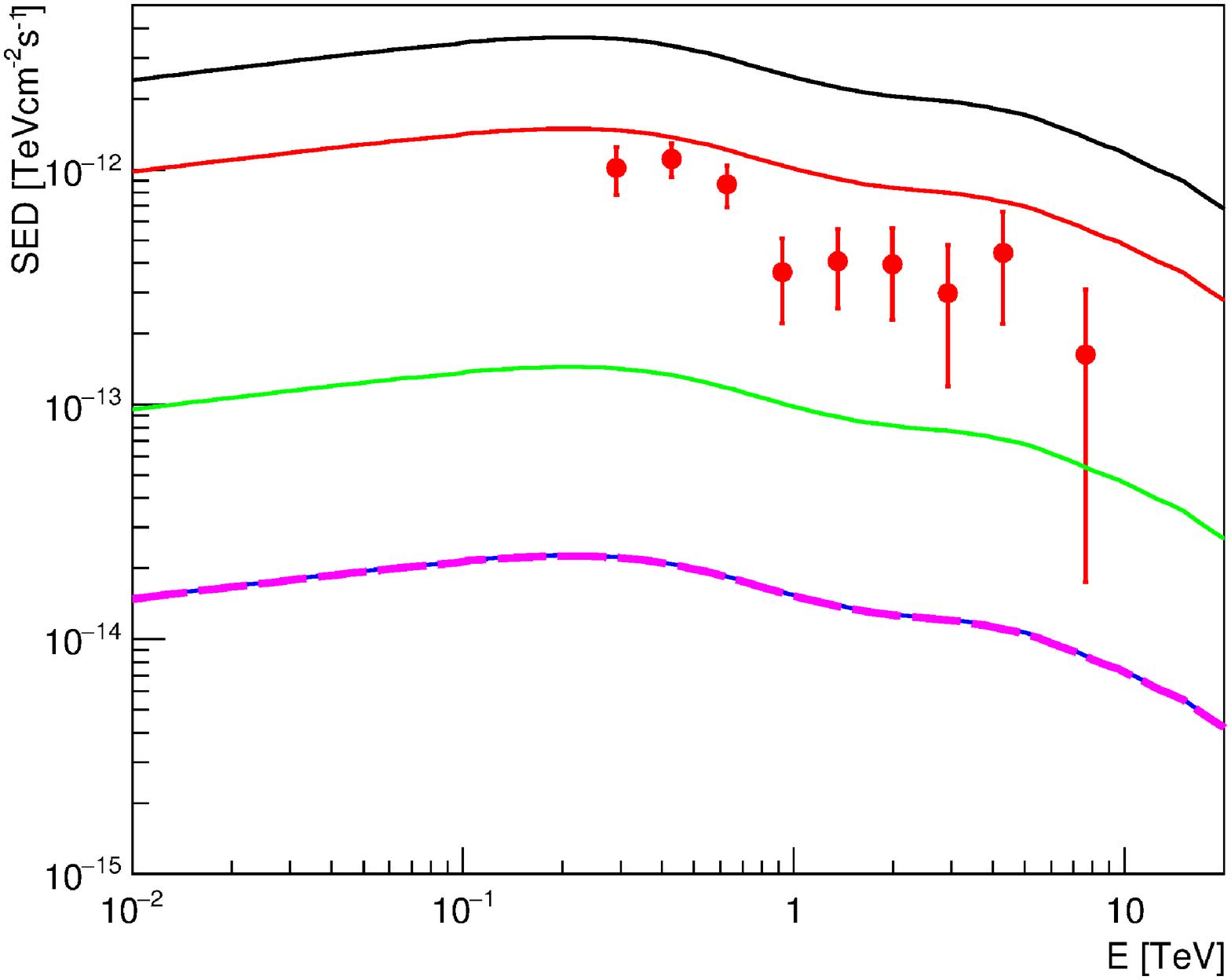}
\caption{Model spectrum for different values of $B_{0}$. Colors denote the same values of $B_{0}$ as in Figure 20.}
\label{fig21}
\end{figure}

In this section we consider the intermediate hadronic cascade model with different values of the $z_{c}$ parameter. We perform absolute normalization of the spectrum using the T14 model. We consider the source 1ES 0229+200 and two options for model parameters presented in T14 --- the one corresponding to $\Gamma$= 10, as well as to $\Gamma$= 30. 1ES 0229+200 is an archetypal extreme $TeV$ blazar that was considered in Essey \& Kusenko (2010b) as a very good candidate for the basic hadronic model. Fig.~\ref{fig21} presents normalized observable gamma-ray spectra for the same values of the $B_{0}$ parameter as in Fig.~\ref{fig20} and $z_{c}= 0$.

\begin{figure}
\centering
\includegraphics[width=8cm]{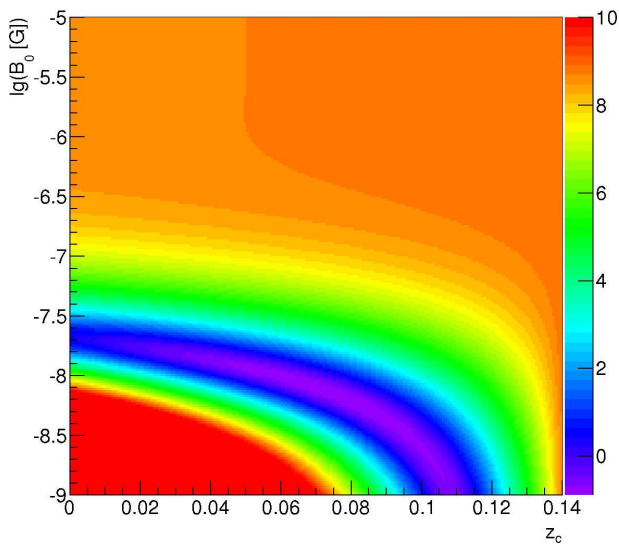}
\caption{Significance ($Z$-value) of the intermediate hadronic model exclusion for $\Gamma$= 10 and blazar emission parameters from T14.}
\label{fig22}
\end{figure}

\begin{figure}
\centering
\includegraphics[width=8cm]{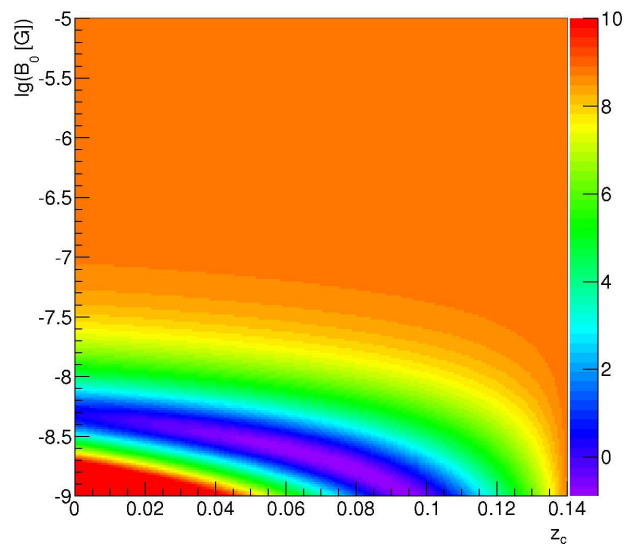}
\caption{Same as in Fig.~\ref{fig22}, but for $\Gamma$= 30.}
\label{fig23}
\end{figure}

Finally, we performed an exhaustive scan over a wide range of parameters $(z_{c},B_{0})$ on a 200$\times$200 grid for two options of the source emission models presented in T14, labelled by the values of $\Gamma$=10 and $\Gamma$=30. Using a simple $\chi^{2}$ goodness-of-fit statistical test, for every version of observable spectrum we calculated a $p$-value according to the prescriptions of Beringer et al. \cite{beringer}. After that, these $p$-values were converted to $Z$-values (i.e. statistical significance) according to the prescriptions of Beringer et al. \cite{beringer}. Fig.~\ref{fig22} shows $Z$-values in color; it was drawn for the case of $\Gamma$= 10, and Fig.~\ref{fig23} --- for the case of $\Gamma$= 30. All values of $Z$>10 $\sigma$ were set to 10 $\sigma$ and, thus, are displayed in red. The lower-left parts of Fig.~\ref{fig22}--Fig.~\ref{fig23}, filled by red color, correspond to the case when the intensity of the observable $\gamma$-rays is much greater than the one required by observations. Other parts of these graphs (violet and blue) denote the range of acceptable models. Finally, the upper-right parts of these graphs display the regions of parameters where the corresponding model is excluded. It is remarkable that all considered configurations with $B_{0}$ above 100 $nG$ are excluded with significance $Z$>7 $\sigma$.

During this calculation, it will be remembered, we have neglected the losses of protons on EBL from the source to the observer. However, there are two main factors that strongly outweigh this one, namely: 1) injection of protons in context of the T14 model is done at an energy $E_{p-inj}$ much lower than 1 $EeV$; therefore, for the case of $E_{p-inj}$= 10 $GeV$ and $\gamma_{p}$=2 the total energy of accelerated protons is 5 times greater than the energy of UHE protons, thus, the modification factor in this case gains an additional factor of 0.2 2) synchrotron and curvature radiation losses of primary UHE protons will futher diminish their total energy. The last factor is particularly strongly constraining for the case of the Blandford-Znajek emission mechanism (Blandford-Znajek \cite{blandford}) due to strong magnetic fields in the acceleration region, see Neronov et al. \cite{neronov09} and Neronov \& Ptitsyna \cite{ptitsyna}.

\section{Summary and conclusions}

In this work we reviewed the main possibilities of how cascades from primary $\gamma$-rays or protons may influence the data interpretation when testing extragalactic $\gamma$-ray propagation models, mainly dealing with the source redshift range 0.1--0.3. We reviewed the main regimes of electromagnetic cascade development: 1) one-generation regime, which holds when $E_{\gamma0}<$10 $TeV$, 2) the universal regime, which is satisfied when $E_{\gamma0}>$100 $TeV$ 3) a possibility of the extreme energy cascade regime at $E_{\gamma0}>$1 $EeV$. 

We developed a fast, simple and suffuciently precise hybrid code that allows to calculate the observable spectrum of $\gamma$-rays for the case of primary proton. We performed calculations of observable spectra for this case and investigated several versions of the hadronic cascade model --- the basic model (all gamma-rays are generated by primary protons), the intermediate model (the same, but the proton beam is dissolved at some $z_c$, the basic model is the sub-set of intermediate models with $z_{c}$= 0), the modified model that includes the primary component. We discussed the signatures of the spectrum for the basic hadronic model for $z_{s}$= 0.186. The signatures at a comparatively low observable energy are almost the same as the ones for the case of a purely electromagnetic cascade --- that is, the $E^{-1.60}$ $dN/dE$ spectrum below $\approx 200$ $MeV$ and $E^{-1.85}$ spectrum from $\approx200$ $MeV$ to $\approx200$ $GeV$. However, at higher energy ($E>$200 $GeV$) the spectrum in the hadronic model is much harder than the universal spectrum for the case of a pure electromagnetic cascade.

We also performed the comparison of two hybrid calculations of observable spectra for different values of $z_{c}$ from 0 up to nearly $z_{s}$. The results of these calculations with two different methods are in very good agreement for the case of the proton primary energy $E_{p0}\le$ 30 $EeV$, i.e. when the pair-production losses are dominant. For the case of $E_{p0}$= 100 $EeV$ the agreement between the two methods is also good for the basic hadronic model and the intermediate hadronic model with comparatively low $z_{c}<z_{s}/2$, but not so good if $z_{c}\ge z_{s}/2$. The reason of this disagreement may be connected to the possibility of the extreme energy cascade regime. New calculations and measurements of the universal radio background are necessary in order to decide which spectrum is more realistic. 

While we confirmed the claim that the shape of the observable spectrum in the basic hadronic model depends only slightly on the primary proton energy, it was found that in context of the intermediate hadronic model the shape of the spectrum strongly depends on the $z_{c}$ parameter in the high energy region. 

We performed a series of fits to the spectra of extreme $TeV$ blazars with various extragalactic $\gamma$-ray propagation models, namely: the absorption-only model, the electromagnetic cascade model, the basic hadronic model, the intermediate hadronic model, the modified hadronic model. Most of the fits presented in this paper are formal best fits. It was found that, as a rule, both the electromagnetic cascade model and the hadronic cascade model are able to provide a reasonable fit to the observed spectral shape. 

For the case of blazar 1ES 0347-121 and the electromagnetic cascade model we performed the calculation of modification factor $K_{B}$ with respect to the absorption-only model. We found that $K_{B}$>1 in the energy region of 1-10 $TeV$. Therefore, the version of the electromagnetic cascade model that was considered in this work for this source predicts that it is possible that the observable intensity is higher than in the absorption-only model for the same level of EBL without any new physics. For the case of the source 1ES 0229+200 we also performed a fit in the framework of an exotic model with $\gamma\rightarrow ALP$ oscillations and a calculation of the modification factor $K_{B}$ vs. voidiness $K_{V}$ dependence in the framework of the electromagnetic cascade model. We found that when $K_{V}$ is lower than 1, the value of $K_{B}$ does not fall at high energies, but, on the contrary, even grows in the energy range of 1-15 $TeV$, thus making the effects induced by electromagnetic cascades even more pronounced.

Finally, using the fact that many blazars are situated in galaxy groups or clusters with comparatively strong magnetic fields (Muriel \cite{muriel}, Oikonomou et al. \cite{oikonomou}), we performed the testing of the emission model of Tavecchio (2014). We found that for $B_{0}$>100 $nG$ this emission model is excluded with significance $Z>$7 $\sigma$. Our results show that the hadronic cascade model experiences significant difficulties. This conclusion is in line with results obtained by Razzaque et al. \cite{razzaque}. Further testing of the hadronic and electromagnetic cascade models may be performed on an event-by-event level with advanced analysis tools such as GammaLib and ctools package (Knoedlseder et al. \cite{knoedlseder}). Such a work is underway in our group.

We conclude that cascades from primary $\gamma$-rays or nuclei may induce an appreciable background for astrophysical searches for $\gamma-ALP$ oscillation that needs to be taken into account. The electromagnetic and hadronic cascade models deserve further study, also in context of searches for other exotic phenomena, such as Lorentz invariance violation (Tavecchio \& Bonnoli \cite{tavecchio16}). The knowledge of EGMF would provide a significant insight into the intergalactic cascade phenomena in the Universe. Further measurements with the CTA array of Cherenkov telescopes (Acharya et al. \cite{acharya}) will likely clarify these issues. Together with existing instruments Fermi LAT (Atwood et al. \cite{atwood}) and AGILE (Rappoldi et al. \cite{rappoldi}), future experiments that have either high sensitivity or good angular resolution, such as GAMMA-400 (Galper et al. \cite{galper}) or the novel balloon-borne emulsion $\gamma$-ray telescope GRAINE (Takahashi et al. \cite{takahashi}), may also prove to be helpful in this task.

\begin{acknowledgements}
      This work was supported by RFBR, Grant 16-32-00823. The authors are grateful to Dr. V.V. Kalegaev for permission to use the SINP MSU space monitoring data center computer cluster and to V.O. Barinova, M.D. Nguen, D.A. Parunakyan for technical support.
\end{acknowledgements}

\newpage

\begin{appendix}
\section{Electromagnetic cascades in the universal regime}

\begin{figure}
\centering
\includegraphics[width=8cm]{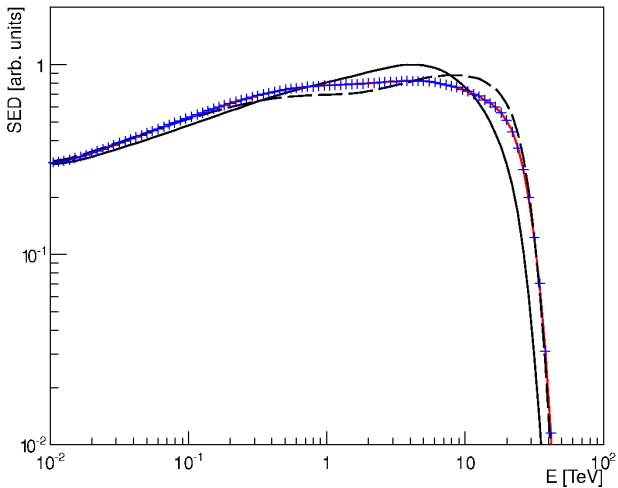}
\caption{Electromagnetic cascade spectra in the universal regime for the case of $z$= 0.02. Black solid --- gamma-ray, 100 $TeV$; black dashed --- electron, E= 100 $TeV$; red solid --- gamma-ray, E= 1 $PeV$, blue crosses --- electron, E= 1 $PeV$.}
\label{figA1}
\end{figure}

\begin{figure}
\centering
\includegraphics[width=8cm]{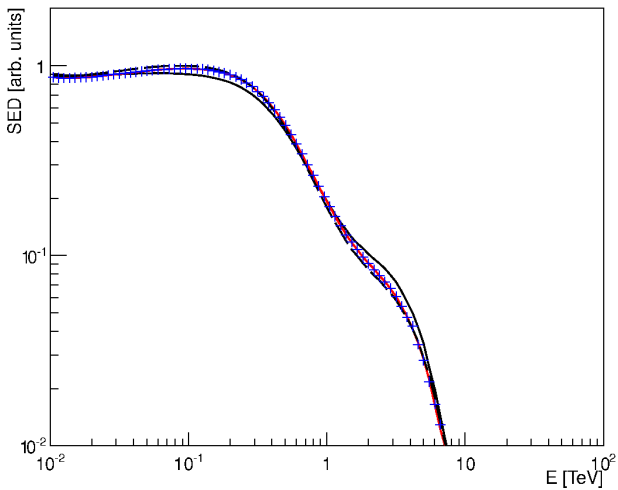}
\caption{Same as in Fig.~\ref{figA1}, but for $z$= 0.186.}
\label{figA2}
\end{figure}

\begin{figure}
\centering
\includegraphics[width=8cm]{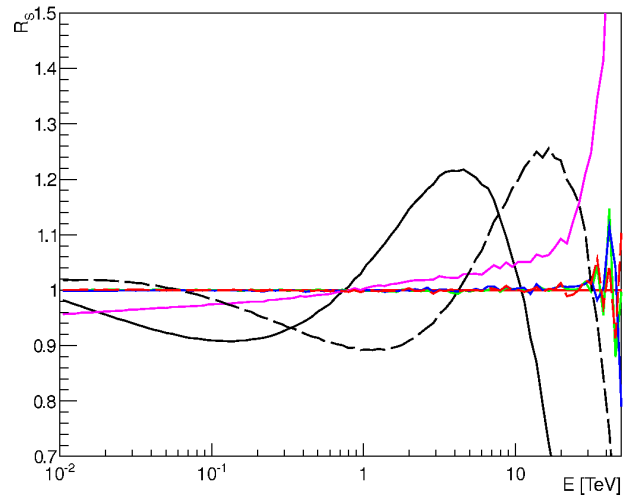}
\caption{Ratio of the spectra to the spectrum of primary gamma-ray with energy 1 $PeV$ for the case of $z$= 0.02. Black solid --- primary gamma-ray, 100 $TeV$, black dashed --- electron, E= 100 $TeV$; red solid --- gamma-ray, E= 1 $PeV$, red dashed --- electron, E= 1 $PeV$, green solid --- gamma-ray, E= 10 $PeV$, blue solid --- gamma-ray, E= 100 $PeV$, magenta --- gamma-ray, E= 1 $EeV$.}
\label{figA3}
\end{figure}

Here we present some additional plots to reveal the borders of the universal regime of EM cascade development. Fig.~\ref{figA1}--\ref{figA2} show the spectra for the case of two different redshifts, different primaries (electron and $\gamma$-ray), and two primary energies. We have checked that for the case of greater energies, 10 $PeV$ and 100 $PeV$, the spectra for both redshifts and both primaries are practically coincident (within the width of the line) with ones shown in these figures for the case of $E_{0}$=1 $PeV$. Fig.~\ref{figA3}--\ref{figA3} show the ratio of the spectra to the spectrum for the case of primary $\gamma$-ray and $E_{0}$= 1 $PeV$ for different primaries and different $E_{0}$, as well for two values of $z_{s}$.

\begin{figure}
\centering
\includegraphics[width=8cm]{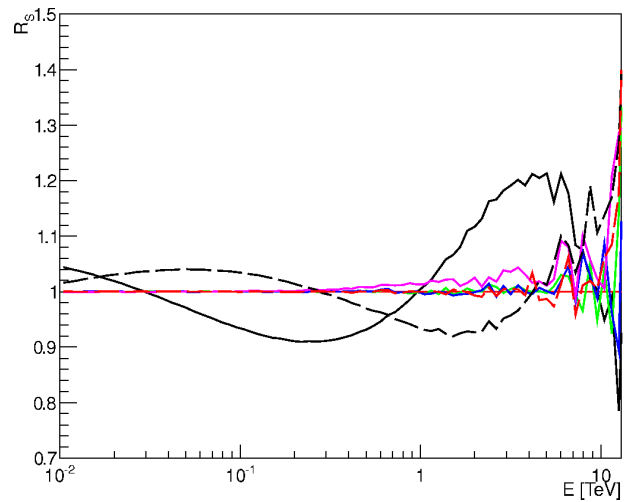}
\caption{Same as in Fig.~\ref{figA3}, but for $z$= 0.186.}
\label{figA4}
\end{figure}

\section{Absorbed and cascade components for various primary spectra of $\gamma$-rays}

\begin{figure}[t]
\centerline{\includegraphics[width=0.50\textwidth]{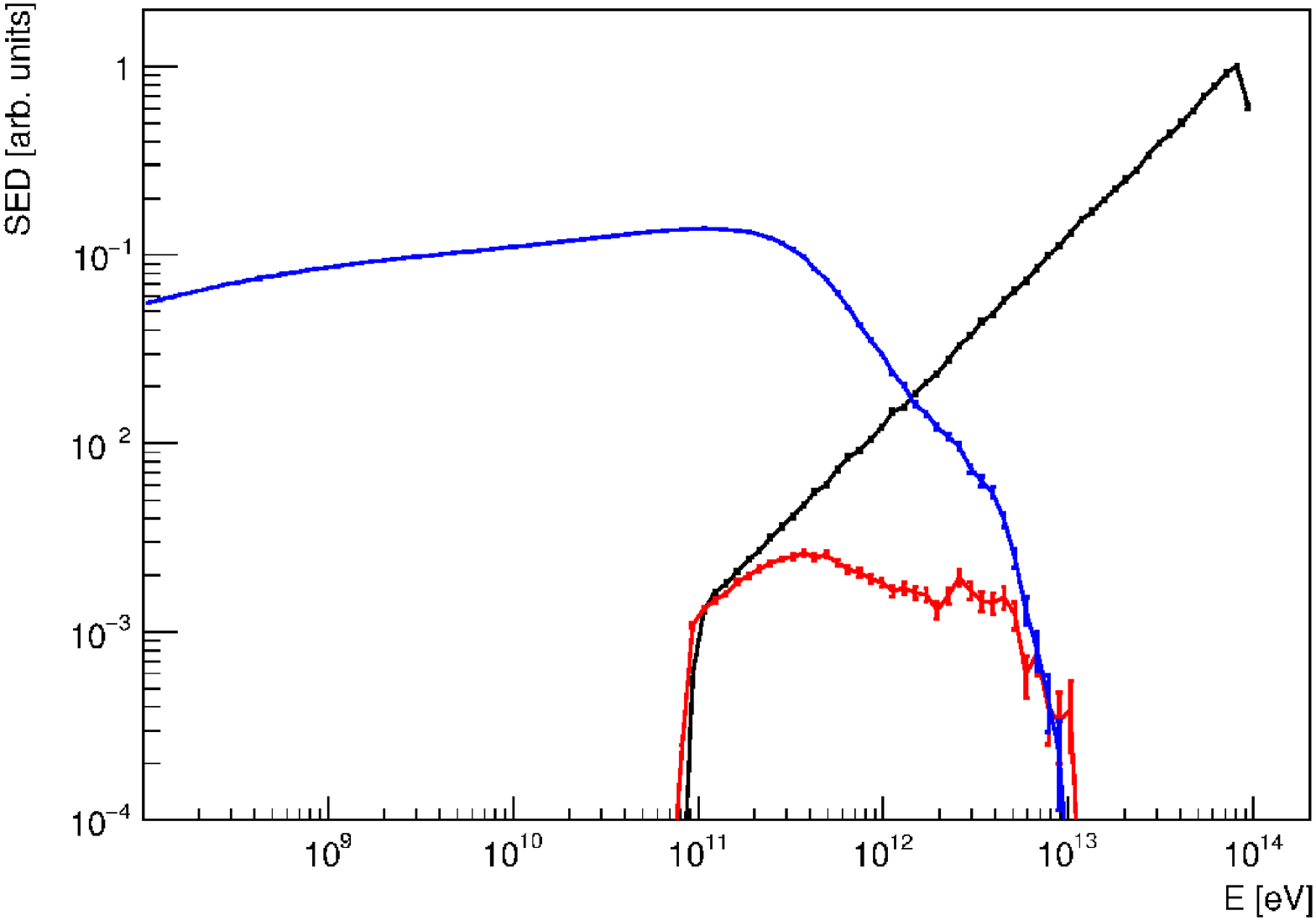}\includegraphics[width=0.50\textwidth]{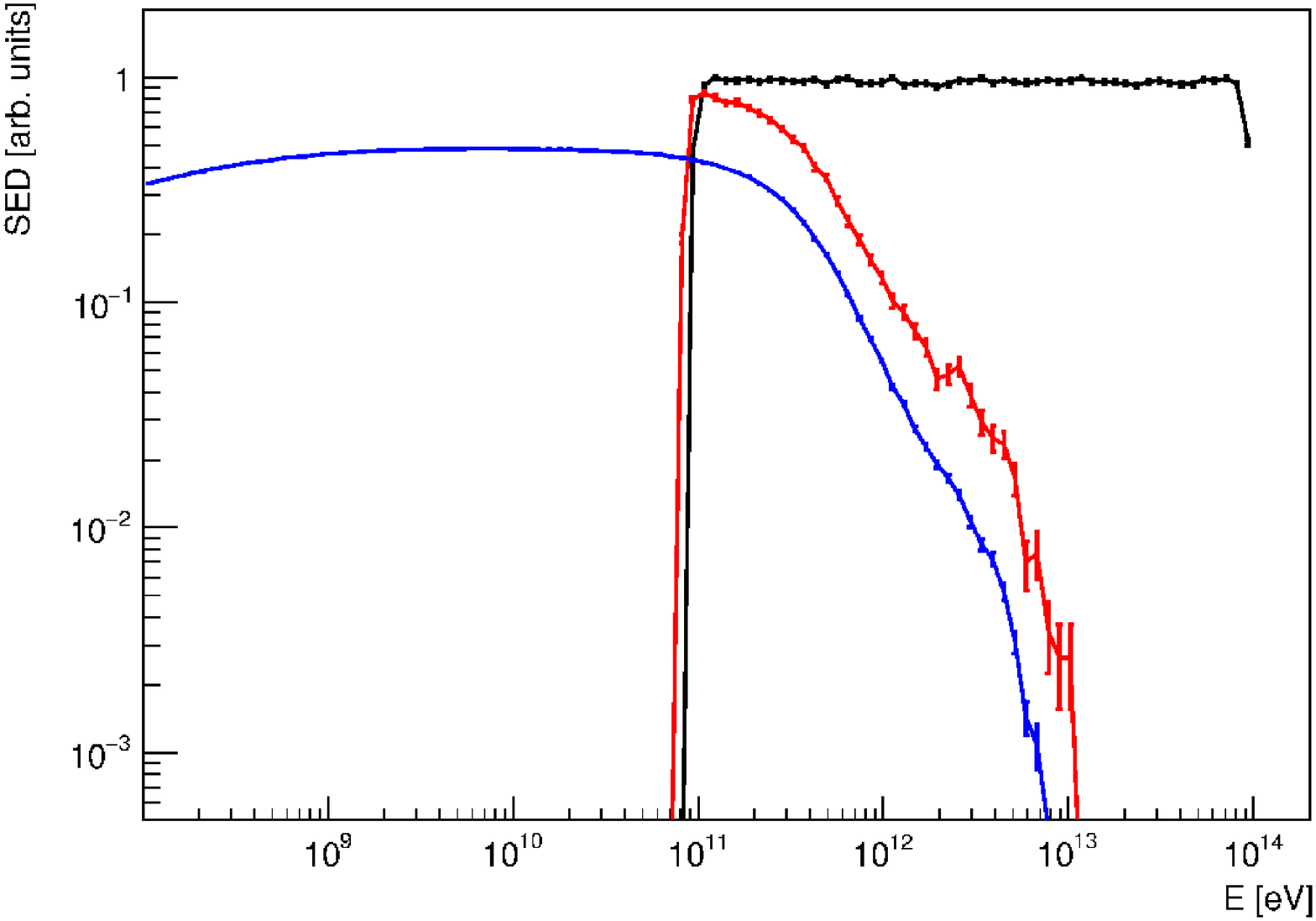}}
\centerline{\includegraphics[width=0.50\textwidth]{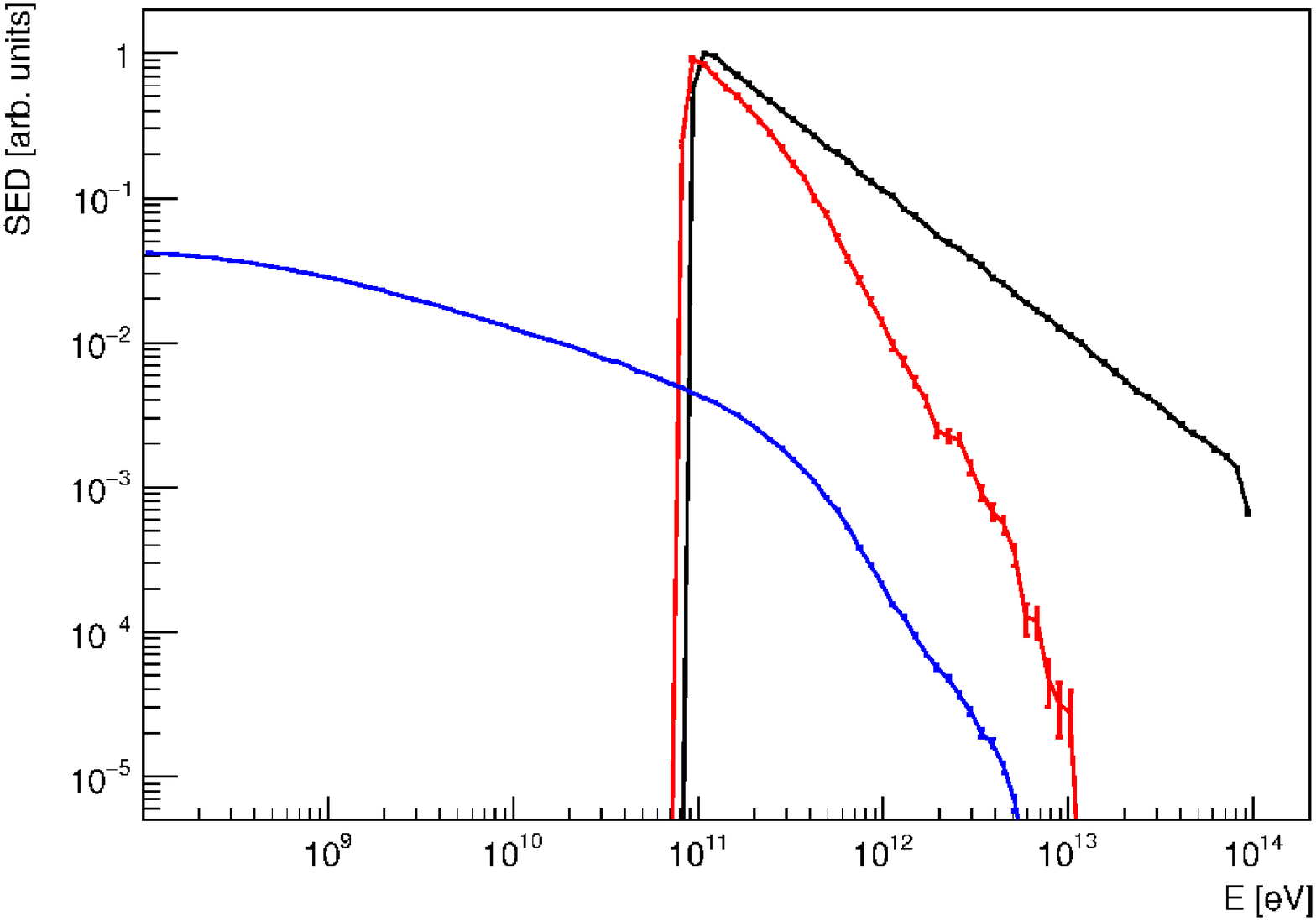}\includegraphics[width=0.50\textwidth]{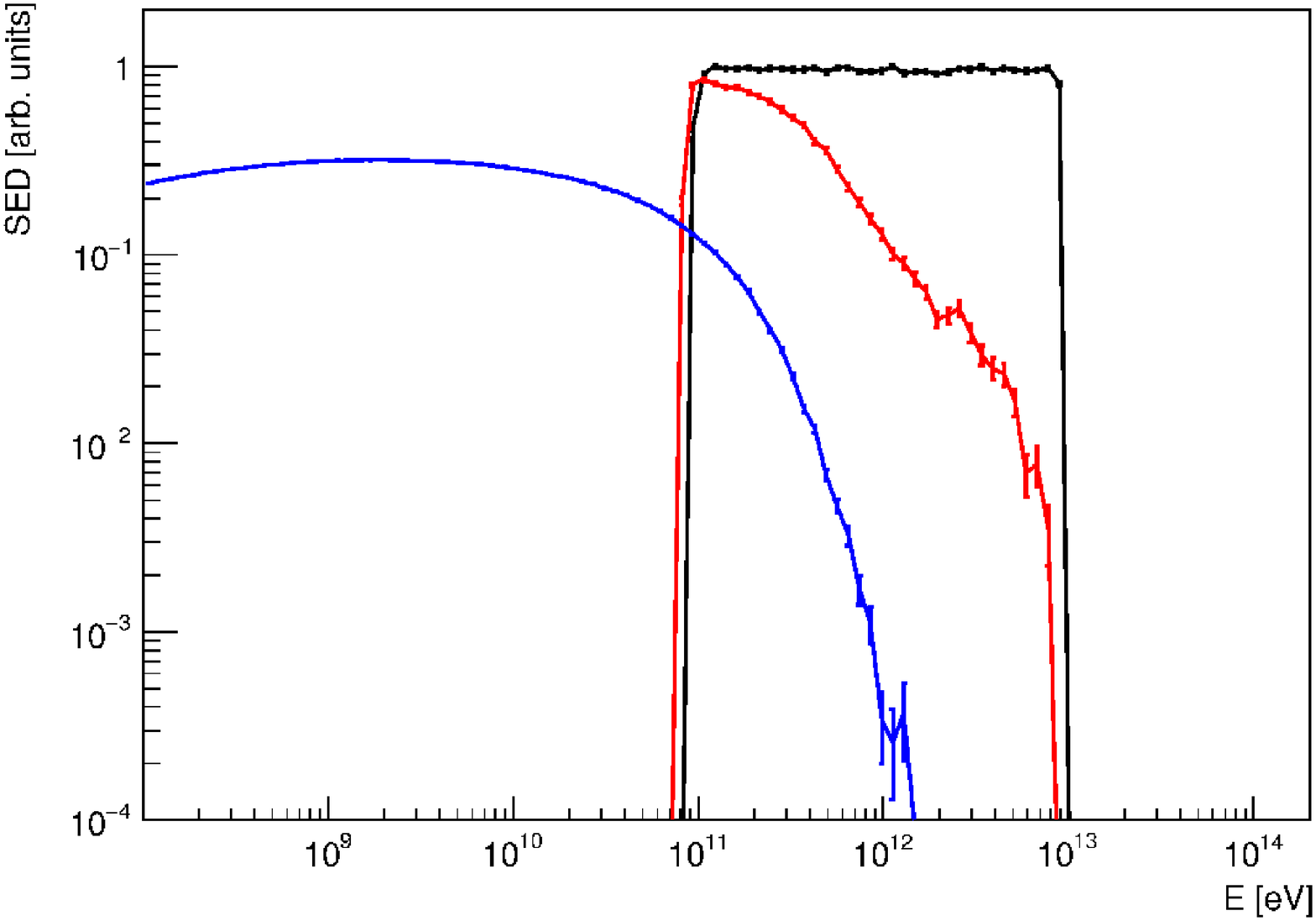}}
\centerline{\includegraphics[width=0.50\textwidth]{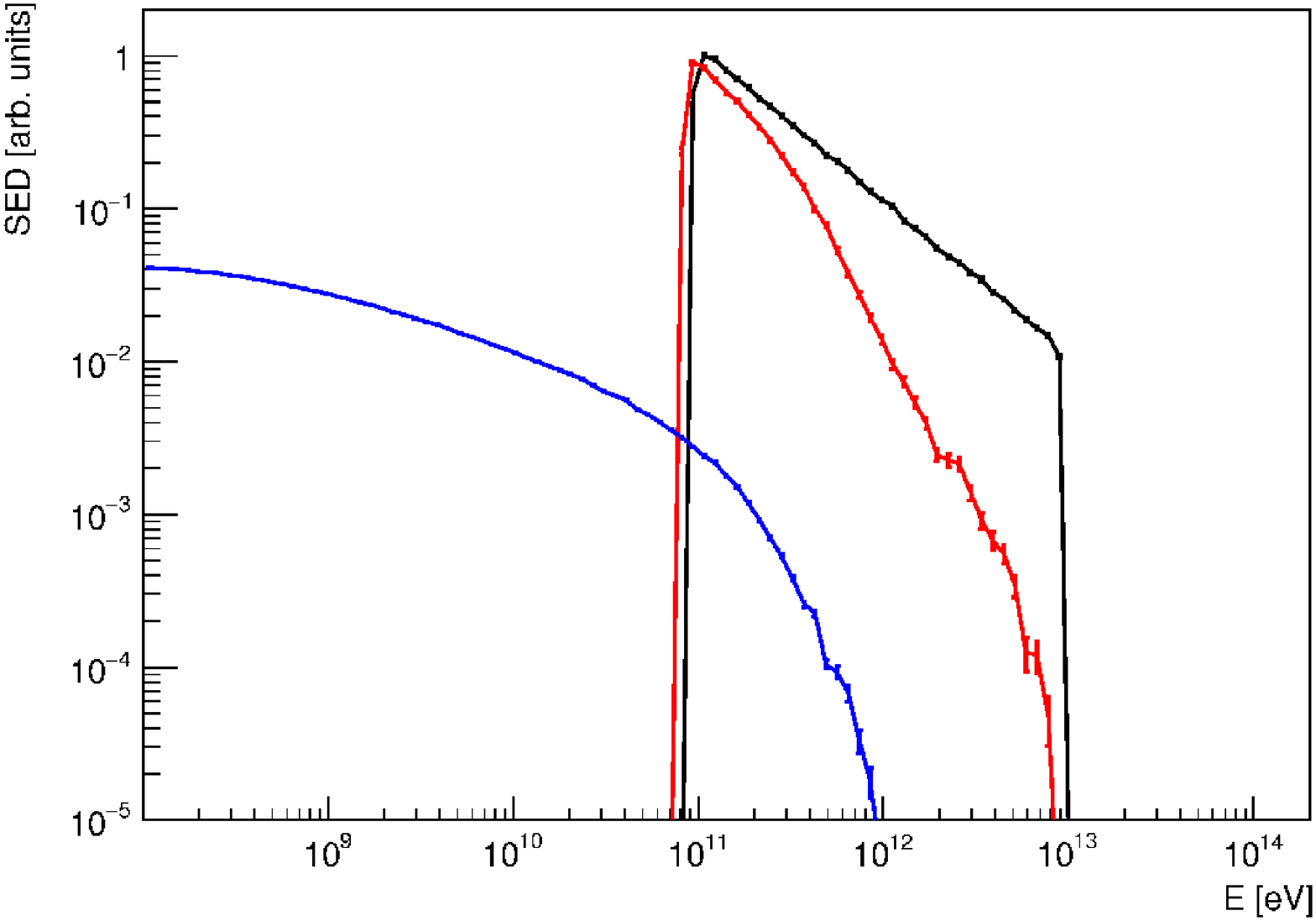}\includegraphics[width=0.50\textwidth]{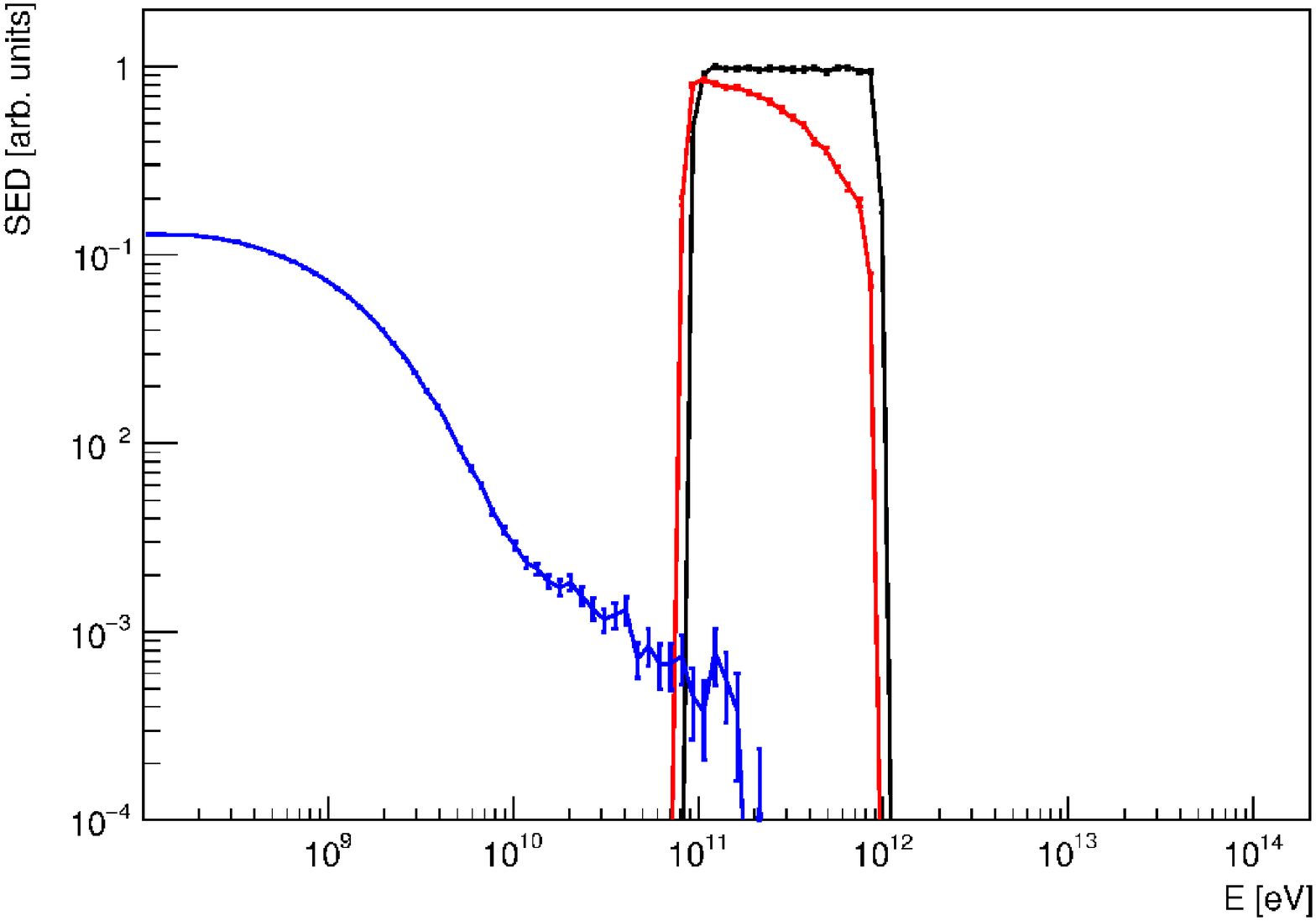}}
\caption{Primary and secondary components (observable spectrum) of $\gamma$-rays from primary $\gamma$-rays with power-law shape and different parameters. Black line --- primary spectrum in the source, red lines --- absorbed and redshifted primary component, blue lines --- cascade component. Bars denote statistical fluctuations.}\label{figB1}
\end{figure}

In Fig.~\ref{figB1} we present several other examples of calculations for the case of a purely electromagnetic cascade and the following primary spectrum:
\begin{equation}
dN/dE_{0} \propto E_{0}^{-\gamma}\cdot\theta(E_{0}-E_{0max}), 
\end{equation}
where $\theta(E_{0}-E_{0max})$= 1 if $E_{0}<E_{0max}$ and 0 otherwise. Fig.~\ref{figB1} top-left graph is drawn for the case of $\gamma$= 1 and $E_{0max}$= 100 $TeV$; Fig.~\ref{figB1} top-right --- for $\gamma$= 2 and $E_{0max}$= 100 $TeV$, Fig.~\ref{figB1} middle-left --- $\gamma$= 3 and $E_{0max}$= 100 $TeV$, Fig.~\ref{figB1} middle-right --- $\gamma$= 2 and $E_{0max}$= 10 $TeV$, Fig.~\ref{figB1} bottom-left --- for $\gamma$= 3 and $E_{0max}$= 10 $TeV$, Fig.~\ref{figB1} bottom-right --- for $\gamma$= 2 and $E_{0max}$= 1 $TeV$.

\onecolumn
\twocolumn

\section{Optical depth $\tau$ in the ELMAG code}

\begin{figure}[t]
\centerline{\includegraphics[width=0.50\textwidth]{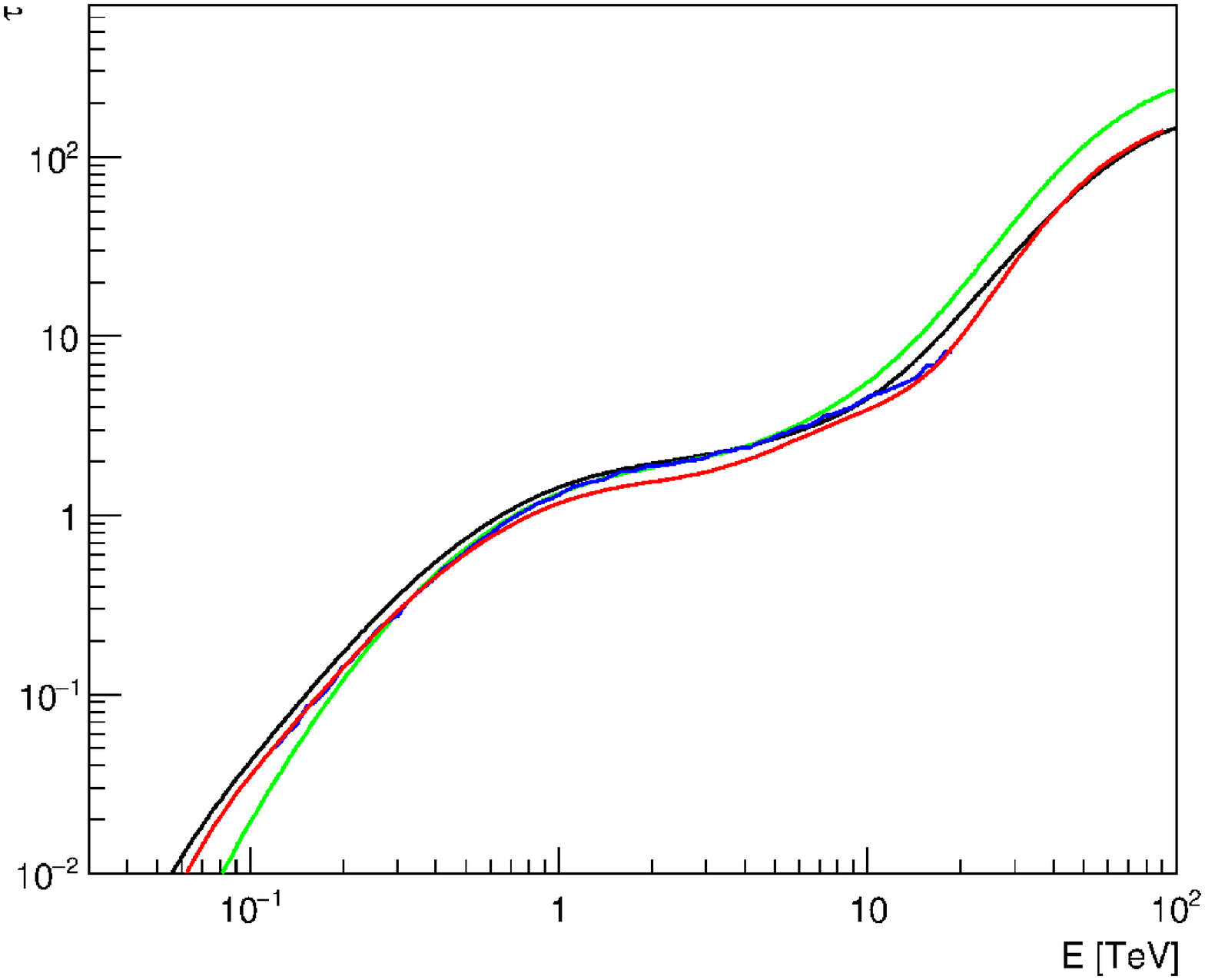}\includegraphics[width=0.50\textwidth]{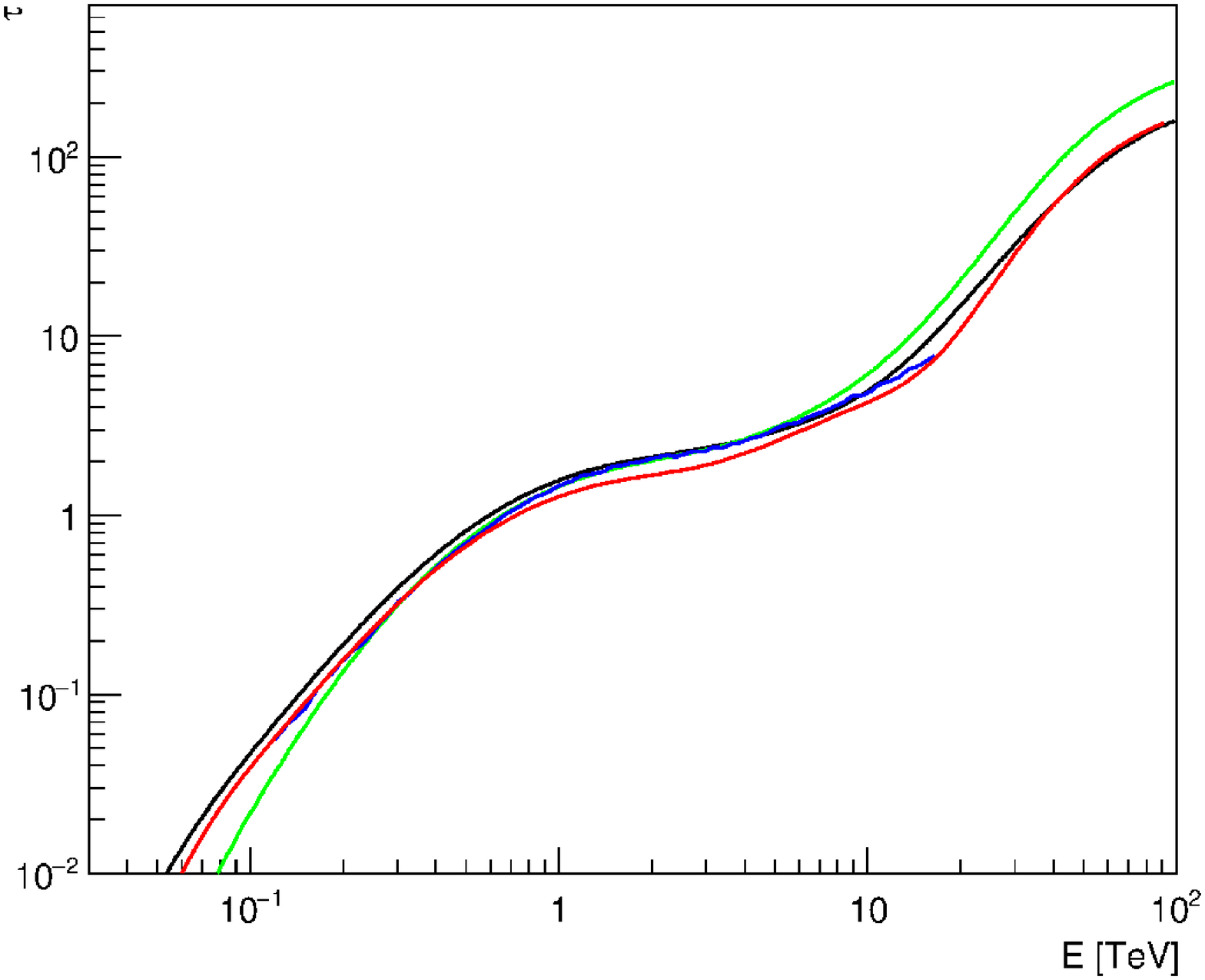}}
\centerline{\includegraphics[width=0.50\textwidth]{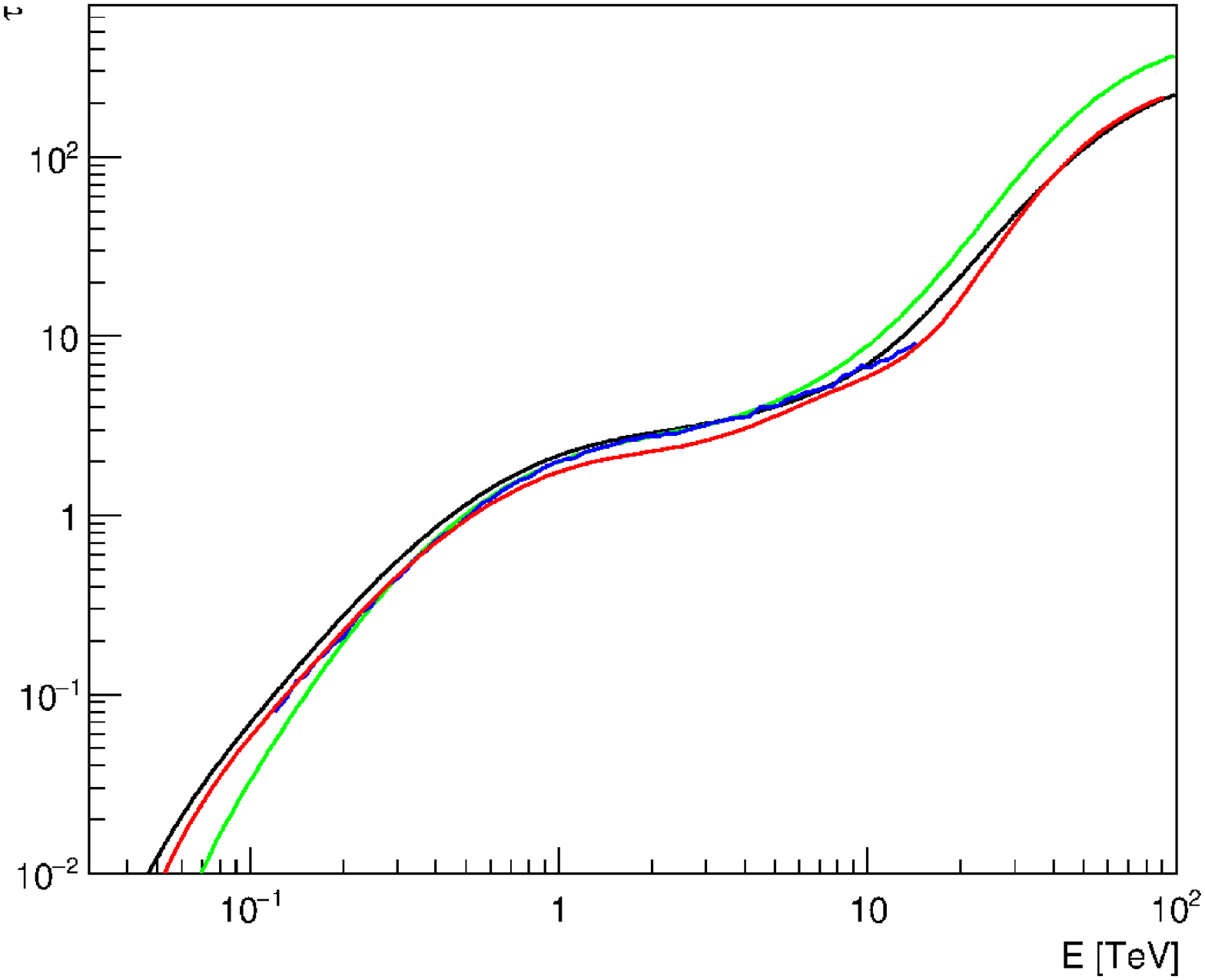}\includegraphics[width=0.50\textwidth]{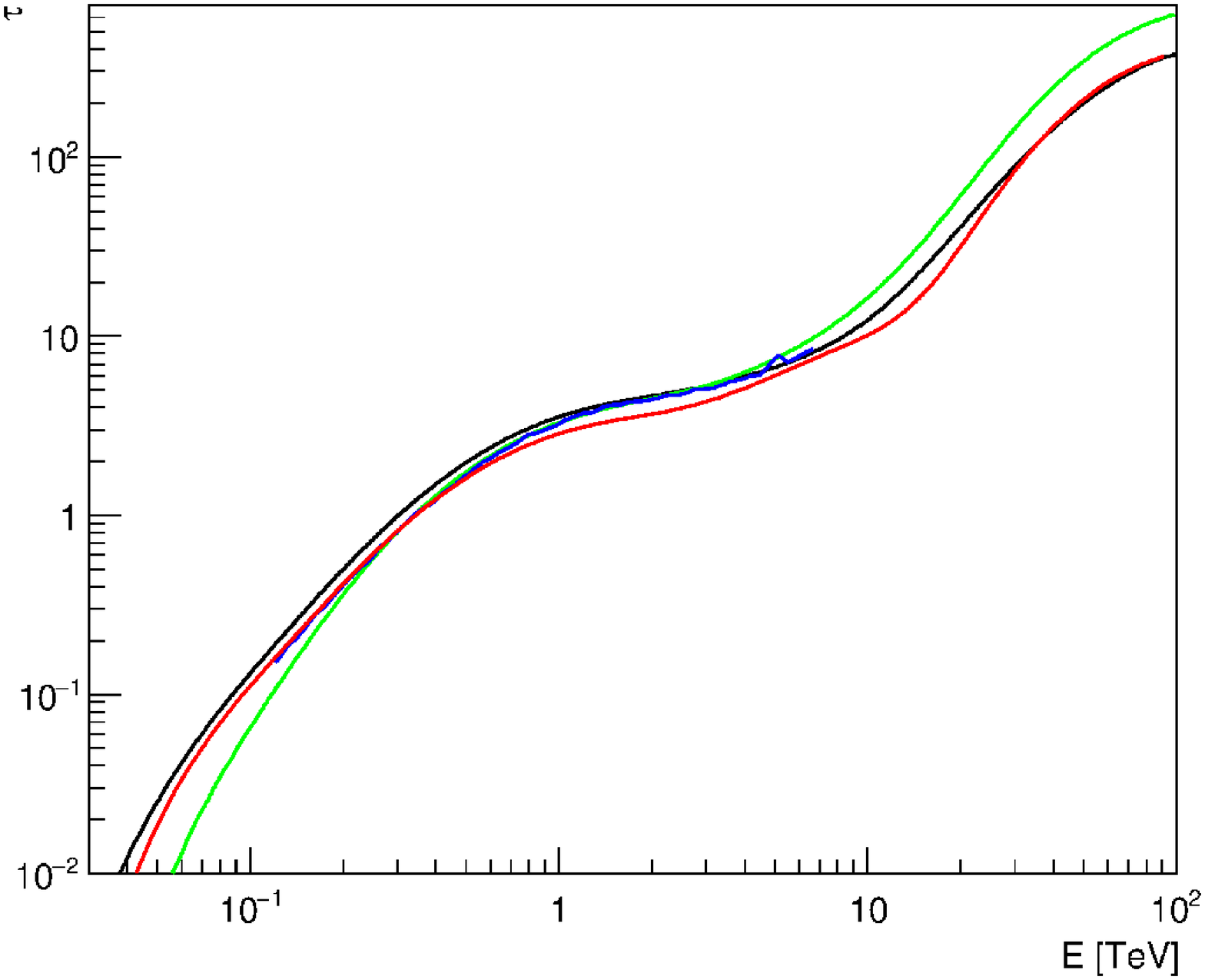}}
\centerline{\includegraphics[width=0.50\textwidth]{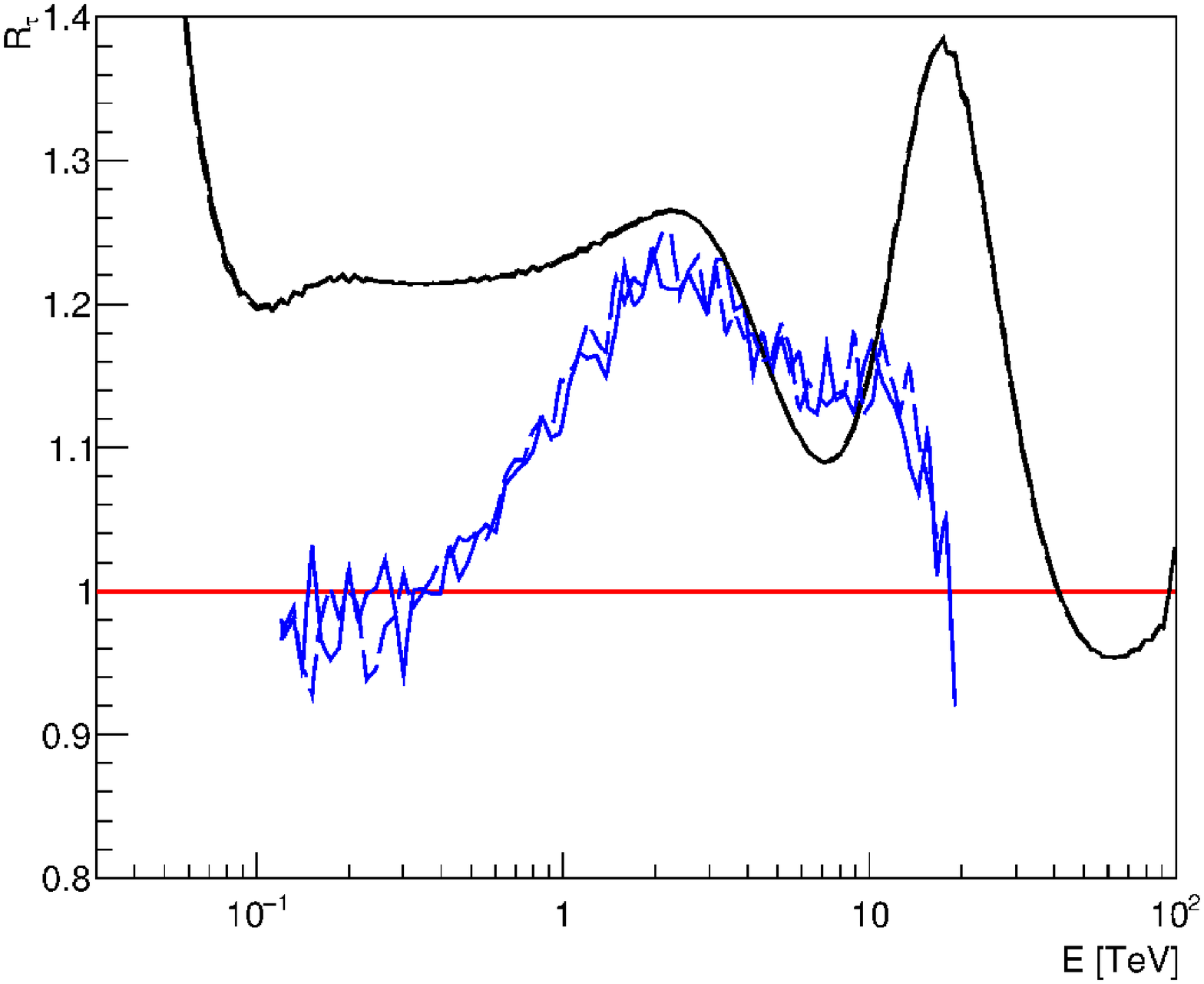}\includegraphics[width=0.50\textwidth]{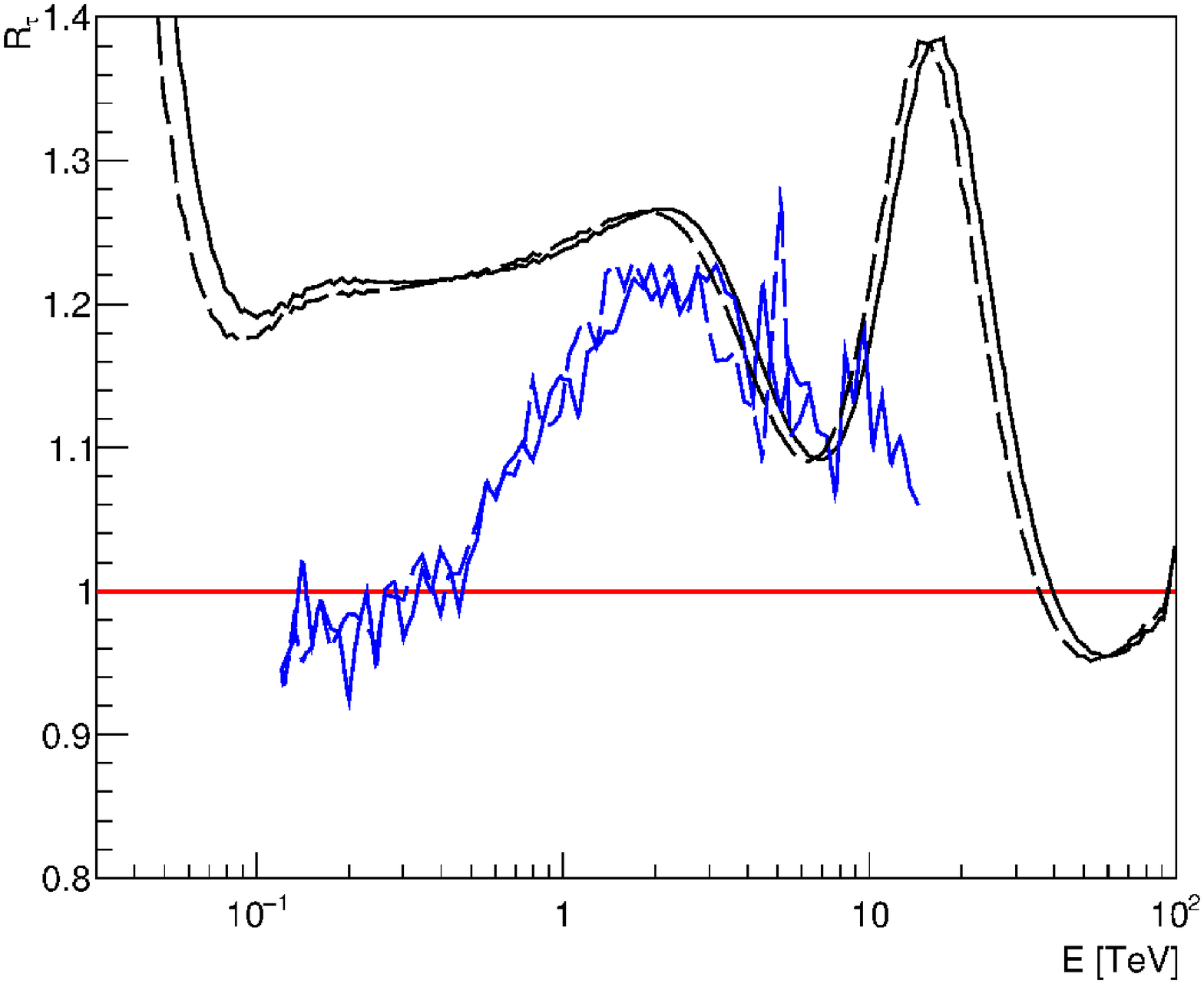}}
\caption{Comparison of $\tau$ vs. energy for G12 (black), F08 (green), KD10 (red), ELMAG KD10 (blue) models for $z= 0.129$ (top left figure), $z= 0.14$ (top right figure), $z= 0.186$ (middle left figure), $z= 0.287$ (middle right figure) and ratios of G12 and ELMAG KD10 to KD10  (bottom left figure: solid lines for $z= 0.129$, dashed --- for $z= 0.14$, bottom right figure: solid lines for $z= 0.186$, dashed --- for $z= 0.287$).}\label{figC1}
\end{figure}

Fig.~\ref{figC1} shows the comparison between several models of optical depth $\tau$ with the version of the KD10 EBL model as implemented in the ELMAG code. Other EBL models include G12, KD10 (original implementation) and F08.

\onecolumn
\twocolumn

\section{Comparison of observable $\gamma$-ray spectra with other studies}

\begin{figure}[h]
\centerline{\includegraphics[width=0.50\textwidth]{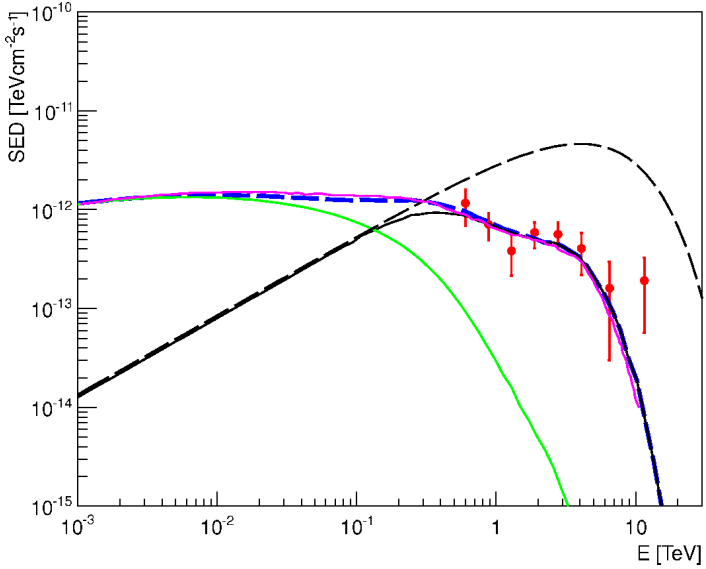}}
\caption{The comparison our calculations with the ELMAG code and the result of Vovk et al. (2012) for $z= 0.14$. Magenta line denotes the total observable spectrum calculated by Vovk et al. (2012), other lines --- our results: dashed black line --- primary spectrum in the source, solid black line --- absorbed primary component, solid green line --- cascade component, dashed thick blue line --- total model spectrum.}\label{figD1}
\end{figure}

\begin{figure}[h]
\centerline{\includegraphics[width=0.50\textwidth]{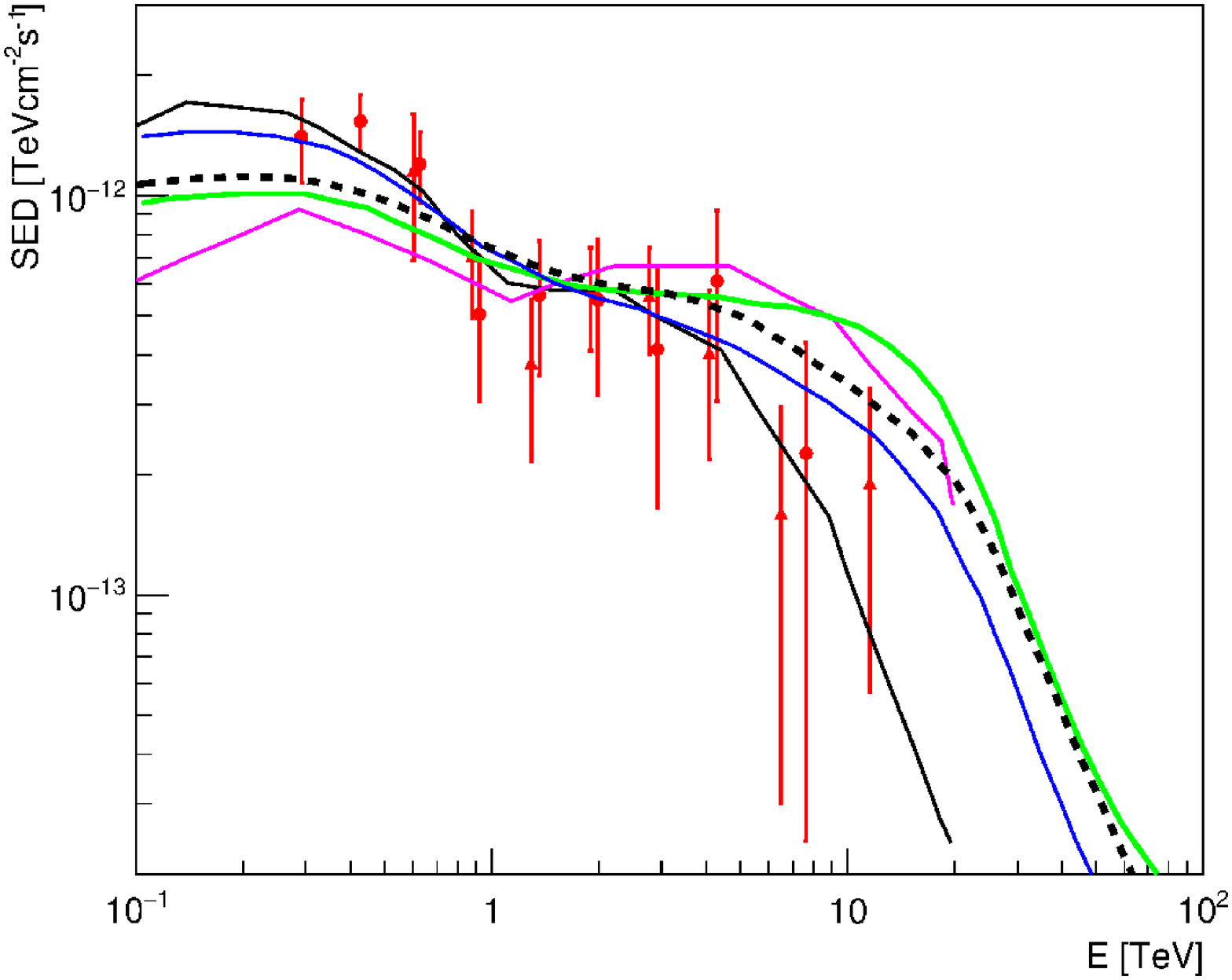}}
\caption{The comparison between hadronic model from our work black- dashed curve (power-law spectrum from 1 EeV to 100 eev, gamma= -2) with other works (solid curves). Black: Essey et al. (2011), high EBL intensity; magenta: Essey et al. (2011), low EBL intensity; green: Murase et al. (2012), KD10 EBL; blue: Murase et al. (2012), Kneiske et al. (2004) best fit EBL.}\label{figD2}
\end{figure}

Here we present the comparison of our test calculations with other works. Fig.~\ref{figD1} shows good agreement between our calculations with the ELMAG code (assuming the KD10 EBL model, as implemented in ELMAG) and the result of Vovk et al. (2012), obtained with the F08 EBL model. Fig.~\ref{figD2} presents the comparison of our result for the case of the basic hadronic cascade model with the ones obtained by Essey et al. (2011) and Murase et al. (2012).

\end{appendix}
\end{document}